\documentclass[fleqn,usenatbib]{mnras}
\usepackage{newtxtext,newtxmath}
\usepackage[T1]{fontenc}
\usepackage{ae,aecompl}
\usepackage{bigints}

\usepackage{graphicx}	
\usepackage{amsmath}	
\usepackage{bigints}
\usepackage{natbib}
\usepackage{caption}
\usepackage{ragged2e}
\usepackage{placeins}
\usepackage{bigints} 
\usepackage{physics}
\usepackage{cleveref}
\usepackage[export]{adjustbox}
\usepackage{longtable}
\usepackage{xtab}
\usepackage{bm}
\usepackage[mathscr]{eucal}
\usepackage{comment}
\usepackage{multirow}
\usepackage{tabularx}
\usepackage[table,xcdraw]{xcolor}
\usepackage{setspace}
\usepackage{longtable,rotating}
\usepackage{threeparttable}
\usepackage{booktabs}
\usepackage{tcolorbox}
\usepackage{mathtools,leftindex,tensor}
\usepackage[version=4]{mhchem}
\usepackage{makecell}
\usepackage{listings}
\DeclareRobustCommand{\ttfamily}{\fontencoding{T1}\fontfamily{lmtt}\selectfont}
\usepackage[normalem]{ulem}
\tcbuselibrary{breakable,xparse,skins}

\usepackage{hyperref}
\hypersetup{
    colorlinks=true,
    linkcolor=blue,
    filecolor=magenta,      
    urlcolor=cyan,
    pdftitle={Misty, patchy, and turbulent CGM with mCC},
    pdfpagemode=FullScreen,
    }

\usepackage{tikz} 
\usetikzlibrary{shapes.geometric,arrows.meta} 

\usepackage{scalerel,tikz}
\usetikzlibrary{svg.path}
\definecolor{orcidlogocol}{HTML}{A6CE39}
\tikzset{orcidlogo/.pic={
 \fill[orcidlogocol] svg{M256,128c0,70.7-57.3,128-128,128C57.3,256,0,198.7,0,128C0,57.3,57.3,0,128,0C198.7,0,256,57.3,256,128z};
 \fill[white] svg{M86.3,186.2H70.9V79.1h15.4v48.4V186.2z}
 svg{M108.9,79.1h41.6c39.6,0,57,28.3,57,53.6c0,27.5-21.5,53.6-56.8,53.6h-41.8V79.1z M124.3,172.4h24.5c34.9,0,42.9-26.5,42.9-39.7c0-21.5-13.7-39.7-43.7-39.7h-23.7V172.4z}
 svg{M88.7,56.8c0,5.5-4.5,10.1-10.1,10.1c-5.6,0-10.1-4.6-10.1-10.1c0-5.6,4.5-10.1,10.1-10.1C84.2,46.7,88.7,51.3,88.7,56.8z};
}}
\newcommand\orcidicon[1]{\href{https://orcid.org/#1}{\mbox{\scalerel*{
\begin{tikzpicture}[yscale=-1,transform shape]
\pic{orcidlogo};
\end{tikzpicture}
}{|}}}}

\defcitealias{Dutta2024}{D24}
\defcitealias{Hummels2024}{HR24}

\title[Misty, patchy, and turbulent CGM with mCC]{Misty, patchy, and turbulent: constraining the cool circumgalactic medium with mCC}

\author[Bisht et al.]{
\newauthor
Mukesh Singh Bisht$^{1\,\orcidicon{0000-0002-1497-4645}}$\thanks{E-mail: msbisht@rrimail.rri.res.in (MSB)},
Prateek Sharma$^{2\,\orcidicon{0000-0003-2635-4643}}$\thanks{E-mail: prateek@iisc.ac.in (PS)},
Alankar Dutta$^{2,3\,\orcidicon{0000-0002-9287-4033}}$, 
Biman B. Nath$^{1\,\orcidicon{0000-0003-1922-9406}}$ 
\\
$^{1}$Raman Research Institute, Bengaluru 560080, India\\
$^{2}$Department of Physics, Indian Institute of Science, Bengaluru 560012, India\\
$^{3}$Max-Planck-Institut f\"ur Astrophysik, Karl-Schwarzschild-Straße 1, 85748 Garching b. M\"unchen, Germany
}

\date{Accepted XXX. Received YYY; in original form ZZZ}

\pubyear{2024}

\begin{document}
\label{firstpage}
\pagerange{\pageref{firstpage}--\pageref{lastpage}}
\maketitle

\begin{abstract}

The circumgalactic medium (CGM) is the largest baryon reservoir around galaxies, but its extent, mass, and temperature distribution remain uncertain. We propose that cool gas ($\sim 10^4$ K) in the CGM resides in clumpy structures referred to as cloud complexes (CCs) rather than uniformly filling the entire CGM volume. Each CC contains a mist of tiny cool cloudlets dispersed in a warm/hot medium ($\sim 10^5$ \hbox{--} $10^6$~K). Modeling CCs in the mist limit (unit area covering fraction within a CC) simplifies the calculation of observables like ion absorption columns, equivalent widths, compared to modeling individual cloudlets from first principles. Through Monte Carlo realizations of CCs, we explore how CC properties affect the observed variation in observables. We find that a power-law distribution of CCs ($dN_{\rm CC}/dR \propto R^{-1}$) with a total of $\sim 10^3$ CCs each with a radius of $\sim 10$ kpc and total cool gas mass of $\sim 10^{10} M_\odot$ reproduces MgII column density and equivalent width distribution trends with impact parameter for the COS-Halos sample (Werk+ 2013). We further show that the area-averaged MgII column density, combined with the area covering fraction, provides a robust proxy for estimating the cool CGM mass, independent of other model parameters. Modeling a larger number of (smaller size) cloudlets within a CC shows that line blending from individual cloudlets results in turbulent broadening on the CC scale. This work presents a practical framework for linking CGM models with observations of a multiphase CGM, illuminating the distribution of cool gas in galaxy halos. 

\end{abstract}




\section{Introduction}
\label{sec:introduction}

The circumgalactic medium (CGM) is the vast reservoir of gas surrounding the stellar disk and the interstellar medium (ISM) of galaxies \citep{Tumlinson2017, Faucher2023}. Various observations and numerical simulations suggest CGM to be multiphase with gas temperature varying by $\sim 2\hbox{-}3$ orders of magnitude ($\sim 10^4$-$10^7$ K for Milky Way-like galaxies). Unlike the intracluster medium (ICM) of massive clusters of galaxies, which has been mapped extensively in X-ray emission (a recent review is \citealt{Donahue2022}), the hot CGM in Milky Way-like galaxies (like X-ray emitting ICM, also likely to be the mass/volume dominant phase) is too dilute and compact to be mapped in emission. The cool/warm gas ($\sim 10^{4-5}$ K) in the CGM of foreground external galaxies is detected in absorption lines of low/intermediate ionization metal ions like MgII, CaII, SiII, CII, CIV, OVI in the spectra of bright background sources (typically quasars; \citealt{Srianand1993, Charlton2003, Chen2010, Bordoloi2011}). Cool gas has recently been observed even in emission (MgII emission; \citealt{Burchett2021, Zabl2021, Guo2023, Pessa2024}), but emission is most sensitive to the densest cool gas.

A fundamental but unconstrained physical property of the CGM (unless qualified, we refer to the CGM of Milky Way-like galaxies) is its mass fraction in the cool phase. Although the cool ion (e.g., Mg II) column density (inferred from absorption lines) directly counts the number of ions along the sightline, the cool CGM mass is highly uncertain because of the limited number of quasars behind most intervening galaxies. Additionally, a significant variation in the inferred column density at similar impact parameters introduces uncertainty in estimating the cool gas mass.
A key result of this paper is that the combination of the average column density and the area covering fraction (both observationally available quantities) provides a robust estimate of the cool gas mass in the CGM ($\sim 10^{10} M_\odot$ for Milky Way-like galaxies). Metal ion absorption is also affected by the elemental abundances, the photoionizing background, non-equilibrium effects, and dust depletion. Since the dispersion measure provides column density of free electrons, which is not affected by these complications, fast radio bursts (FRBs) are promising probes of the total CGM mass (\citealt{Prochaska2019, Cook2023}). However, we cannot distinguish the various thermal phases of the CGM from FRB dispersion measures.

The cool gas in the CGM likely has various formation channels (\citealt{Decataldo2024}), including, e.g, cooling from hot gas in the CGM due to thermal instabilities (\citealt{McCourt2012, Voit2017}), cosmological accretion onto the galaxy (\citealt{Rahmani2018, Afruni2019}), stellar and AGN outflows (\citealt{Augustin2021, Burchett2021}), and stripping from satellite galaxies (\citealt{Rubin2012, Roy2024}).
Cool gas clouds are also disrupted by shear instabilities, potentially leading to their evaporation in the presence of a surrounding hot medium (\citealt{Lan2019}). Several theoretical and semi-analytical studies have attempted to investigate the structure and distribution of cool gas within the CGM (\citealt{Stern2016, Hummels2024, Faerman2023, Yang2025}).
The models presented in this paper are agnostic to the specific cool gas formation and disruption mechanisms but provide a tool for building phenomenological models of the mass and radial distribution of the cool gas in the CGM.
  
The cool gas is observed to have an area covering fraction of order unity (\citealt{Dutta2020, Huang2021}), while the volume fraction is expected to be $\lesssim 1 \%$ (\citealt{Faerman2023, Dutta2024}). This suggests that the cool gas is not uniformly distributed in the CGM, unlike the hot gas, but rather distributed in clumpy structures dispersed in the hot CGM.  

Recently, \citealt{Dutta2024} (hereafter  \citetalias{Dutta2024}) modeled the cool gas with a three-phase, one-zone CGM model. This model fills all the cool CGM gas uniformly as a mist of numerous cloudlets, with a small volume filling fraction but a unity area covering fraction, in the entire CGM. We refer to this as the {\it mCGM} (misty CGM) model. They modeled the cool gas in one of the {\sc TNG50-1} halos and predicted the baseline column density of MgII to be $\sim 10^{13} \, \rm cm^{-2}$ (the solid cyan line in their Fig. 11). This is smaller than the typical observed column density, despite the large scatter in observed values and several stringent upper limits, suggesting that the cool gas does not block all sightlines uniformly. In the {\it mCGM}  model, the cool gas is distributed throughout the CGM volume, resulting in a low number density and therefore a low column density.  
To explain the empty sightlines and the observed scatter in the column density, the cool gas ought to be distributed in a smaller volume than the entire CGM. This will result in a higher particle number density and a larger column density for sightlines passing through cool clouds and non-detection along the empty sightlines. One can distribute the cool gas in several cloud complexes (CCs) rather than uniformly throughout the entire CGM and better explain the observations. In this paper, we extend the work of \citetalias{Dutta2024}, moving beyond a completely misty CGM to a more realistic CGM filled with misty cloud complexes. We refer to our model based on CCs as {\it mCC} model, which stands for misty cloud complex.

Another closely related work to ours is the {\it CloudFlex} model (\citealt{Hummels2024}; hereafter \citetalias{Hummels2024}). They generated numerous ($\gtrsim 10^6$) cloudlets within a CC, modeled absorption profiles, and based on their analytic halo scale model for cool gas mass distribution, predicted the equivalent width (EW) distribution of the MgII line.
The CCs in the CGM are analogous to terrestrial clouds that are made up of an astronomical number of tiny water droplets. 
The CGM observations of individual absorption components at high-velocity resolution suggest the cloudlets to be $\lesssim 1$ pc  (e.g., Fig. 17 of \citealt{Sameer2024} and Fig. 3 of \citealt{Chen2023}). It is impossible to computationally model such an enormous number of cloudlets within a CC from first principles\footnote{If 1 pc is the typical cloudlet size, we would need $\sim 10^{13}$ cloudlets to make up $10^{10} M_\odot$ in the cool CGM.}. Instead, in this work (section \ref{sec:analytical}) we present a computationally tractable approach complementary to \citetalias{Hummels2024}, treating the cloudlets within a CC analytically in the mist limit and generating Monte Carlo realizations of a manageable number ($10^{3-4}$) of CCs throughout the CGM.

We put our {\it mCC} model in the context of {\it mCGM} (\citetalias{Dutta2024}) and {\it CloudFlex} (\citetalias{Hummels2024}) models. Following \citetalias{Hummels2024}, we create Monte Carlo realizations of cloudlets within a single CC in section \ref{sec:turb}, with an increasing number of smaller cloudlets to show that the misty CC approximation provides a satisfactory description of the otherwise computationally intractable realization of an enormous number of cloudlets within a CC. Later in section \ref{sec:realistic}, we generate Monte Carlo realizations of not only individual cloudlets within a single CC but also the distribution of CCs across the CGM. In this computationally prohibitive model, we cannot reach the mist limit, but we compare the distribution of EWs and obtain the best model parameters. To conclude, our {\it mCC} model not only acts as a tractable bridge between the simple misty CGM model of \citetalias{Dutta2024} and the realistic but computationally expensive {\it CloudFlex} model (\citetalias{Hummels2024}), but also provides an intuitive understanding of the complex distribution of cool gas in the CGM.

The MgII doublet metal transition $\lambda\lambda \, 2796,2803$ is well observed and studied in the absorption of quasar light from the intervening CGM of external galaxies. In this work, we focus on MgII $\lambda 2796$, but our model can also be used to analyze other cool ions like SiII, CaII, and CII, and with simple extension, also intermediate ions such as CIV and OVI (c.f. section \ref{sec:intermediate}). Starting with MgII observational data in section \ref{sec:observation}, we present our misty cloud complex ({\em mCC}) model in section \ref{sec:analytical}. In section \ref{sec:turb}, we zoom in on a single CC to verify the mist limit, particularly in velocity space. In section \ref{sec:realistic} we consider the distribution of cloudlets within CCs (rather than assuming the mist limit), distributed non-uniformly in the CGM. We end with a discussion and summary in sections \ref{sec:discussion} and \ref{sec:summary}, respectively.

\begin{figure}
    \centering
    \includegraphics[width=\columnwidth]{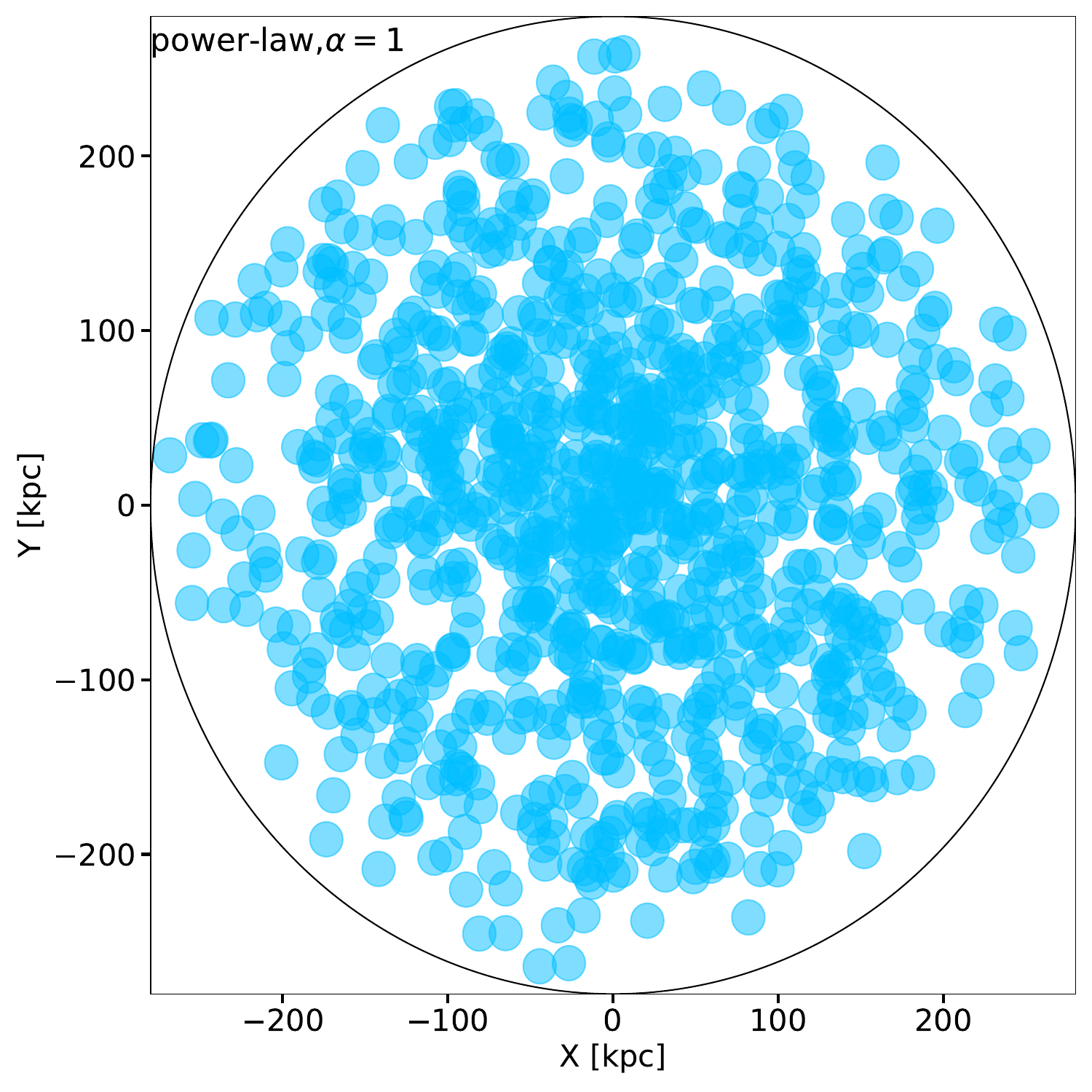}
    \caption{The line of sight (LOS) projected distribution of $10^3$ CCs each of radius $10$ kpc in the CGM of radius $280$ kpc with a power-law CC distribution of index $\alpha = 1$ (Eq. \ref{eq:pl_dist}). Notice that there are numerous empty regions towards the outer regions of CGM and comparatively fewer empty regions in the center. The LOSs passing through CGM outskirts will, therefore, not produce strong ion absorption in contrast with the central sightlines intersecting multiple CCs. }
    \label{fig:cc_cartoon}
\end{figure}

\section{Observational Data}
\label{sec:observation}

Several surveys have been conducted to study cool gas in the CGM in the last $\sim 2\hbox{-}3$ decades over several redshift and stellar mass ranges, eg. MAG${\textsc{ii}}$CAT (\citealt{Nielsen2013}), MAGG (\citealt{Dutta2020}), SDSS DR16 (\citealt{Anand2021}), CUBS (\citealt{Qu2023}), and MEGAFLOW (\citealt{Cherrey2025}), etc. For uniformity in redshift and stellar masses, in our work, we compare our model results with the COS-Halos (\citealt{Werk2012}) survey and the Magellan MagE Mg II (M3) Halo project (\citealt{Huang2021}).
The COS-Halos survey comprises 44 $\sim L^*$ galaxies with stellar mass in the range $10^{9.5}\hbox{-}10^{11.5} \, M_\odot$ with a median of $10^{10.5} \, M_\odot$ at the impact parameter of $15\hbox{-}160$ kpc. The redshift range of these galaxies is $0.1\hbox{-}0.4$ with a median redshift of $0.22$. We obtained the MgII column density and equivalent width data from \citealt{Werk2013} and the corresponding normalized impact parameter values from \citealt{Werk2014}.
The M3 Halo project comprises $211$ isolated galaxies with an impact parameter of $9-497$ kpc, with a median of $94$ kpc. The redshift range is $0.1-0.48$ with a median of $0.21$. These galaxies span a wide range in stellar mass of $2\times 10^8 - 4\times 10^{11} \, M_{\odot}$ with a median stellar mass of $4\times 10^{10} \, M_{\odot}$. They obtained the best-fit log-log relation between the MgII equivalent width $W_{2796}$ and the impact parameter ($R_\perp$) for the isolated galaxy sample as,
\begin{eqnarray}
\nonumber
    \mathrm{log} \, W_{2796}(\mathring{\rm A}) & =  (1.35\pm0.25) - (1.05\pm0.17) \mathrm{log} \, R_\perp  \\
    & + (0.21\pm 0.08)\times (\rm {log} \, M^*_\odot - 10.3).
\end{eqnarray}
They have also estimated the covering fraction at various impact parameter ranges with various equivalent width thresholds, which we will compare with our model prediction later.

Note that the two surveys have similar median stellar mass and redshift. Therefore, we use the observational data from both surveys to compare the MgII column density, EW, and covering fraction with our model predictions. Since the median redshift of these surveys is $\sim 0.2$, we compute the MgII ion fraction at the redshift of $0.2$ for our model. Our model can also be compared with observations from other surveys with different redshift ranges, provided that the ion fraction is computed at that redshift.

\section{Misty cloud complexes in the CGM}
\label{sec:analytical}

Motivated by observations and simulations (area covering fraction and volume fraction), we assume that the cool gas resides in cloudy structures rather than filling the entire CGM volume (\citetalias{Dutta2024}). We refer to these structures as `cloud complexes' (CCs) following \citetalias{Hummels2024}. These CCs are filled with tiny cool gas cloudlets. One of the caveats of the misty CGM  model of \citetalias{Dutta2024} is that it predicts cool gas along all LOSs, which is inconsistent with observations. If, instead, the cool gas is spread non-uniformly in these CCs, there will be sightlines not covered by cool gas. 
For illustration, in Figure \ref{fig:cc_cartoon}, we show the line of sight (LOS) projected distribution of CCs in the CGM where the number density of CCs as a function of distance from the center follows a power-law profile with a power-law index of $1$. There are $10^3$ CCs in total, each of radius $10$ kpc, distributed in the CGM of radius $280$ kpc. There are numerous empty regions without any CCs. The LOSs passing through these regions will result in no detection of cool gas (corresponding to the upper limits on the MgII column). These empty regions are everywhere in the CGM (fewer in the central region than in the outskirts). The LOSs passing through multiple CCs will result in a higher column density of cool ions. Therefore, an intrinsic scatter in the column density at a similar impact parameter is a natural outcome of the distribution of cool gas in CCs rather than uniformly in the entire CGM.

We consider the CCs to be misty (the mist limit is attained for numerous enough tiny cloudlets; see Appendix \ref{sec:mist_limit}), with the area covering fraction of cloudlets within a CC equal to unity ($f_A^{\rm cl} =  1$). The area covering fraction of CCs ($f_A^{\rm CC}$ different from $f_A^{\rm cl}$) is the fraction of the CGM area occupied by CCs. This cannot exceed unity if we do not separately count the areas of overlapping CCs; otherwise, it may exceed unity. 
Assuming the mist limit within each CC, we distribute the CCs in the CGM, compute the observables (column density, equivalent width distributions, covering fraction), and compare them with observations. 
For concreteness, we compare average quantities for uniform and a power-law distribution of CCs. 

If $M_{\rm cool}$ and $N_{\rm CC}$ are the total cool gas mass and the total number of CCs in the CGM, respectively, then the mass of each CC is $M_{\rm CC} = M_{\rm cool}/N_{\rm CC}$, assuming (for simplicity) that all the CCs have the same mass. The average number density of cool gas in each CC (distinct from the physical density of the cool gas) is then $\langle n_{\rm gas}\rangle = 3 M_{\rm CC}/(4 \ \pi R_{\rm CC}^3 \ \mu m_{\rm p})$, where $R_{\rm CC}$ is the radius of CC and $\mu m_p$ is the mean molecular weight which we assume to be $0.6 m_p$ ($m_p$ is proton mass) since hydrogen is mostly ionized even in cool CGM. We fix the mass and radius of CCs and refer to this model as the `basic' model. Later, in Sec. \ref{sec:advanced}, we consider the size and mass distribution of CCs in the CGM and refer to it as the `advanced' model.

\subsection{Analytical estimates}
Before examining the Monte Carlo realization of CCs distributed across the CGM, it is helpful to provide analytic estimates of the average number of CCs and the mean ion column density along a LOS to facilitate a better understanding of the model.

\begin{table}
    \centering
    {\renewcommand{\arraystretch}{1.2}
    \begin{tabular}{c|c}
    \hline
    Symbol & Meaning \\
    \hline
    $R_{\rm CGM}$ & Radius of CGM  \\
    $R_{\rm CC}$ & Radius of a cloud complex \\
    $R_\perp$ & Impact parameter \\
    $M_{\rm cool}$ & Total cool gas mass in the CGM \\
    $M_{\rm CC}$ & Mass of a cloud complex \\
    $N_{\rm CC}$ & Number of cloud complexes in the CGM \\
    $\alpha$ & Power-law index of CC radial distribution \\
    $N_{\rm cl}$ & Number of cloudlets in a CC \\
    $r_{\rm cl}$ & Radius of a cloudlet \\
    $f_A^{\rm CC}$ & Area covering fraction of CCs in CGM \\
    $f_A^{\rm cl}$ & Area covering fraction of cloudlets in a CC \\
    $\langle N_{\rm CC,LOS} \rangle$ & Average number of CCs intersected along a sightline\\
    $\langle N_{\rm ion} \rangle$ & Average column density of an ion along a sightline \\
    $f_{\rm V}$ & Volume filling fraction of cool gas in a CC\\
    $f_{\rm V}^{\rm CC}$ & Volume filling fraction of CCs in CGM\\
    $n_{\rm cool}$ & Physical number density of cool gas \\
    $\langle n_{\rm gas} \rangle$ & Average density of gas in a misty CC \\
    $\langle n_{\rm CC} \rangle$ & Average number density of CCs \\
    $\langle n_{\rm ion} \rangle$ & Average number density of an ion \\
    $f_{\rm ion}$ & Ion fraction \\
    $b_{\rm thermal}$ & Thermal broadening \\
    $b_{\rm turb,CC}$ & Turbulent broadening across a CC \\
    $b_{\rm tot}$ & Total broadening \\
    $W_{2796}$ & Equivalent Width of MgII \\
    $v_{\rm LOS}$ & LOS velocity \\
    \hline
    \end{tabular}}
    \caption{Symbols used in this paper and their description. Our {\it fiducial parameters} are $R_{\rm CGM}=280$ kpc, $R_{\rm CC} = 10$ kpc, $M_{\rm cool} = 10^{10} M_\odot$, $N_{\rm CC} = 10^3$, and $\alpha=1$.}
    \label{tab:tab1}
\end{table}

\subsubsection{Uniform distribution of cloud complexes}

We begin with a uniform distribution of CCs in the CGM. So, the average number density of CCs is $\langle n_{\rm CC} \rangle = 3 N_{\rm CC}/(4 \ \pi R_{\rm CGM}^3)$,\footnote{Note that we do not consider the inner $10$ kpc region of the galaxy. The average number density of CCs ($\langle n_{\rm CC} \rangle$) does not vary significantly if we consider only the CGM volume excluding the ISM, since $R_{\rm CGM}^3 \gg (10 {\rm~kpc})^3$.} where $R_{\rm CGM}$ is the CGM radius.
The average number of CCs intersected along a given LOS (see Table \ref{tab:tab1} for a list of commonly used symbols) at an impact parameter $R_\perp$ is 
\begin{equation}
    \langle N_{\rm CC,LOS} \rangle (R_\perp) = 2 \int_0^{s_{\rm max}}\langle n_{\rm CC}\rangle \pi R_{\rm CC}^2 ds,
\end{equation}
where $ds$ is the line element along the LOS and $s_{\rm max}=\sqrt{R_{\rm CGM}^2 - R_\perp^2}$. This trivial integral gives the average number of CCs encountered along a LOS, 
\begin{eqnarray}
    \langle N_{\rm CC,LOS} \rangle (R_\perp) = \frac{3N_{\rm CC}R_{\rm CC}^2}{2R_{\rm CGM}^3}\sqrt{R_{\rm CGM}^2 - R_\perp^2}.
    \label{eq:no_CC}
\end{eqnarray}
The area covering fraction of CCs in a CGM starts to approach unity when the average number of CCs along a LOS exceeds unity, i.e., $\langle N_{\rm CC,LOS} \rangle \gtrsim 1$. For fiducial parameters ($R_{\rm CC} = 10$ kpc, $R_{\rm CGM} = 280$ kpc), this happens when $N_{\rm CC} \gtrsim 500$. The average column density of an ion is
\begin{equation}
    \langle N_{\rm ion} \rangle (R_\perp) =  2\int_0^{ s_{\rm max}} (\langle n_{\rm CC}\rangle  \pi R_{\rm CC}^2 ds)(\langle n_{\rm ion} \rangle (R) \langle L \rangle), 
\end{equation}
where the terms in the first bracket count the average number of CCs intersected along $ds$ and the terms in the second bracket represent the average column density of an ion from a single CC. Here $\langle n_{\rm ion} \rangle (R)$ is the average number density of an ion in a CC at a distance $R$ from the center, which equals $\langle n_{\rm gas} \rangle \times f_{\rm ion}(R)$, where $f_{\rm ion}(R)$ is the ion fraction that depends on the physical number density of the cool gas for a given redshift and metallicity and $\langle n_{\rm gas} \rangle$ is the average number density of cool gas in a CC (distinct from physical density of cool gas)\footnote{Physical density is the true or the local density of the cool gas whereas the average density is the global average density taking into account the full available volume (see Figure 2 of \citetalias{Dutta2024}).}, $\langle L \rangle$ is the area-averaged path length across a CC, which is equal to,
\begin{equation}
    \langle L \rangle = \frac{\int_0^{R_{\rm CC}} 2\pi b db \, 2\sqrt{R_{\rm CC}^2 - b^2}}{\int_0^{R_{\rm CC}} 2\pi b db} = \frac{4}{3}R_{\rm CC}.
\label{eq:mean_l}    
\end{equation}
Therefore, we get the following expression for the average column density of an ion along the LOS at an impact parameter of $R_{\perp}$, 
\begin{equation}
    \langle N_{\rm ion} \rangle (R_\perp) = 2 N_{\rm CC}  \left(\frac{R_{\rm CC}}{R_{\rm CGM}}\right)^3 \langle n_{\rm gas}\rangle \int_{R_{\perp}}^{R_{\rm CGM}} f_{\rm ion}(R) \frac{R dR} {\sqrt{R^2 - R_\perp^2}}.
\label{eq:col_u}
\end{equation}

\subsubsection{Power-law distribution of cloud complexes}

Next, we use a power-law distribution of the number of CCs as a function of $r$ of the form
\begin{equation}
    \langle n_{\rm CC} \rangle (R) = n_0 \left( \frac{R}{R_{\rm CGM}} \right)^{-\alpha},
\label{eq:pl_dist}
\end{equation}
where $\alpha$ is the power-law index and $n_0$ is obtained using the normalization condition $N_{\rm CC} = \int \langle n_{\rm CC} \rangle (R) 4 \pi R^2 dR$. We get the following expressions for the average number of CCs intersected and the column density of an ion along a LOS at an impact parameter $R_\perp$,
\begin{equation}
   \langle N_{\rm CC,LOS} \rangle (R_\perp) = 2n_0 \pi R_{\rm CC}^2 R_{\rm CGM}^\alpha \int_{R_\perp}^{R_{\rm CGM}} \frac{R^{-\alpha +1}}{\sqrt{R^2-R_\perp^2}} \, dR,
\end{equation}    
and
\begin{equation}
    \langle N_{\rm ion} \rangle (R_\perp) = \frac{8}{3} n_0 \pi R_{\rm CC}^3 R_{\rm CGM}^\alpha \langle n_{\rm gas} \rangle  \int_{R_\perp}^{R_{\rm CGM}} f_{\rm ion}(R) \frac{R^{-\alpha +1}}{\sqrt{R^2-R_\perp^2}} \, dR.
\label{eq:col_p}
\end{equation}

In addition to the average column density, we also calculate the standard deviation using estimates derived in Appendix \ref{app:std}.

\subsubsection{Volume \& mass fraction of cool gas in CCs}
The mass fraction of CGM in the cool ($\sim 10^4$ K) phase is subdominant compared to the volume-filling hot phase (e.g., \citetalias{Dutta2024}, \citealt{Faerman2023}). The volume fraction is even smaller since cool gas is denser. Table 3 in \citetalias{Dutta2024} quotes a cool mass fraction of 11.9\% and a volume fraction of 0.16\% for a Milky Way-like TNG50 halo. Since in our CC model, cool gas is confined only within CCs, we expect the volume and mass fraction of cool gas within a CC to be higher than the whole CGM.

The volume fraction of cool gas within a CC is
\begin{eqnarray}
\nonumber
    f_V &=& \frac{M_{\rm cool}}{N_{\rm CC} \mu m_p n_{\rm cool} (4 \pi R_{\rm CC}^3/3) } = 0.016 \left(\frac{M_{\rm cool}}{10^{10}\ M_\odot}\right) \\ 
    && \times \left( \frac{N_{\rm CC}}{10^3}\right)^{-1} \left(\frac{n_{\rm cool}}{10^{-2} \, \rm cm^{-3}}\right)^{-1} \left(\frac{R_{\rm CC}}{10 \, \rm kpc}\right)^{-3},
    \label{eq:fV_cloudlet}
\end{eqnarray}
and the mass fraction is,
\begin{equation}
    f_M = \frac{M_{\rm CC}}{M_{\rm CC} + (1 - f_V) \mu m_p n_{\rm hot} (4 \pi R_{\rm CC}^3/3)},
\end{equation}
which equals $0.62$ for our fiducial parameters (assuming $n_{\rm hot} = 10^{-4}$ cm$^{-3}$ for hot gas in pressure equilibrium with cool clouds). As expected, the volume and mass fraction of cool gas within CCs are higher than in the full CGM. The volume fraction of CCs within the CGM is
\begin{equation}
f_V^{\rm CC} = N_{\rm CC} \left ( \frac{R_{\rm CC}}{R_{\rm CGM}} \right)^3 = 0.05 \left ( \frac{N_{\rm CC}}{10^3}\right) \left ( \frac{R_{\rm CC}}{10 \rm \, kpc}\right)^3 \left ( \frac{R_{\rm CGM}}{280 \rm \, kpc}\right)^{-3}. 
\end{equation}
However, the volume fraction of cool gas in the CGM as a whole would be tiny, $f_V^{\rm CC} \times f_V \approx 0.08$ \%, consistent with Table 3 of \citetalias{Dutta2024} based on a Milky Way-like TNG50 halo.
The volume fraction of CCs is typically small, yet the area covering fraction can be substantial because of projection along the LOS (e.g., see Figure \ref{fig:cc_cartoon}). This also means that the probability for the overlap of two CCs is small ($\sim 5$\%), implying that such overlapping CCs do not affect our statistics. Therefore, we do not bother about ensuring that CCs do not overlap. 

\subsection{Physical number density profile of cool gas}
\label{sec:cool_density}
The pressure of the cool gas is expected to be higher in the central regions than in the outskirts. This pressure variation will result in density variation across the CGM, with higher density in the center and lower density in the outskirts. Motivated by \citealt{Stern2016}, who modeled the cool photoionized CGM gas 
and found that the mean cool gas density scales as $R^{-1\pm 0.3}$ (see also, \citealt{Faerman2025}), we assume the following cool gas density profile,
\begin{equation}
\label{eq:ncool_PL}
n_{\rm cool}(R)= 10^{-3} \times \left( \frac{R}{R_{\rm CGM}} \right)^{-1} \, \rm cm^{-3}.
\end{equation}
A similar density scaling with radius is also found by \citealt{Werk2014} for the COS-Halos sample. The CGM is expected to have a shallower density profile compared to the dark matter density because of feedback (e.g., see Fig. 2 in \citealt{Sharma2012} for the hot gas density profiles inferred in groups and clusters), and we try this specific form to test the impact of varying $n_{\rm cool}$ with radius. The variation in the physical density of cool gas only changes the ion fraction ($f_{\rm ion}$; see Appendix \ref{app:con_factor}) as a function of radius at a given redshift and for a given metallicity.

\subsection{Understanding analytical results}

Using the analytical estimates and the physical density of cool gas, we calculate and compare the mean MgII column density for various parameters. We adopt $M_{\rm cool}=10^{10} \, M_\odot$, $R_{\rm CC}=10$ kpc, $N_{\rm CC}=10^3$, and $R_{\rm CGM}=280$ kpc as our fiducial values. There are several observational studies of the coherence length of the metal-bearing cool gas in the CGM. \citealt{Afruni2023} and \citealt{Shaban2025} found the coherence length of MgII to be $\sim 5$ and $2.7$ kpc respectively while \citealt{Duttar2024} estimated it to be $\approx 10$ kpc. 
Therefore, we choose $10$ kpc as our fiducial value for the radius of CC. To obtain the MgII ion fraction, we consider Eq. \ref{eq:ncool_PL} for the cool gas density profile. We compute the MgII ion fraction using {\sc CLOUDY} (\citealt{Ferland2017}) at a redshift of $0.2$ and a temperature of $10^4$ K. We choose KS18 (\citealt{Khaire2019}) as the background radiation field (see Figure \ref{fig:ion_frac}). We also consider a constant metallicity of $0.3$ solar (\citealt{Prochaska2017}) to compute the MgII column density.

\begin{figure}
    \centering
    \includegraphics[width=\columnwidth]{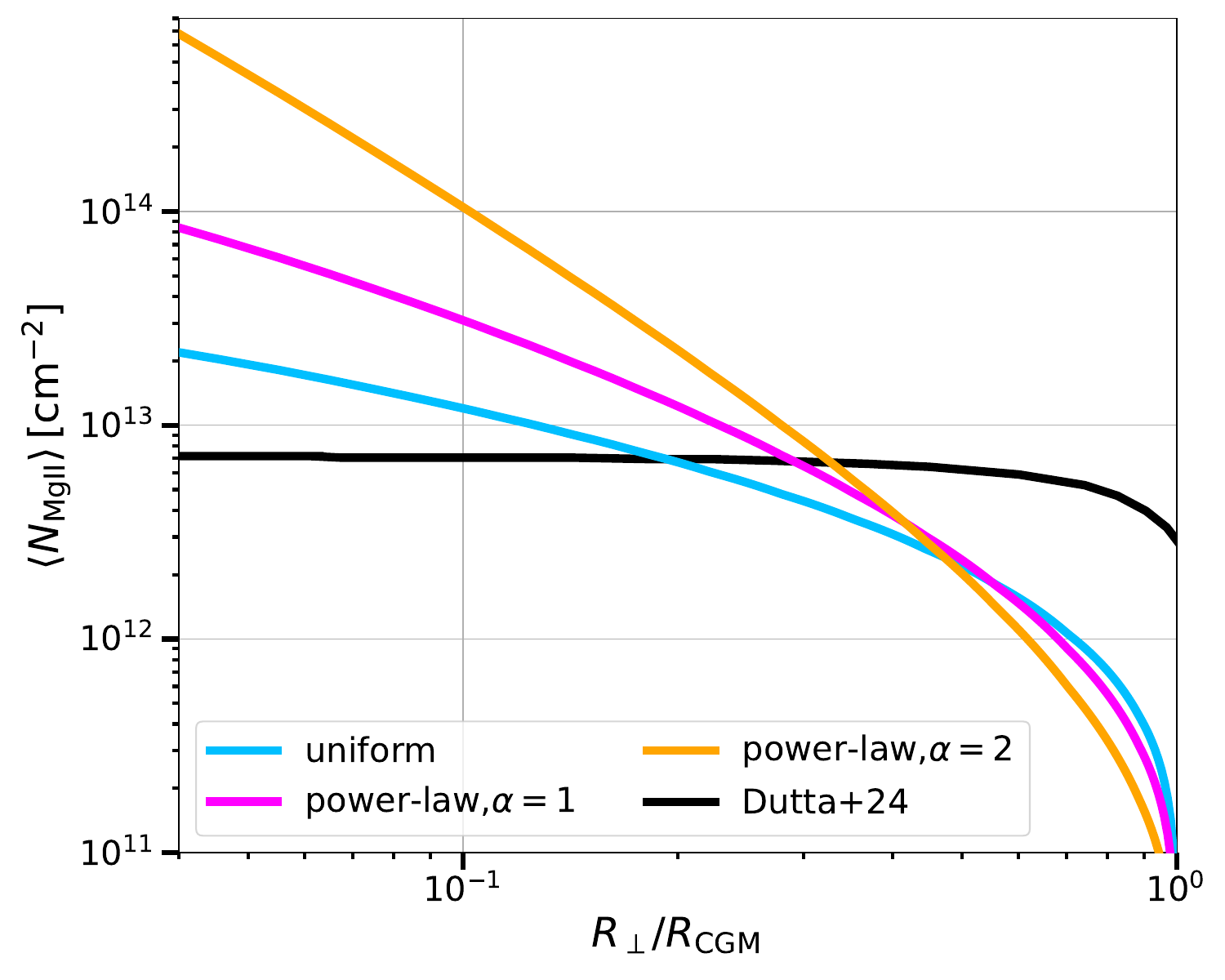}
    \includegraphics[width=\columnwidth]{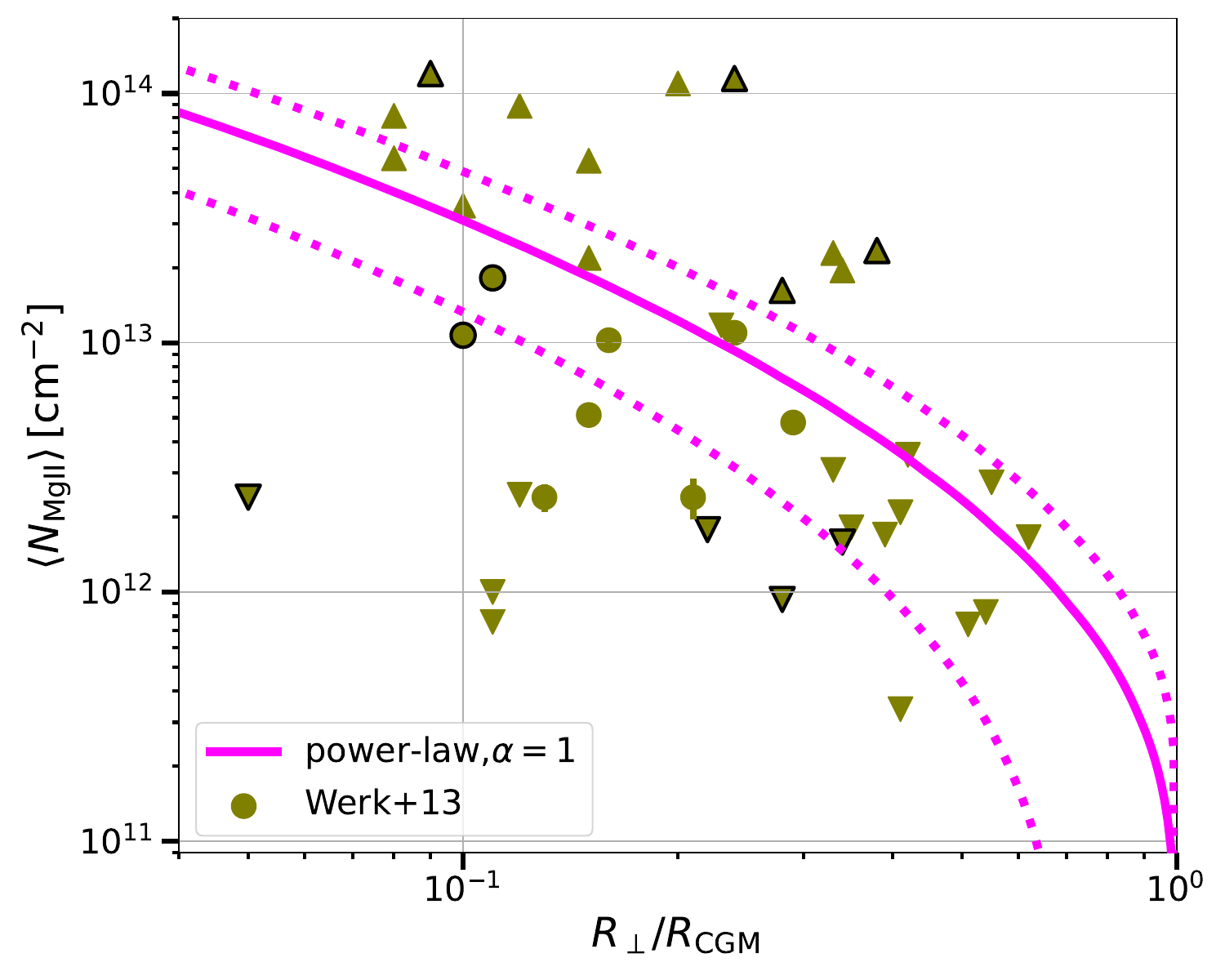}
    \caption{The \textit{top panel} shows the mean MgII column density as a function of normalized impact parameter. The black line shows the baseline column density from \citetalias{Dutta2024}. The blue line shows the mean MgII column density for uniform distribution, while magenta and orange lines show the same for power-law distribution of CCs in the CGM with power-law index of $1$ and $2$ respectively (Eqs. \ref{eq:col_u}, \ref{eq:col_p}). The baseline column density predicted by \citetalias{Dutta2024} shows no trend with the impact parameter, while the uniform and power-law distributions show a declining trend, which is observationally observed. However, the column density values for the power-law index of $\alpha =1$ (magenta line) better match the typical values from observations. So we choose a power-law distribution with index $\alpha=1$ as our fiducial parameter.
    The solid magenta line in the \textit{bottom panel} shows the mean MgII column density for 
    the fiducial value of $\alpha=1$. The dotted magenta lines show the expected standard deviation (see Appendix \ref{app:std}) from the average. The data points are from the COS-Halos survey (\citealt{Werk2013}). Even the data with smaller redshift ($0.15-0.25$) and stellar mass bins ($10.5-11$ in $\log_{10}$; encircled with black color) show intrinsic scatter in the column density data.}
    \label{fig:variation_an}
\end{figure}

In the \textit{top panel} of Figure \ref{fig:variation_an}, we show the average MgII column density as a function of the normalized impact parameter for various models. The black line shows the baseline MgII column density from the misty CGM model of \citetalias{Dutta2024}. We choose the parameters similar to (but not identical to; in particular, our CGM extent of $280$ kpc is smaller than their choice of $333$ kpc) what \citetalias{Dutta2024} found for one of the Milky Way-like TNG50 halos. The blue line shows the average MgII column density for uniform distribution (Eq. \ref{eq:col_u}) while magenta and orange lines show the column density for power-law distribution with power-law index of $\alpha=1$ and $2$ respectively (Eq. \ref{eq:col_p}). The column density profile from \citetalias{Dutta2024} shows no trend with impact parameter,\footnote{This is due to constant ion fraction values assumed at all radii. In Fig. 11 of  \citetalias{Dutta2024}, they assume a constant $f_{\rm MgII} = n_{\rm MgII}/n_{\rm Mg}$, which gives a flatter MgII column density compared to a self-consistent model that allows $f_{\rm MgII}$ to vary with pressure (e.g., see Fig. 6 of \citealt{Faerman2023}).} while uniform and power-law distributions of CCs in CGM show a declining trend with the impact parameter, which is what is observed. This declining trend has contributions from both the power-law distribution of CCs and the variable ion fraction with radius. Clearly, uniform and power-law distributions with $\alpha = 2$, respectively, under- and over-predict the typical MgII column density values. Therefore, we choose a power-law distribution with index $\alpha = 1$ as our fiducial model.

In the \textit{bottom panel}, the solid magenta line shows the mean MgII column density for our fiducial model ($\alpha=1$).
The dotted magenta lines show the expected spread (standard deviation) in column density (see Eq. \ref{eq:err_n}), which is computed by taking care of the variation in the number of CCs along a LOS and the deviation in the intersected length of a CC. To estimate the standard deviation, we choose the MgII ion fraction value at the smallest 3D distance (equal to the impact parameter) at a given impact parameter.  
Also plotted are the observations from the COS-Halos (\citealt{Werk2013}) survey. The circles show detections, while the upper and lower triangles show the lower and upper limits, respectively. The data points encircled with black color are the sub-samples with a narrower redshift range of $0.15-0.25$ and stellar mass range of $10.5-11.0$ (in log10). The observationally inferred column densities are typically the total column density along a LOS, summing over the individually detected components. 

Notice a larger scatter in the observed MgII column density values at similar impact parameters. The large scatter can arise due to various possibilities like a broad range in the halo/stellar masses of galaxies, a range in redshifts, different galaxy types (quiescent/star-forming), and orientation. However, the sub-samples (points encircled with black color) with a narrow range in stellar mass and redshift also show significant scatter. Therefore, there must be an intrinsic or true scatter in the column densities of ions tracing the cool CGM. 
The true scatter most likely arises due to the clumpy nature of the cool gas, projection effects, and intrinsic scatter in the cool gas number density at similar impact parameters (\citealt{Yang2025}). 
Using absorption line measurements in the spectra of multiple background quasars at different impact parameters, \citealt{Rao2013} and \citealt{Lehner2020} mapped the CGM of M31 and found a clear scatter in the column densities of ions tracing its cool gas.
By combining the observed column densities of SiII, SiIII, and SiIV, \citealt{Lehner2020} showed that the total silicon column density also exhibits scatter, indicating the clumpy nature of the cool gas.
The true scatter would be clearer from future observations/surveys (e.g., \citealt{Ng2025}) with a larger sample of galaxies having similar stellar mass, redshift, and types. 
Our analytical expressions (Eqs. \ref{eq:col_u}, \ref{eq:col_p}) pass through the general scatter of the observations. The average column density does not provide stringent constraints on the nature and distribution of CCs because of a large spread and scattered upper/lower limits in the observationally inferred MgII column density. These observations motivate our Misty CC model.

\subsection{Monte Carlo realizations of CCs in the CGM}
\label{sec:MCMC}
The analytical estimates of the mean column density and its deviation provide an excellent framework to compare the trends in observed column density. However, to compare the possible intrinsic spread in observed MgII column densities and EW with the model prediction, we need to create a Monte Carlo realization of CCs in the CGM. 
The intrinsic scatter in the observed column density values at similar impact parameters can be reproduced by limiting the cool gas to misty CCs rather than filling the entire CGM by cool mist (as in \citetalias{Dutta2024}).

To generate a Monte Carlo realization of CCs in the CGM, we populate $N_{\rm CC}$ number of CCs each of radius $R_{\rm CC}$ in the CGM for power-law distributions with index $\alpha=1$. We first sample the radial distance $r$ of the center of CC from the power-law distribution (see Eq. \ref{eq:pl_dist}). We then draw a uniform deviate in $\phi$ ranging from $0$ to $2\pi$ and in $\cos \theta$ from $-1$ to $1$. Using these $r,\phi$ and $\theta$, we compute the $x,y,z$ coordinates. We ensure that the CCs are not populated beyond the CGM boundary ($R_{\rm CGM}$) by recomputing the CC coordinates in case they are. 

After generating the centers of CCs, we shoot $10^4$ LOSs (spaced uniformly in $\log_{10} R_\perp$) through the CGM in the $z$ direction. 
We then compute the $x,y$ coordinates of the LOS by drawing a random value of $\phi$ uniformly between $0$ and $2\pi$ and calculate the number of CCs intersected and the corresponding intersected lengths along each LOS. We calculate the total column density for each intersected CC using the intersected length, the average cool gas density in the CC ($\langle n_{\rm gas} \rangle$), and the MgII ion fraction. To calculate the MgII column density, we calculate the MgII ion fraction using the physical cool gas density (Eq. \ref{eq:ncool_PL}) and the ion fraction at a redshift of $0.2$, assuming a cool gas temperature of $10^4$ K, metallicity of $0.3$, and KS18 as background radiation field (see Appendix \ref{app:con_factor}). Recall that our CCs are assumed to be uniform and in the mist limit. In the observations, column density is typically derived from the Equivalent Width (EW) based on Voigt profile modeling of various absorption components in the background quasar continuum produced by the intervening CGM. In our model, we first calculate the column density and then the corresponding EW, assuming a velocity field in the CGM and turbulent broadening across a CC, which we discuss below in detail.

\subsubsection{Computing Equivalent Width}
\label{sec:EW}

\begin{figure}
    \centering
    \includegraphics[width=1\columnwidth]{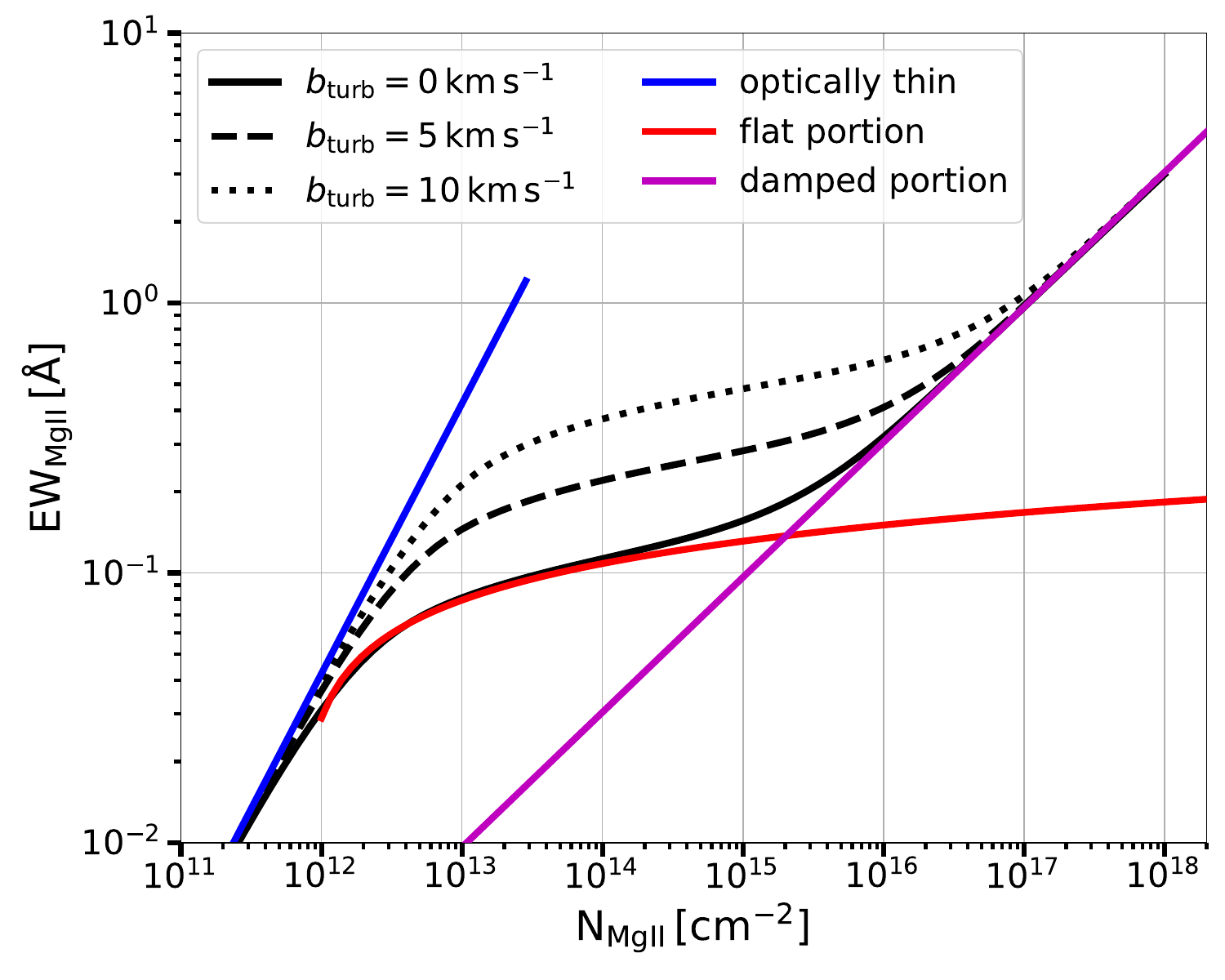}
    \caption{Variation in MgII EW with MgII column density. The black lines show the EW for various $b_{\rm turb}$ values. The solid blue, red, and magenta lines show the analytical relation between EW and column density in the optically thin limit, and in the flat and damped portions of the curve of growth. These analytic relations are from \citealt{Draine2011} (see Chapter 9). We assume thermal broadening at $T=10^4$ K for MgII, which gives $b_{\rm thermal}=2.6 \, \rm km \, s^{-1}$. In the flat portion (applicable to most of our CCs, see Fig. \ref{fig:variation_an}), the EW increases with increasing turbulent broadening.}
    \label{fig:eq_col_den}
\end{figure}

We can compute the EW using the column density estimated above using the curve of growth and assuming a reasonable $b$-parameter (that characterizes the thermal and turbulent broadening of the absorption line; \citealt{Draine2011}). For a MgII column density $\gtrsim 10^{13}$ cm$^{-2}$, the EW will be lower than the linear extrapolation of the flat portion of the curve of growth (see Figure \ref{fig:eq_col_den}). For example, if we have $10$ CCs (coinciding in LOS and turbulent velocities so that they produce a single absorption feature) along a LOS, each with a MgII column density $\sim 10^{13} \, \rm cm^{-2}$, then the EW computed using the total column density will be lower than the sum of EWs computed for each CC individually. However, summing the EWs is appropriate when the LOS CCs are kinematically non-overlapping. The correct EW along a LOS lies between these two extremes and depends on the details of the LOS and the turbulent velocities of cloud complexes.

\subsubsection{Ansatz for turbulent broadening across a CC} 
\label{sec:turb_ansatz}

To compute EW, we consider both thermal and turbulent broadening across a CC. We assume the temperature of the cool gas to be $10^4$ K, which gives thermal broadening of $b_{\rm thermal} = \sqrt{2k_B T/(A m_p)} = 2.6 \rm \, km \, s^{-1}$ for MgII. We also consider the contribution of turbulent broadening ($b_{\rm turb,CC}$; across a CC) for each CC. We compute $b_{\rm turb,CC}$ in as follows. Assuming that turbulence is driven at CGM global scales with velocity dispersion $\sigma_{\rm turb, CGM}$ that cascades down to CC scale without loss,\footnote{In Kolmogorov-like turbulence, kinetic energy cascades without loss across inertial scales with the turbulent velocity at scales $l$ and $L$ related as $v_l = v_L(l/L)^{1/3}$, down to the small viscous scale at which it dissipates (\citealt{Kolmogorov1991}).} the velocity dispersion at CC scale is given by $\sigma_{\rm turb,CC} = \sigma_{\rm turb, CGM} \times (R_{\rm CC}/R_{\rm CGM})^{1/3}$. Therefore, the total broadening is $b_{\rm tot}=\sqrt{b_{\rm thermal}^2 + b_{\rm turb,CC}^2}$ (in the micro-turbulent limit; \citealt{Mihalas1978}), where $b_{\rm turb,CC}=\sqrt{2}\sigma_{\rm turb,CC}$. To estimate $\sigma_{\rm turb, CGM}$, we assume the turbulent Mach number of the hot ($\sim 2 \times 10^6$ K), volume-filling CGM to be $\sim 0.5$ (e.g.,  \citealt{Schmidt2021,Mohapatra2022}).\footnote{Direct observations of turbulence in the hot CGM are not available for Milky Way mass halos. The CGM is expected to have a larger turbulent Mach number than the intracluster medium, for which the hot phase turbulence is subsonic with a turbulent Mach number $\sim 0.2$ (\citealt{Hitomi2016}).} These assumptions give $\sigma_{\rm 3D turb, CGM} \approx 107 \rm \, km \, s^{-1}$. For $R_{\rm CC}=10$ kpc and $R_{\rm CGM}=280$ kpc, we get $\sigma_{\rm 3D turb,CC} = 35 \rm \, km \, s^{-1}$. Thus, the 1D turbulent dispersion is $\sigma_{\rm turb,CC} = \sigma_{\rm 3D turb, CC}/\sqrt{3} = 20 \rm \, km \, s^{-1}$.
For simplicity, we choose the turbulent broadening parameter to be the same for all intersected lengths across the CC. 

\begin{figure}
    \centering
    \includegraphics[width=\columnwidth]{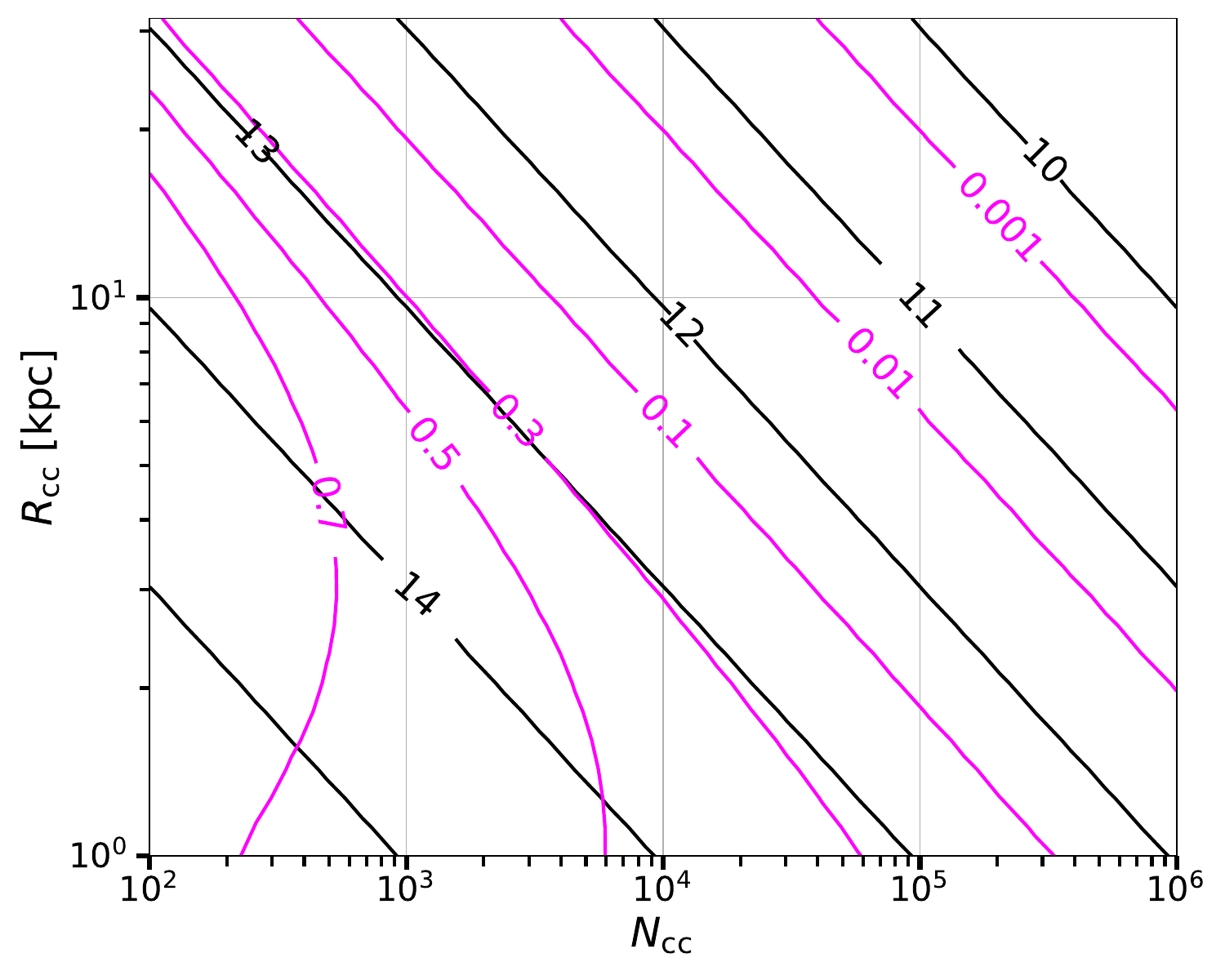}
    \caption{Contours of equal (in log$_{10}$) maximum MgII column density (corresponding to LOS passing through the center of a CC; black lines) and equivalent width (magenta lines, in units of $\rm \mathring{A}$) in the parameter space of $N_{\rm CC}$ and $R_{\rm CC}$ for a single cloud complex with $M_{\rm cool} = 10^{10} M_\odot$. For MgII column densities $\gtrsim 10^{13}$ cm$^{-2}$, the EW saturates in the flat part of the curve of growth (see Figure \ref{fig:eq_col_den}) and the curves of constant EW are no longer parallel to the constant column density contours.  The EW increases with $R_{\rm CC}$ for a fixed MgII column density because of a larger $b_{\rm turb, CC}$ (see Fig. \ref{fig:eq_col_den}).
    }
    \label{fig:contour}
\end{figure}

Figure  \ref{fig:contour} shows the contour plot of the maximum MgII column density (black lines indicate ${\rm log}_{10} N_{\rm MgII}$; corresponding to the LOS passing through the center of a CC) and equivalent width (in $\rm \mathring{A}$; magenta lines) for a single CC in the $N_{\rm CC}-R_{\rm CC}$ parameter space for a fixed total cool gas mass $M_{\rm cool}=10^{10} \, \rm M_{\odot}$. The MgII column density for a LOS passing through the center of a CC is $\langle n_{\rm MgII} \rangle \times 2R_{\rm cc}$, which simplifies to
\begin{equation}
    N_{\rm MgII, max} =  9.2 \times 10^{12} \ {\rm cm^{-2}} \left(\frac{M_{\rm cool}}{10^{10} \ M_{\odot}} \right) \left( \frac{N_{\rm CC}}{10^3} \right)^{-1} \left( \frac{R_{\rm CC}}{10 \, \rm kpc} \right)^{-2}.
    \label{eq:max_col}
\end{equation}

To compute the MgII column density above, for simplicity, we assume the MgII ion fraction to be $0.2$. Note that MgII fraction of $0.2$ correspond to the gas density of $\sim 10^{-2} \, \rm cm^{-3}$ at redshift of $\sim 0.2$ (see Fig. \ref{fig:ion_frac}). We first generate the normalized absorption profile ($I_\nu=\exp[-\tau_\nu]$, where $\tau_\nu \propto N_{\rm MgII}$ is the frequency-dependent optical depth; see Eq. 9.5 in \citealt{Draine2011}) and then compute the area under the normalized absorption profile as the definition of EW. 
In the linear regime of the curve of growth ($N_{\rm MgII}\lesssim 10^{12} \, \rm cm^{-2}$; see Fig. \ref{fig:contour}), the EW and column density contours run parallel to each other, which suggests a single value of EW for a given value of column density (Eq. \ref{eq:max_col}). At higher column densities, $N_{\rm MgII}\gtrsim 10^{12} \, \rm cm^{-2}$, for a fixed column density the EW increases with an increase in $R_{\rm CC}$ as $b_{\rm turb,CC}$ is larger for a larger CC according to our ansatz for CGM turbulence (see Figure \ref{fig:eq_col_den}). This regime lies in the flat part of the curve of growth. Therefore, to obtain a higher column density and EW per CC, one needs larger and fewer CCs. Thus, the observed column density and EW distributions can constrain the cool CGM mass and the CC numbers and sizes.

\subsubsection{Ansatz for LOS velocity of CCs} 

To compute the absorption spectrum along a LOS, we require the LOS velocities of the intersected CCs. We assign a 3D velocity field with a Gaussian distribution and a Kolmogorov power spectrum\footnote{\citealt{Chen2023} found that the Kolmogorov scaling is consistent with the observations for clouds with sizes $\lesssim$ 1 kpc. The same scaling holds at a much larger scale in extended QSO nubulae \citep{Chen2024}}. ($v_l \propto l^{1/3}$) on a $1800^3$ grid across the entire CGM. Every CC is assigned the velocity from the closest grid cell. 
The turbulent velocity field has zero mean and standard deviation $\sigma_{\rm 3Dturb,CGM} = 107 \rm \, km \, s^{-1}$ (assuming our ansatz for global CGM turbulence). In addition to this bulk velocity, each CC also has a Gaussian spread in its internal velocities ($\sigma_{\rm turb, CC}$), as mentioned earlier. To generate the absorption profile along each LOS, we consider the blending of absorption profiles from all intersected CCs. The optical depth along a LOS is computed as the sum of the optical depths from each intersected CC, $\tau(\nu)=\sum_i \tau_i(\nu)$, where the summation goes over all the intersected CCs. Thus, the normalized absorption profile is given as $I(\nu)=\exp[-\tau(\nu)]$ and the area under it gives the total EW along a LOS.

\begin{figure*}
    \centering
    \includegraphics[width=\textwidth]{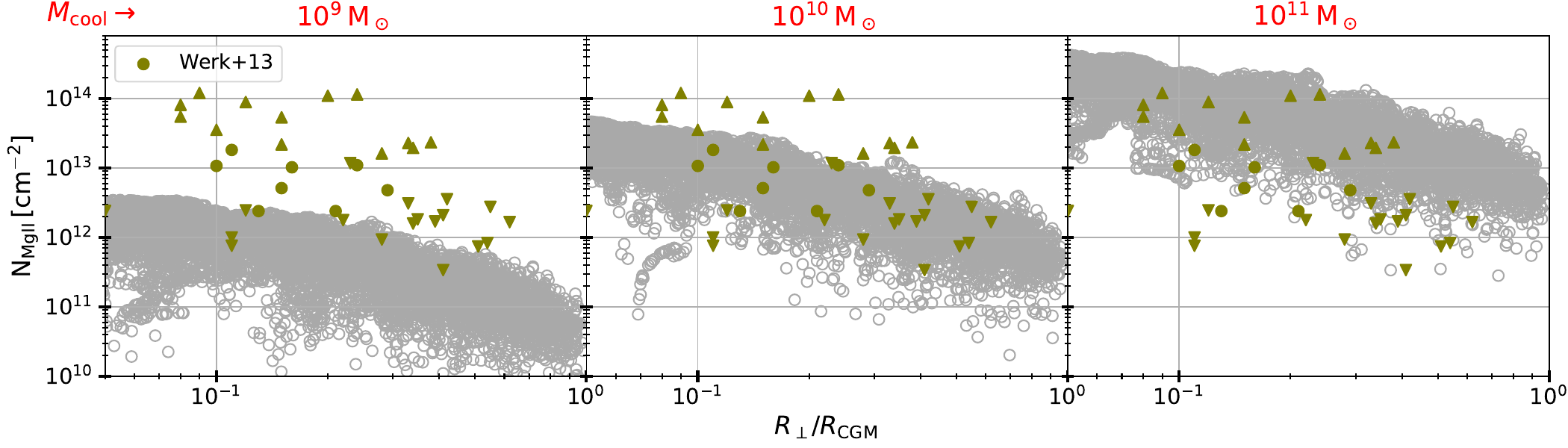}
    \includegraphics[width=\textwidth]{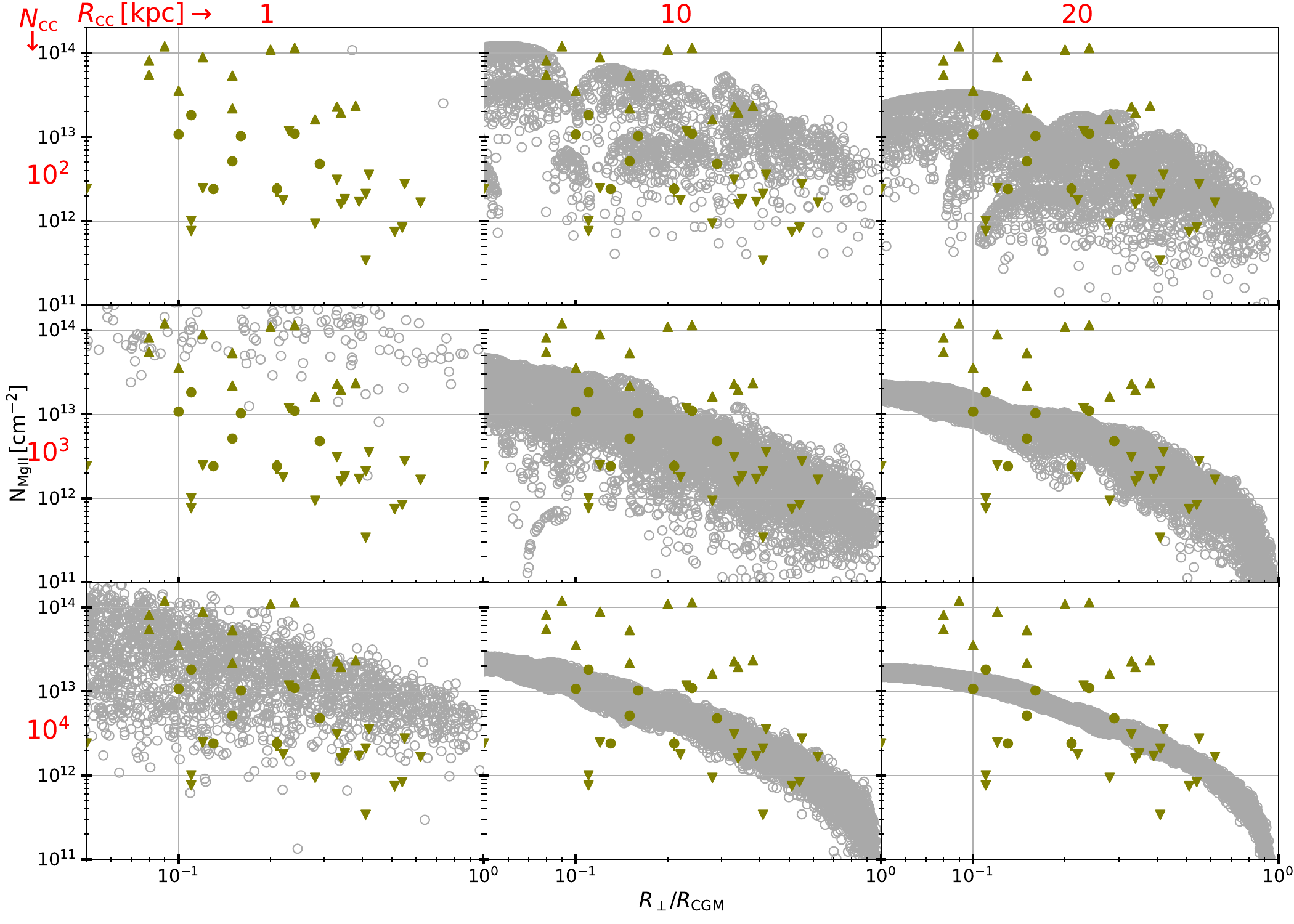}
    \caption{Variation in the MgII column density along $10^4$ sightlines with the cool gas mass (top panels), and number and radius of CCs (bottom $3\times 3$ panels) for a power-law ($\alpha=1$) distribution of CCs in the CGM. Only non-zero values are shown with circles, resulting in a smaller number of them when the covering fraction is small. The data points shown in green color are from the COS-Halos survey (\citealt{Werk2013}). It is evident from the top panels that cool CGM mass of $\sim 10^{10} \, M_\odot$ best matches the observed data.
    The bottom panels show the variation in MgII column density with $R_{\rm CC}$ and $N_{\rm CC}$ for a fixed cool gas mass of $10^{10} \, M_\odot$. Notice that the increase in the size of CC results in less scatter in the column density because with larger CCs, the volume fraction of CCs is larger, and a similar number of CCs is intersected along different sightlines. A similar trend is observed for a larger number of CCs in the CGM. 
    For the range of parameters we explored, $N_{\rm CC} = 10^3$ CCs with $R_{\rm CC} = 10$ kpc and $M_{\rm cool} = 10^{10} M_\odot$ best explain the observed MgII column density distribution. } 
    \label{fig:col_var}
\end{figure*}

\subsection{Comparing MgII column density with observations}

The top panels of Figure \ref{fig:col_var} show the column density distribution along 10$^4$ sightlines (empty sightlines are not shown) across the CGM with $10^3$ CCs and a CC radius $R_{\rm CC} = 10$ kpc, but with the cool gas mass of $10^9$, $10^{10}$ and $10^{11} M_\odot$; we choose the power-law index $\alpha=1$. The data points shown in green color are from the COS-Halos survey (\citealt{Werk2013}). The average MgII column density varies linearly with the cool gas mass, which suggests that the column density distribution of cool gas tracers (like MgII) is a quantitative indicator of the cool gas mass in the CGM. The trend and average MgII column density are most consistent with a cool CGM mass $\sim 10^{10} M_\odot$. The distribution of MgII column density from our Monte Carlo-generated LOSs is too small (large) for $10^9(10^{11}) M_\odot$ in the cool CGM.

The bottom $3\times 3$ panels of Figure \ref{fig:col_var} show the variation in MgII column density for cool CGM mass of $10^{10} M_\odot$ with a variation of the number of CCs ($N_{\rm CC}$ across rows) and the CC radius ($R_{\rm CC}$ across columns). For small $R_{\rm CC}$ and $N_{\rm CC}$, the detected column densities are much larger, but the covering fraction is small since most sightlines do not encounter cool gas. Since we only show detections, the cases with a small covering fraction will have a smaller number of grey circles. For larger $R_{\rm CC}$, the covering fraction increases, and the individual LOS column densities are smaller. With an increasing number of CCs and CC sizes (bottom right panels), the scatter in MgII column density decreases. The observed spread in the inferred column densities of cool/warm gas can thus help us constrain cool/warm CGM parameters such as $M_{\rm cool}$, $N_{\rm CC}$, and $R_{\rm CC}$. The observed column density distribution is most consistent with the simulated distribution for $N_{\rm CC} =10^{3}$, $R_{\rm CC} = 10$ kpc, $M_{\rm cool} = 10^{10} M_\odot$, and with the power-law index of $\alpha=1$ (see Fig. \ref{fig:variation_an}), so we adopt these as our fiducial values.

\subsection{Comparing MgII Equivalent Width and covering fraction with observations}

For each LOS, we quote a single EW, corresponding to the area under the absorption spectrum across velocities, and do not separate different absorption components. Most of the observational works quote a single EW along a LOS, adding EWs of multiple absorption components if present, implicitly assuming a single absorption component in the curve of growth modeling. The advancement of high-resolution spectrographs and precise modeling will make it possible to model the total absorption on a component-by-component basis, a method currently being applied in a few limited studies (\citealt{Sameer2024}).

\begin{figure}
    \centering
    \includegraphics[width=\columnwidth]{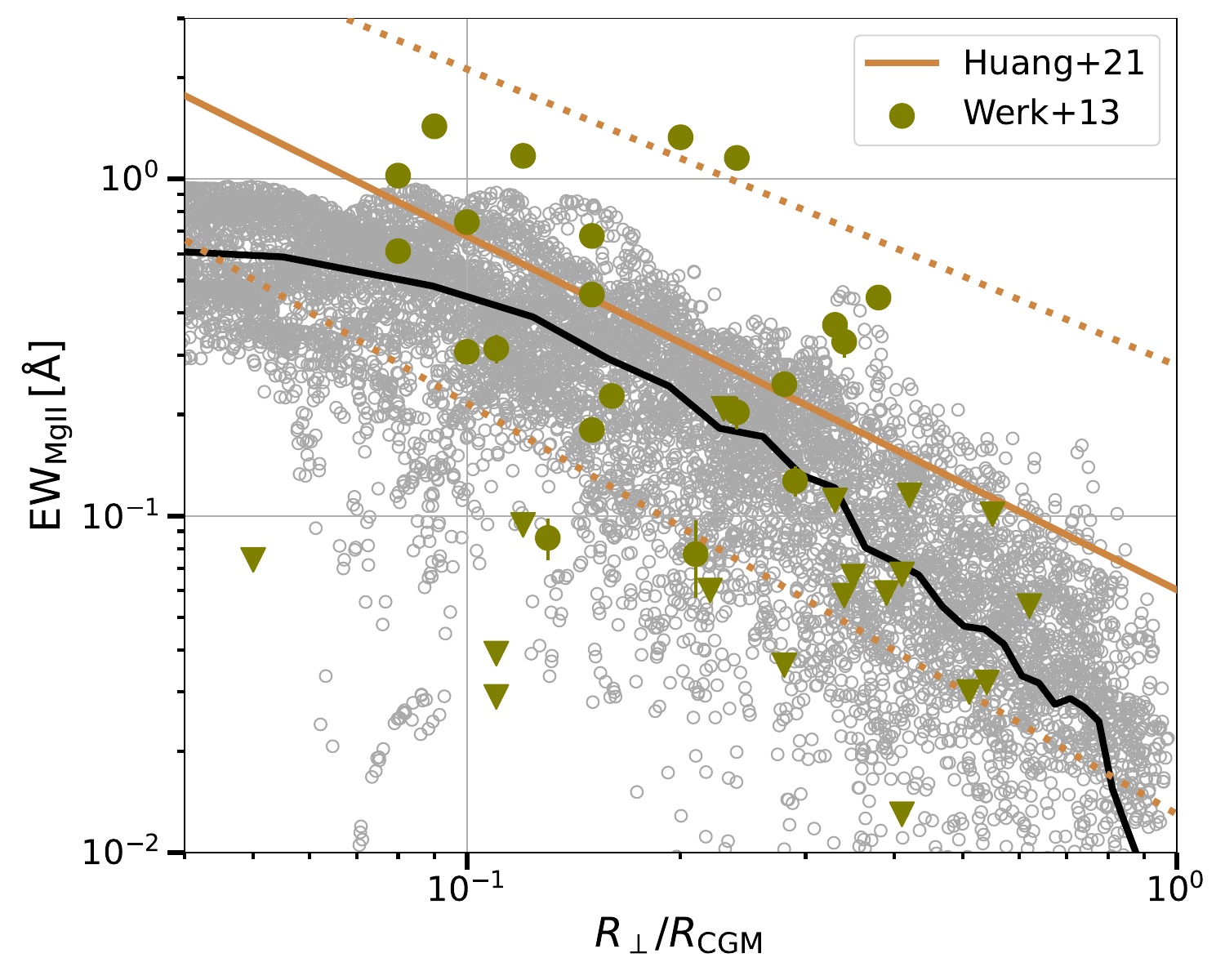}
    \includegraphics[width=\columnwidth]{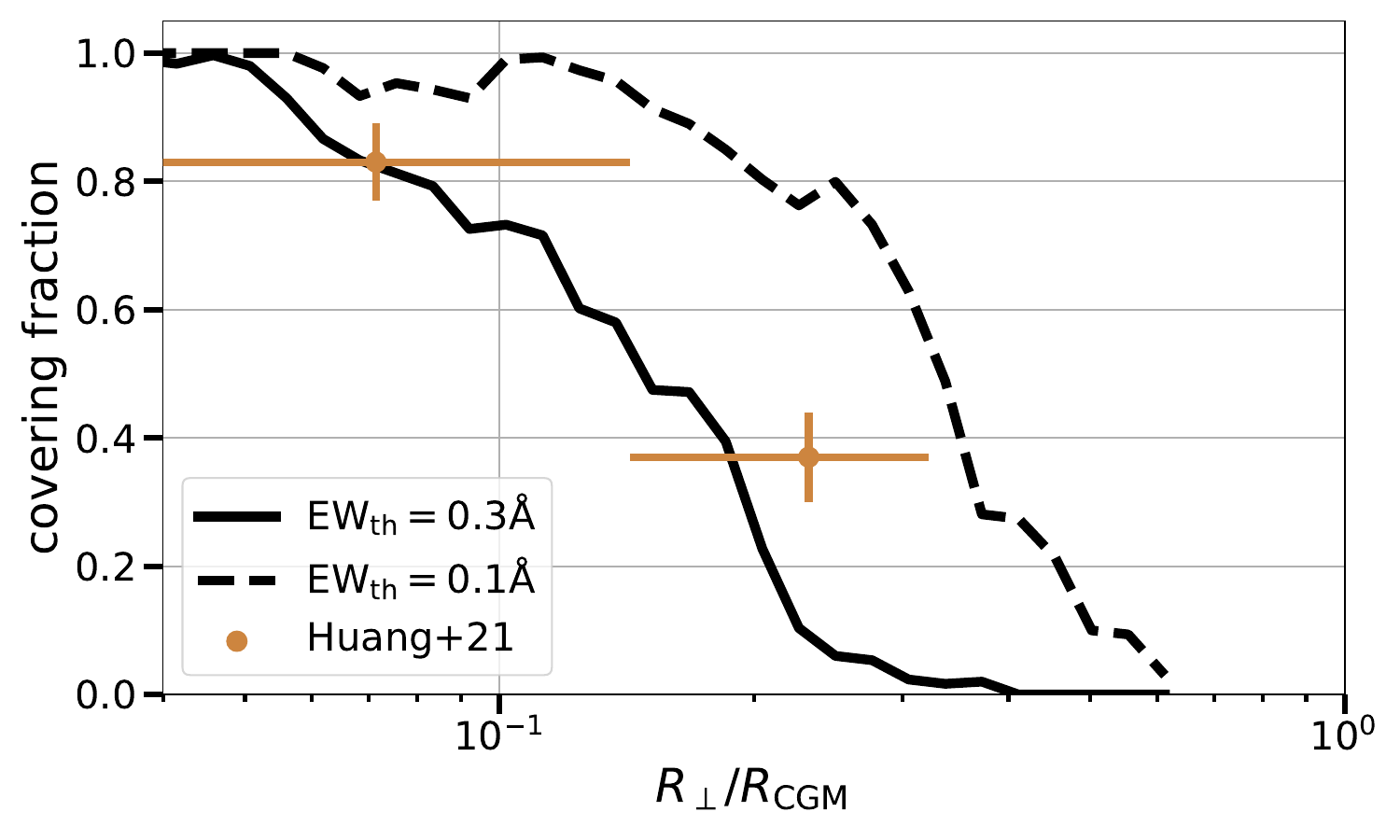}
    \caption{The \textit{top panel} shows the MgII EW distribution as a function of the normalized impact parameter for our fiducial {\it mCC} model. The grey points (note that empty sightlines are not shown) show the total MgII EW as a function of the impact parameter for $10^4$ sightlines. 
    The black solid line shows the mean EW. The solid and dotted orange lines show the best-fit relation and $1 \sigma$ spread based on observations from \citealt{Huang2021}. The green points are the EW data points from the COS-Halos survey (\citealt{Werk2013}). The \textit{bottom panel} shows the covering fraction as a function of the normalized impact parameter: the solid black line for an EW threshold of $0.3$ \r{A} and the dashed black line for $0.1$ \r{A}. The orange points show the covering fraction values from \citealt{Huang2021} for the EW threshold of $0.3$ \r{A}. Our mCC model with the fiducial parameters reproduces the EW distribution and covering fractions. }
    \label{fig:cov_eq}
\end{figure}

The \textit{top panel} of figure \ref{fig:cov_eq} shows the MgII EW for our fiducial {\it mCC} model. The grey scatter points show the EW distribution from Monte Carlo realizations of CCs along various LOSs; empty sightlines are not shown, and the black line shows the mean EW. The observed relation of EW with impact parameter ($R_\perp$) from \citealt{Huang2021} is shown using an orange line, with the dotted lines showing the $1\sigma$ uncertainty. The green data points are from the COS-Halos survey (\citealt{Werk2013}). The EW distribution from our model matches the COS-Halos samples (though there is a large scatter in the observed EWs). Even though our model predicts EW values slightly less than the fits from \citealt{Huang2021}, the values are consistent within the $1\sigma$ limit. Moreover, observational distribution is likely to be biased toward higher values because of the difficulty in detecting weak absorption features. The log-log EW-impact parameter relation from \citealt{Huang2021} predicts larger EW, especially at lower and higher impact parameters. However, adopting a log-linear relation between the EW and impact parameter 
will result in the mean EW distribution shape, which resembles our model (see figure $12$ of \citealt{Dutta2020}).
In the \textit{bottom panel}, the solid and dashed black lines show the MgII covering fraction as a function of the normalized impact parameter from our model for the EW thresholds of $0.3$ \r{A} and $0.1$ \r{A}, respectively. The orange data points are from \citealt{Huang2021} with an EW threshold of $0.3$ \r{A}. 
Our misty CC model with fiducial parameters effectively reproduces the covering fraction and equivalent width (EW) distributions.

\subsection{Estimating cool gas mass in the CGM}

One of the fundamental physical properties of the CGM is its mass distribution across different temperature phases. While the volume-filling hot phase is difficult to observe, quasar absorption lines from cool and warm CGM ions are commonly observed. However, there are large variations in the inferred column densities and several upper and lower limits (e.g., see Fig. \ref{fig:variation_an}). Because of these wild variations, it is hard to estimate the cool and warm gas masses. The large variations and scatter in the column density carry important information that can help us infer the physical properties of the cool/warm CGM.

\subsubsection{Relation between average column density \& covering fraction}

\begin{figure}
    \centering
    \includegraphics[width=\columnwidth]{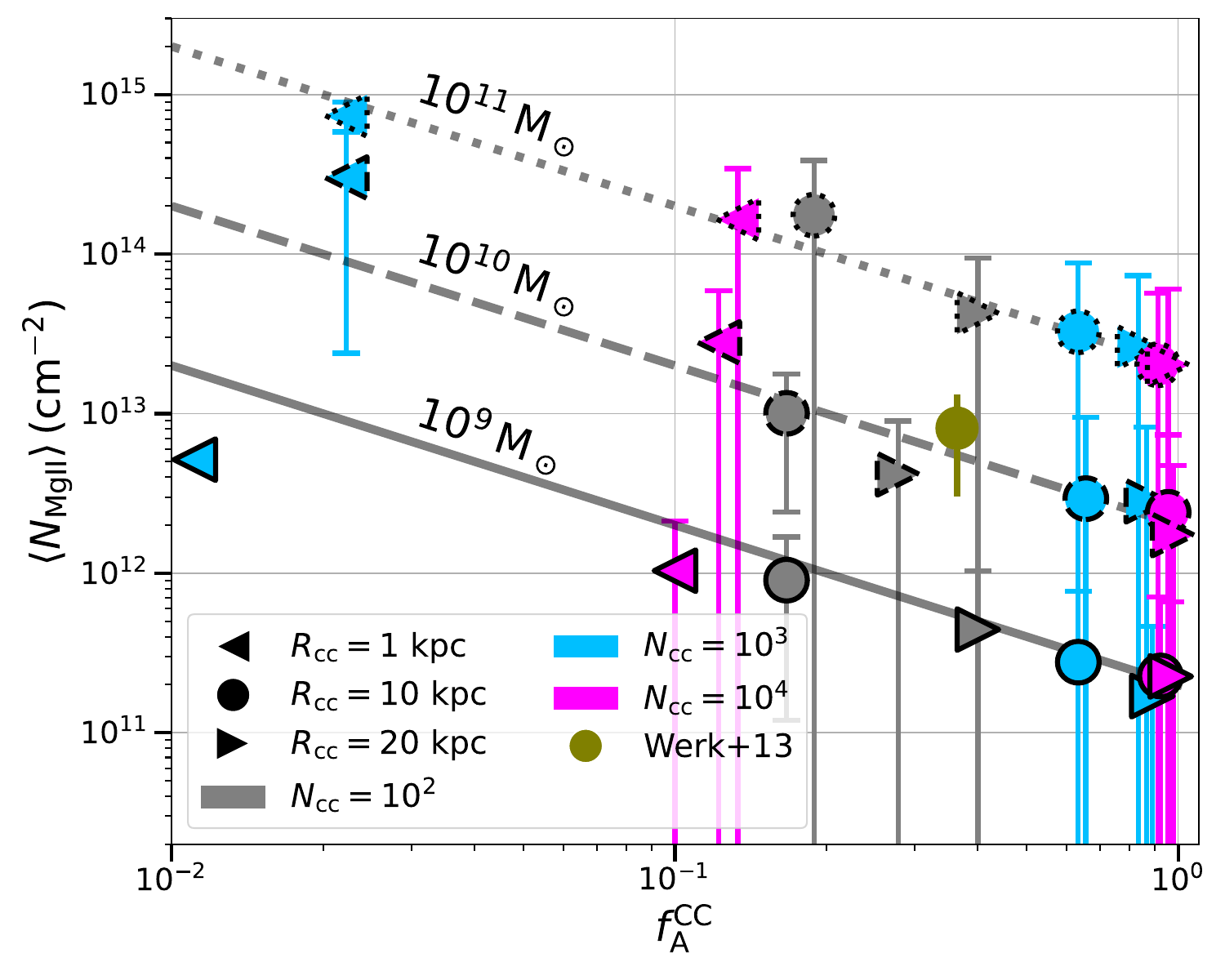}
    \caption{The average MgII column density as a function of the area covering fraction of CCs for various combinations of the cool gas mass, $N_{\rm CC}$ and $R_{\rm CC}$ (see Figure \ref{fig:col_var}). The colored points with solid, dashed, and dotted black lines on their borders are for a cool gas mass of  $10^9$, $10^{10}$, and $10^{11} \, M_\odot$ respectively. The error bars on each point show the standard deviation on the column densities. The dotted, dashed, and solid black lines show the contours of constant average MgII column density times the covering fraction for three different values $\langle N_{\rm mgII} \rangle \times f_{\rm A}^{\rm CC} = 10^{11.3}, 10^{12.3}$, and $10^{13.3}$ cm$^{-2}$, respectively. The product of these two quantities closely tracks the cool CGM mass and hence can be used as an observable proxy for the cool CGM mass. To compute the above quantities, we selected $100$ random sightlines (so that the probability of selection is proportional to the projected CGM area) out of $10^4$ to be statistically consistent with the typical size of observational samples. The green point represents the result from the COS-Halos (\citealt{Werk2013}) survey, the position of which indicates a cool gas mass of $\sim 10^{10} \, \rm M_{\odot}$ for the COS-Halos galaxies.}
    \label{fig:col_cov_frac}
\end{figure}

For the same cool CGM mass, the CCs can be arranged in different physical configurations -- compact CCs with smaller $R_{\rm CC}$ or fewer CCs. With compact CCs and fewer of them, we expect to encounter several empty sightlines, resulting in a small covering fraction. But whenever a LOS encounters a CC, in this case, it is with a large column density (e.g., compare the second-left column of Figure \ref{fig:col_var} with the bottom-left column). One can easily verify that the product of the average column density (averaged over detections) and the covering fraction is roughly the same for the same cool CGM mass. 
Figure \ref{fig:col_cov_frac} shows the correlation between the average MgII column density and the covering fraction of CCs in the CGM for various combinations of $R_{\rm CC}$, $N_{\rm CC}$, and $M_{\rm cool}$. We chose $100$ random samples (with selection probability proportional to the projected area 
from a total of $10^4$ (see Figure \ref{fig:col_var}) to be statistically consistent with observations and compute the average MgII column density and covering fraction (defined as the ratio of non-empty sightlines and the total number of sightlines [100]) for these.\footnote{We bin the projected CGM into $10$ radial annuli of equal areas and then select $10$ LOSs randomly from each annulus, resulting in a total of $100$ LOSs. This ensures that the probability of the selection of a LOS is proportional to the area of the annulus to which it belongs.} The colored points with solid, dashed, and dotted black lines on their borders correspond to the cool gas mass of $10^9, 10^{10}$, and $10^{11} \, \rm M_\odot$, respectively. The error bar shows the standard deviation in the column densities. The dotted, dashed, and solid black lines show the contours of a constant average column density times the covering fraction for the three cases with $\langle N_{\rm mgII} \rangle \times f_{\rm A}^{\rm CC} = 10^{11.3}, 10^{12.3}$, and $10^{13.3}$ cm$^2$. The points corresponding to the same cool CGM mass have a similar value of $\langle N_{\rm mgII} \rangle \times f_{\rm A}^{\rm CC}$, regardless of $R_{\rm CC}$ and $N_{\rm CC}$. 
The green point shows the result from the COS-Halos (\citealt{Werk2013}) survey. The positions of the data point indicate a cool gas mass of approximately $\sim 10^{10} \, \rm M_{\odot}$ for COS-Halos galaxies. We calculate the covering fraction from COS-Halos data by dividing the number of detected LOSs by the total number of observed sightlines, excluding lower limits. A more accurate statistical analysis of the data is beyond the scope of our paper. Thus, we have shown that the combination of the covering fraction and average column density can constrain the mass of the cool CGM. Moreover, the intrinsic spread in column density distribution constrains the number of CCs and their sizes.

\subsection{`Advanced' Model: varying size and mass of CCs}
\label{sec:advanced}

So far, we have considered the ‘basic’ model where we consider a fixed radius and cool gas mass of cloud complexes.
However, cool clouds in the CGM are observed to have a range of sizes and masses, from $\sim$ pc size (\citealt{Hsu2011}) to $\sim$ kpc size (\citealt{Duttar2024}) and mass varying from $\sim 10^4$ to $\sim 10^8 \, M_\odot$. A large variation in size and mass of cool clouds is also seen in cosmological simulations (\citealt{Nelson2020,Ramesh2024,Ramesh2025}) and idealized simulations (\citealt{Gronke2022,Das2024,Tan2024}). Galactic feedback processes also affect the distribution of clouds in the inner CGM, making the radial distribution profiles shallower in the inner region of the CGM, and even a turnover in the index can occur at lower radii (\citealt{Nelson2020,Augustin2025}). 
Therefore, in this subsection, we move beyond our `basic' model with fixed size and mass of CCs in the CGM, and introduce a size and mass distribution of CCs, examining the impact of these modifications on the observables. Along with these modifications, we also incorporate a realistic radial distribution of CCs in the CGM to mimic the effect of galactic feedback processes. We refer to this modified model as the `advanced' model.

Motivated by the observations and simulations mentioned above, we adopt a power-law distribution of the radius of spherical CCs in the CGM as.
\begin{equation}
    \frac{dN_{\rm CC}}{dR_{\rm CC}} \propto R_{\rm CC}^{-\eta}
    \label{eq:size_dist}
\end{equation}
\citealt{Gronke2022,Das2024,Tan2024} in their idealized simulation consistently found that the volume distribution of cool clouds follows a power-law behaviour $V^{-1}$, where $V$ is the volume of cool clouds. Therefore the size distribution, assuming $r\propto V^{1/3}$, also follows a power-law distribution with $r^{-1}$.
The power-law behaviour of the size distribution of cool clouds is also found in cosmological simulations with index $\sim 1.5$ (as can be inferred from Figs. $10$ and $3$ of \citealt{Ramesh2024} and \citealt{Ramesh2025} respectively).
Therefore, we choose the fiducial value of the power-law index $\eta$ as $1$. Apart from the power-law behavior, simulations show a turnover at smaller sizes (Figs. $10$ and $3$ of \citealt{Ramesh2024} and \citealt{Ramesh2025} respectively, and Fig. $12$ of \citealt{Gronke2022}), likely due to the resolution limit. Therefore, we choose the lower and higher cutoffs of $100$ pc and $10$ kpc on the size distribution of CCs. Similarly, we consider a power-law mass distribution of the CCs.
\begin{equation}
    \frac{dN_{\rm CC}}{dM_{\rm CC}} \propto M_{\rm CC}^{-\zeta}
    \label{eq:mass_dist}
\end{equation}
Like the size distribution, the idealized simulations \citep{Gronke2022,Das2024,Tan2024} show a power-law mass distribution of cool clouds with index $2$, while the cosmological simulations show a power-law with index $\sim 1$ (Fig. $9$ of \citealt{Ramesh2024}).
Therefore, we adopt the fiducial value of the power-law index $\zeta$ as $2$ and lower and upper cutoffs of $10^5$ and $10^7 \, M_\odot$. For the radial distribution of CCs in the CGM, we adopt a broken power-law profile to mimic the effect of galactic feedback processes on the distribution of clouds in the following form.
\begin{eqnarray}
\nonumber
    \frac{dN_{\rm CC}}{dR} & \propto R^{-\alpha_1}; R\leq R_0 \\
    & \propto R^{-\alpha_2}; R>R_0
    \label{eq:cc_real_dist}
\end{eqnarray}
The broken power-law distribution is seen in simulations (Fig. 9 in \citealt{Nelson2020} and Fig. 11 in \citealt{Ramesh2024}).
We adopt the shallower (inner regions) and steeper (outer regions) indices to be $\alpha_1=0.1$ and $\alpha_2=1$, respectively. \citealt{Morgan2025} analyzed the isolated galaxies in the {\sc TNG100} simulation and found a characteristic scale of $0.2$ times the virial radius, corresponding to the size of outflows. Motivated by this result, we choose the break point of the broken power-law to be $R_0=50$ kpc.

With these modifications, we generate CCs in the CGM so that the total CC mass reaches the total cool gas mass in the CGM, which we have taken to be $10^{10} \, M_\odot$. Note that in this `advanced' model with the fiducial parameters, we generate a larger number of CCs ($21,622$) than $10^3$, our fiducial value in the `basic' model. We also ensure that the average number density of cool gas in each CC is less than the physical density (see Eq. \ref{eq:ncool_PL}) at the radial location of the CC by regenerating the mass of the CC, where the average/global density is $3M_{\rm CC}/(4\pi R_{\rm CC}^3 \mu m_{\rm p})$. The above condition also ensures that the cool gas volume fraction within each CC is less than unity. After generating the CCs of variable size and mass, we follow similar steps as in our `basic' model to compute the observables.

\begin{figure}
    \centering
    \includegraphics[width=\columnwidth]{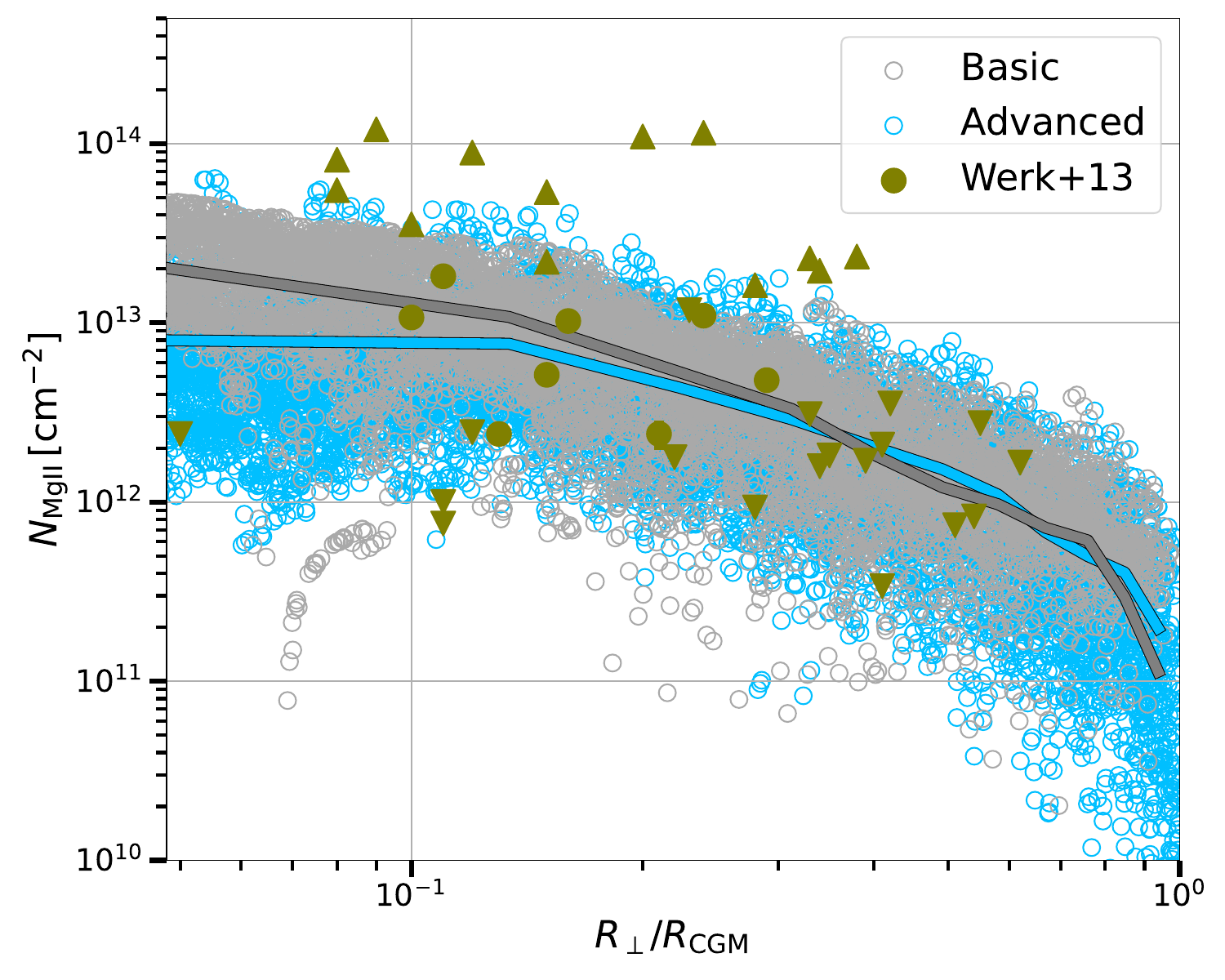}
    \caption{The grey and blue points show the MgII column density for the `basic' and `advanced' (considering the size, mass, and realistic radial distribution of CCs in the CGM) models, respectively, along a total of $10^4$ sightlines (empty sightlines are not shown). The solid lines in the respective colors show the average MgII column density. The green data points are from the COS-Halos survey (\citealt{Werk2013}). Notice a slight decrement in the average column density for the `advanced' model and therefore a larger covering fraction, since the product of average column density and covering fraction is fixed for the fixed total cool gas mass in the CGM. The scatter is larger in the case of the `advanced' model due to the size and mass variation of the CCs.}
    \label{fig:col_advanced}
\end{figure}

In Fig. \ref{fig:col_advanced}, we show the MgII column density as a function of normalized impact parameter for both `basic' (grey points) and `advanced' (blue points) models for $10^4$ sightlines (empty sightlines are not shown). The green data points are from the COS-Halos survey (\citealt{Werk2013}). The solid grey and blue lines show the average MgII column density for the `basic' and `advanced' models, respectively. 
The average column density shows a similar declining trend in both the `basic' and `advanced' models. The `advanced' model shows a larger scatter in column density values compared to the `basic' model due to variation in size and mass of CCs, which effectively generates a larger number of CCs. The overall covering fraction is also higher in the case of the `advanced' model due to a smaller size (larger number) of CCs, the overall average column density is lower, such that the product of average column density and covering fraction is constant for the fixed cool gas mass in the CGM as compared to the `basic' model.

The highest column density is also larger for the `advanced' model due to these variations. Varying the upper/lower limits and slope of the CCs' size, mass, and radial distribution has a similar impact on the column density distribution as in the case of the `basic' model (see Fig. \ref{fig:col_var}). 
The parameter variation that produces a larger number of CCs (larger index and smaller lower/upper cutoff on size and mass distribution), results in the larger covering fraction, and smaller scatter in the column density and smaller average column density value, since the area covering fraction times the average column density is constant for the fixed cool gas mass in the CGM. 
For example, changing the power-law index of mass distribution to $3/1$ from the fiducial value of $2$ results in a total of $50,789/4,165$ CCs, which results in a smaller/larger scatter and smaller/larger average MgII column density.
Similarly, increasing the broken power-law index to $2$ (outer regions) from $1$ results in a larger number of CCs in the central region and fewer in the outskirts. This results in a smaller scatter and covering fraction at inner regions and thereby a larger average column density. 
Changing the break point from $50$ kpc to $100$ kpc does not affect the distribution significantly. Although the above results are generally valid, noticeable variation in column density occurs only when there is a substantial change in these parameters.

\section{Line blending with smaller cloudlets}
\label{sec:turb}

In the last section, we analyzed the distribution of misty  CCs in the CGM. Mist limit refers to a unit area covering fraction within a CC with an enormous number ($N_{\rm cl} \rightarrow \infty$) of tiny ($r_{\rm cl} \rightarrow 0$) cloudlets. In this section, we investigate the mist limit kinematically using the MgII absorption profile by generating a large number of cloudlets (of small size) within a CC. Therefore, we look for the convergence of the MgII absorption profile from tiny cloudlets within a CC and compare it with the misty CC model in the previous section. To do so, following \citetalias{Hummels2024}, we zoom in on a single CC, generate spherical cloudlets within it, and assign velocity to each cloudlet. We assume turbulent broadening across a cloudlet (in a similar manner as done in the previous section) along with thermal broadening, even though turbulent broadening across a $\sim$ parsec size cloudlet is smaller than thermal broadening and the LOS velocity spread of cloudlets across the CC. 
We show that turbulent broadening across a CC emerges due to the blending of absorption profiles of individual cloudlets intersected along a LOS, which are typically closely spaced in the LOS velocity space. 

Instead of generating cloudlets throughout the entire CC ($M_{\rm CC}=10^7 \, M_\odot$, $R_{\rm CC} = 10$ kpc), which would yield an excessively large number of cloudlets ($\sim 10^{10}$ for $r_{\rm cl}=1$ pc), we focus on a smaller cylindrical region along the LOS to keep the total cloudlet count computationally manageable. 
We select a cylinder with a radius of $40$ pc and a height of $20$ kpc, positioning it at the center of the CC. This allows us to generate and analyze fewer cloudlets in a reduced volume.
We uniformly populate cloudlets in the cylindrical volume and ensure that the cloudlets are within the cylinder by regenerating the coordinates of the cloudlets that lie outside it.
We vary the size of spherical cloudlets with $r_{\rm cl}=10,1$, and $0.1$ pc and a fixed $n_{\rm cool}=10^{-2} \, \rm cm^{-3}$. Once we generate the coordinates of all cloudlets, we create a 3D realization of the velocity field following the Kolmogorov spectrum at the CC scale over a grid of $200^3$. Each cloudlet is assigned this background velocity field based on the nearest grid to the cloudlet. The velocity field has zero mean and standard deviation $\sigma_{\rm 3Dturb,CC} = \sigma_{\rm 3Dturb, CGM} \times (R_{\rm CC}/R_{\rm CGM})^{1/3} = 35 \rm \, km \, s^{-1}$ for $R_{\rm CC}=10$ kpc and $R_{\rm CGM}=280$ kpc with $\sigma_{\rm 3Dturb,CGM}= 107 \, \rm km \, s^{-1}$ estimated in the previous section. We then shoot a LOS through the center of the cylinder/CC and compute the number of intersected cloudlets, the respective column densities, the LOS velocities ($z$ component of velocity), and the absorption profile.

Note that we also consider turbulent broadening across a cloudlet following Section \ref{sec:turb_ansatz}, which further broadens the absorption profile compared to purely thermal broadening. Though for smaller size cloudlets ($\lesssim$ pc), this effect is negligible, and thermal broadening dominates over turbulent broadening. We assume the temperature of the cool gas to be $10^4$ K, which corresponds to $b_{\rm thermal}=2.6 \, \rm km \, s^{-1}$ for MgII. To obtain the column density of MgII from the total column density (calculated using the intercepted cloudlet length) of the individual intersected cloudlets, we adopt $f_{\rm Mg_{II}}$ of $0.2$ assuming a $0.3$ solar metallicity (see Appendix \ref{app:con_factor} and figure \ref{fig:ion_frac}).

\begin{figure}
\centering
    \includegraphics[width=\columnwidth]{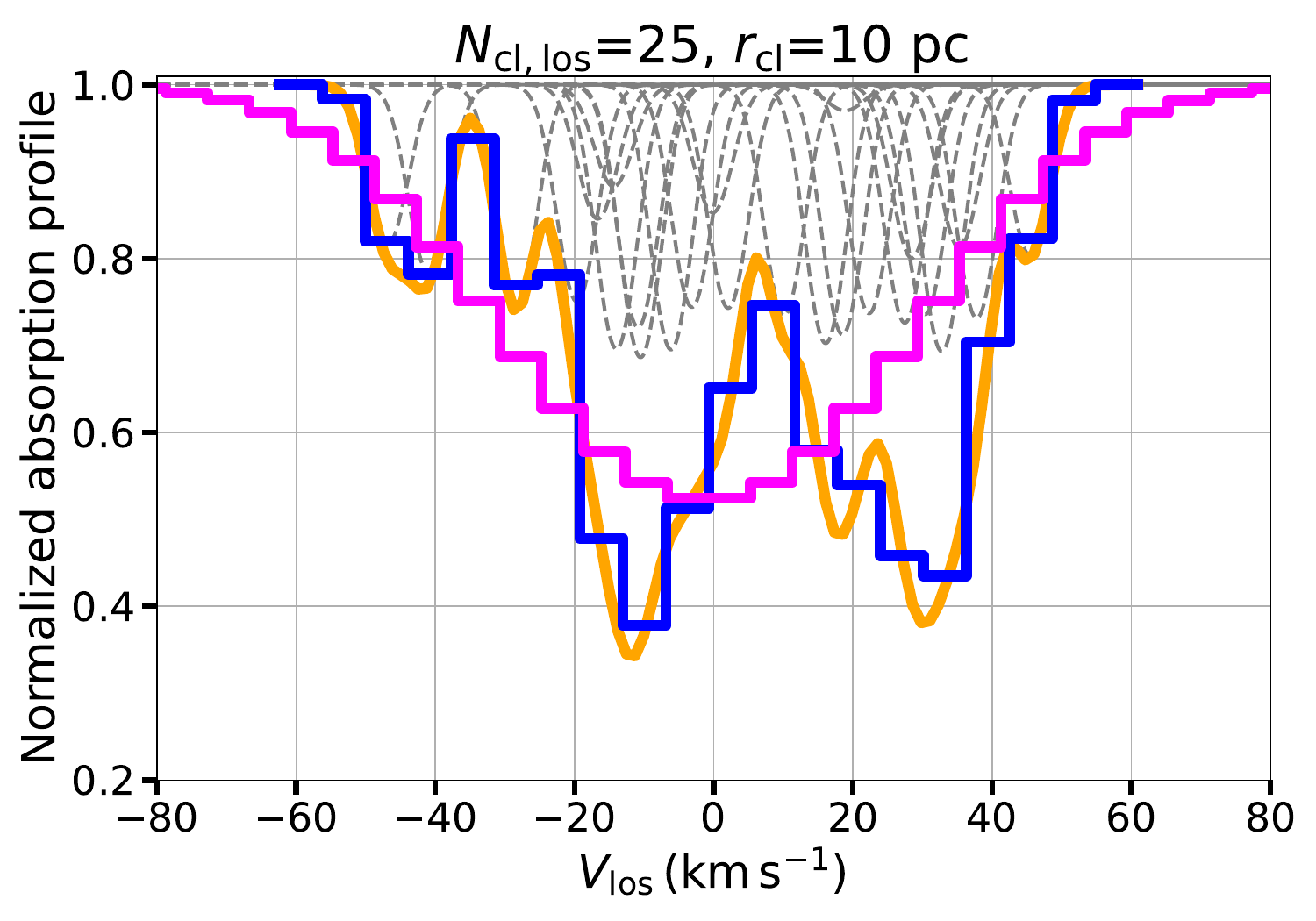}
    \includegraphics[width=\columnwidth]{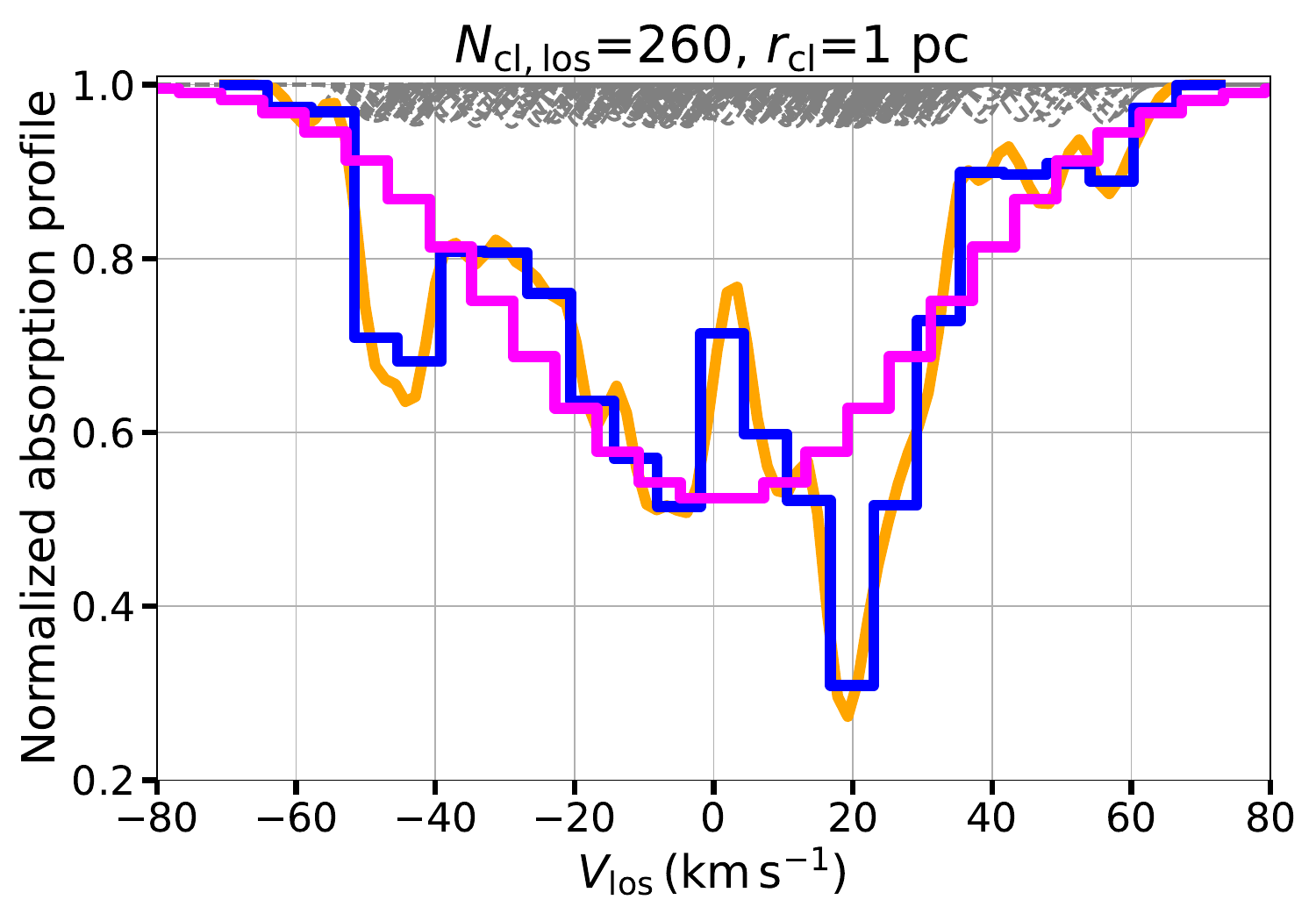}
    \includegraphics[width=\columnwidth]{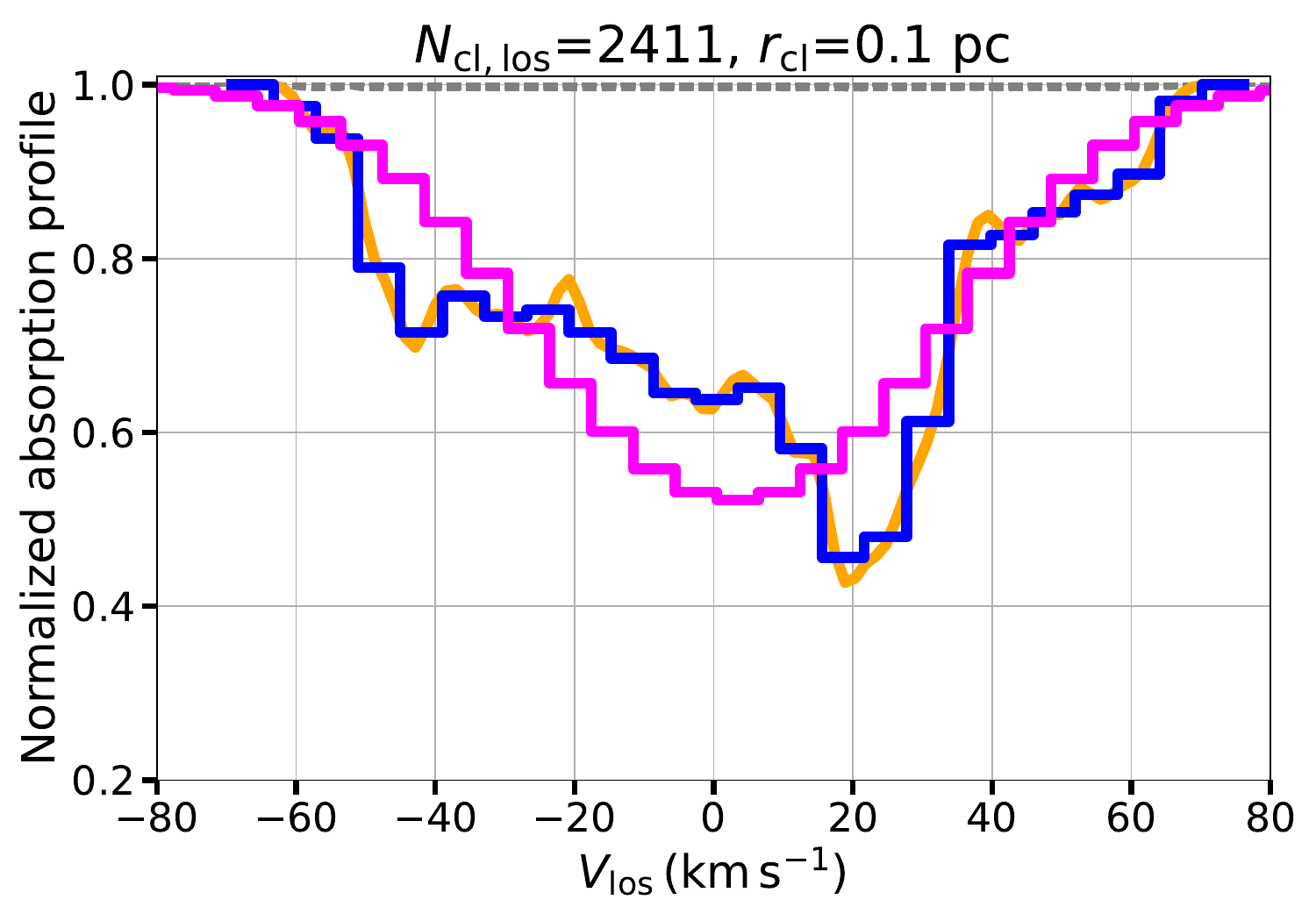}
    \caption{Normalized MgII absorption profile for increasing (decreasing) number (size) of cloudlets in a CC along a LOS. The dashed grey lines show the absorption profile from individual intersected cloudlets. The orange lines show the overall absorption profiles from the intersected cloudlets along the LOS. The blue lines show the overall absorption profile smoothed with a velocity resolution of $6 \, \rm km \, s^{-1}$. The magenta lines show the absorption profile for the misty CC model along the sightline passing through the CC center smoothed with a velocity resolution of $6 \, \rm km \, s^{-1}$ (see section \ref{sec:MCMC} on how to calculate the CC absorption profile in the mist limit).
    As the size of cloudlets decreases from $10$ to $0.1$ pc (top to bottom panels), more and more cloudlets are intersected along the LOS. With an increase in the intersected cloudlets, the overall absorption profiles become broader with fewer absorbing components. For cloudlet size $\lesssim 0.1$ pc, our misty CC model (magenta line, smoothed to a resolution of $6$ km s$^{-1}$) is roughly matching the total absorption profile.}
    \label{fig:line_blended}
\end{figure}

\begin{figure}
    \centering
    \includegraphics[width=\columnwidth]{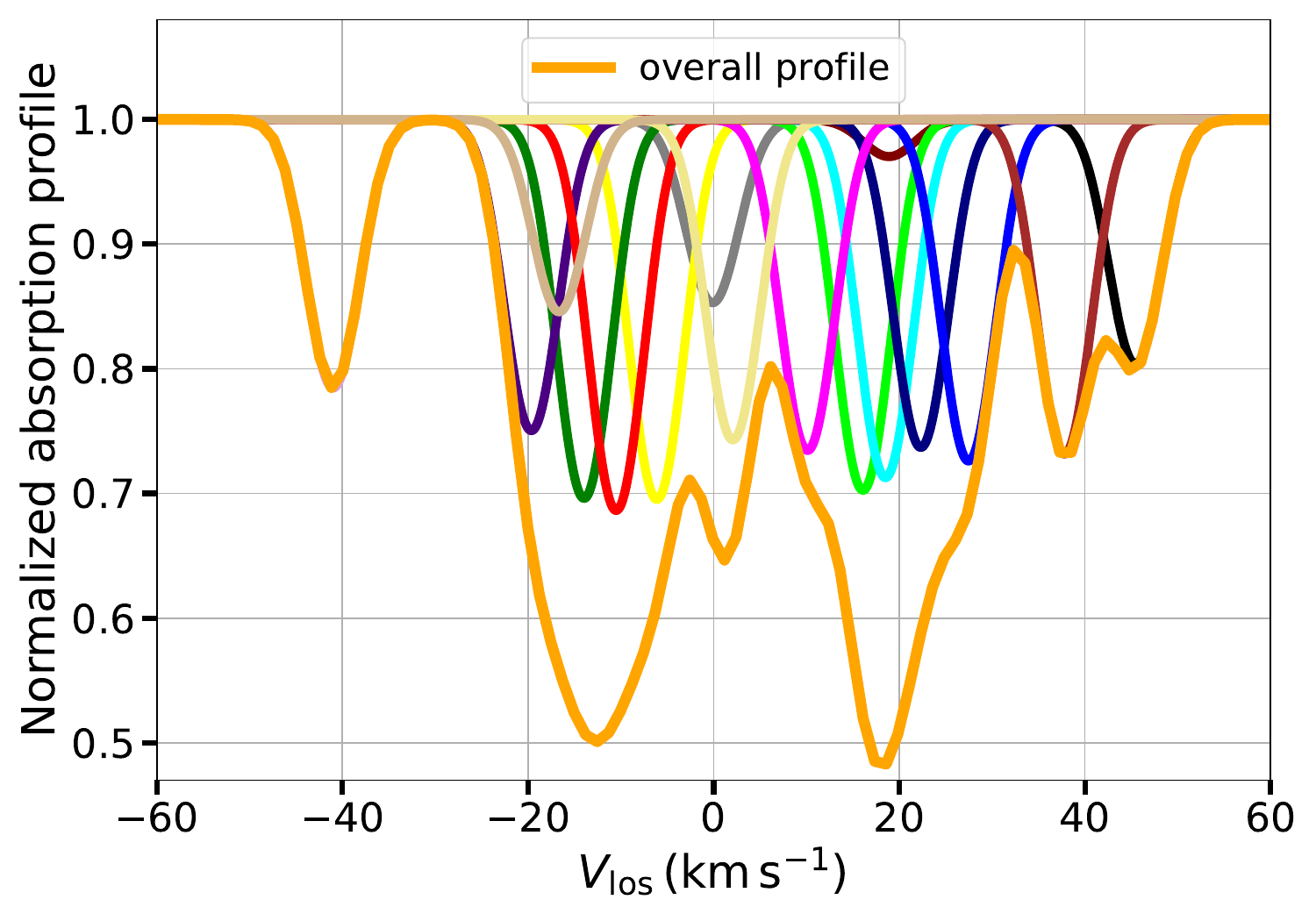}
    \includegraphics[width=\columnwidth]{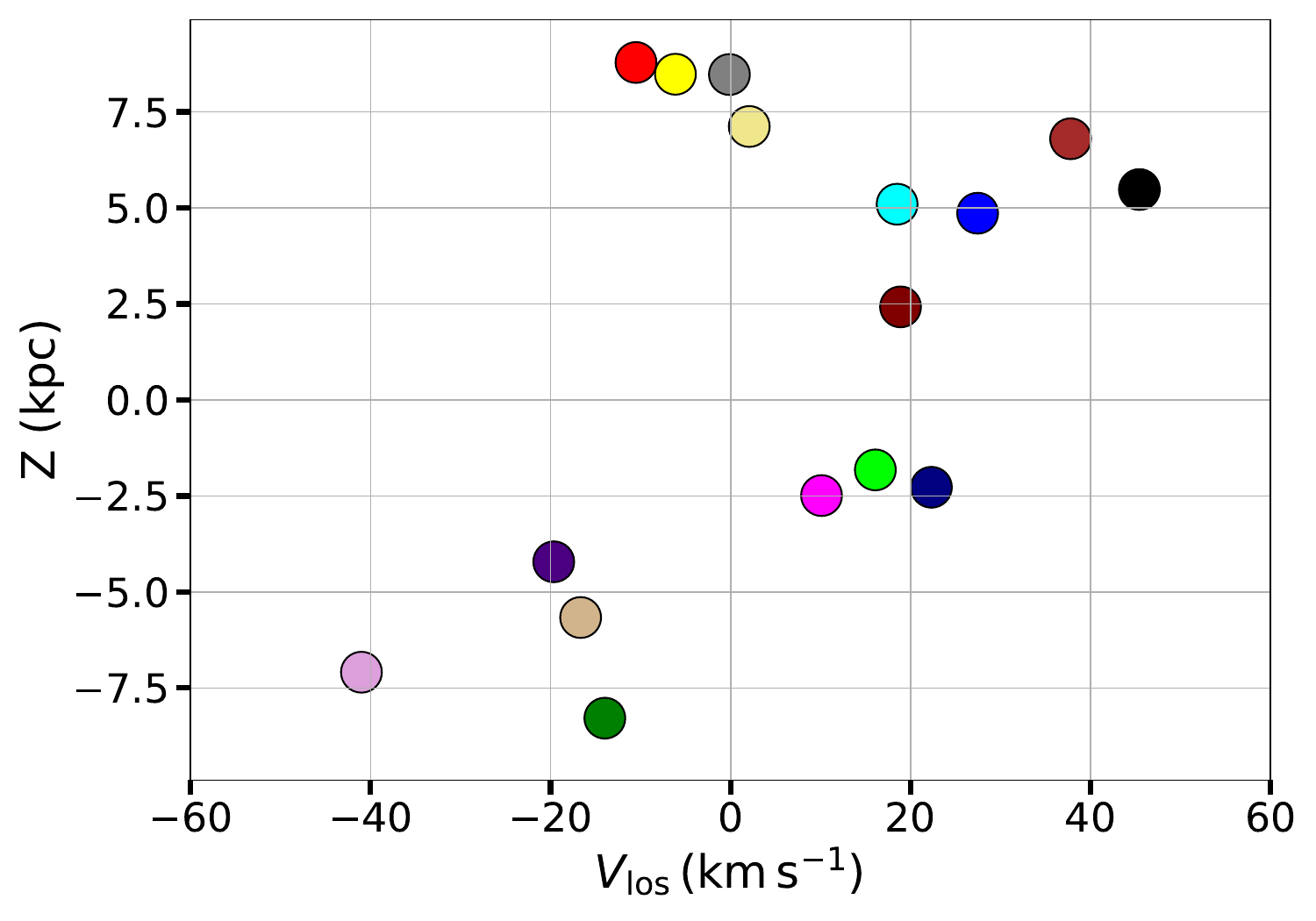}
    \caption{In the \textit{top panel}, the lines in different colors (total $16$) show the MgII absorption profile from individual cloudlets intersected along a LOS through a CC (same as top panel of Fig \ref{fig:line_blended} except that we only consider $16$ out of $25$ cloudlets for clarity). The solid orange line shows the overall/total absorption profile from all the intersected cloudlets. Due to the blending of absorption profiles, only $8$ absorbing components are seen in the overall absorption profile. In the \textit{bottom panel}, we show the $z$ coordinate and the LOS velocity of the intersected cloudlets. The color of the points corresponding to individual cloudlets is consistent with the top panel. Based on the position and velocity distribution, the cloudlets will be blended in the absorption spectrum. This illustration shows that cloudlets that are far apart in physical space may be blended due to similar LOS velocity, whereas cloudlets that are very close in physical space may have disparate LOS velocities, which makes it very difficult to infer 3D information from the 2D observations.}
    \label{fig:line_fit}
\end{figure}

Figure \ref{fig:line_blended} shows the impact of smaller cloudlets and, therefore, a larger number of cloudlets in a CC on the overall absorption profile along a LOS. 
Using total broadening (thermal and turbulent), MgII column density and LOS velocity ($z$ component of velocity) of the individual intersected cloudlets, we generate the absorption profiles of intersected cloudlets as shown with dashed grey lines in Figure \ref{fig:line_blended}. The orange line shows the total absorption profile from all the intersected cloudlets, calculated similarly as done in the previous section for misty CCs across the whole CGM.
The blue line shows the overall absorption profile smoothed with a velocity resolution of $6$ km s$^{-1}$. The magenta line shows the absorption profile from our fiducial misty CC model with $M_{\rm CC}=10^7 \, M_\odot$, $R_{\rm CC}=10$ kpc, and $n_{\rm cool}=10^{-2} \, \rm cm^{-3}$ along a LOS passing through the center of CC smoothed with a velocity resolution of $6$ km s$^{-1}$. 

As the cloudlet size decreases, more cloudlets are generated to maintain the same cool gas mass in a CC. This results in more intersected cloudlets along a LOS ($25,260$, and $2411$ cloudlets for $r_{\rm cl}=10,1$, and $0.1$ pc). The larger number of intersected cloudlets results in an overall broad absorption profile. The overall absorption profiles for $\lesssim 0.1$ pc size cloudlets (bottom panel of Fig. \ref{fig:line_blended}) roughly match the absorption profile predicted by our misty CC model (magenta line). Decreasing the radius of cloudlets further will result in lesser and fewer components in the blue absorption line and a closer match with the mist-limit. This signifies that the turbulent velocity of individual tiny cloudlets results in the turbulent broadening across a CC, well modeled by our misty CC ansatz.  

Figure \ref{fig:line_fit} shows the impact of position and LOS velocity distribution of the cloudlets on the blending of the absorption profiles along a LOS. The lines in the top panel with different colors show the individual MgII absorption profiles along a LOS similar to the top panel of Fig. \ref{fig:line_blended}. Note that we only show $16$ cloudlets out of $25$ intersected cloudlets for clarity. The solid orange line shows the overall absorption profile from all the intersected cloudlets. Depending on the LOS velocity of the intersected cloudlets, the absorption profiles may be blended. In the bottom panel, we show the $z$ coordinate of the intersected cloudlets on the y-axis and the LOS velocity on the x-axis (same as in the top panel). The color of the points in the bottom panel is consistent with the absorption profile of the cloudlets in the top panel. The cloudlet at the leftmost side on the LOS velocity space with $z \approx \hbox{-} 7.5$ kpc and $v_{\rm LOS} \approx \hbox{-} 40 \, \rm km \, s^{-1}$ is not blended with other cloudlets as can be seen with the absorption profile in the top panel.
The two cloudlets around $v_{\rm LOS} \approx -15  \, \rm km \, s^{-1}$ (green and red points in the bottom panel) are blended, as can be seen with the overlapping absorption profiles in the top panel, but are separated by $\approx 16$ kpc in physical space. This signifies that \textit{the cloudlets far away in physical space can be blended due to similar LOS velocity.} This makes it extremely challenging to extract the 3D information of the cool cloud components from 2D observational information.
In contrast, the two cloudlets, which are around $v_{\rm LOS} \approx \hbox{-} 10$ km s$^{-1}$ (red and yellow points in the bottom panel) are close in both velocity and physical space and are also blended as seen in the absorption profile in the top panel.

\section{Direct modeling of cloudlets \& CCs across CGM}
\label{sec:realistic}

In Section \ref{sec:analytical}, we assumed that CCs are misty and focused on the distribution of CCs in the CGM. 
In this section, rather than assuming the mist limit, we extend the work further and distribute the cool cloudlets explicitly within all the CCs across the entire CGM. We analyze the distribution of cloudlets within CCs and CCs across the CGM and check for the parameters that best match the observations within computational limitations. 
The exercise in section \ref{sec:turb} allowed us to investigate the impact of minuscule cloudlets on the observed absorption spectrum. Because of the computational cost, such small cloudlets cannot be populated across the entire CGM. This section takes an approach similar to {\it CloudFlex} (\citetalias{Hummels2024}) and similarly cannot easily reach the mist limit because of computational limitations. In fact, our approach in this section is even more expensive because the distribution functions of various observables from a single CC were combined analytically in \citetalias{Hummels2024}, but here we also treat the distribution of CCs from first principles. The range of investigations in this work allows us to understand the relations between these various related approaches, namely, {\it mCGM}, {\it mCC}, and {\it CloudFlex}.

\begin{table*}
\centering
{\renewcommand{\arraystretch}{1.2}
\begin{tabular}{|c|c|c|c|c|}
\hline
   & Parameter & Fiducial Value  & Tested Range & Description    \\ 
   \hline
   
\multirow{2}{*}{CGM}    & $ R_{\rm CGM}$ (kpc)    & $280$  & \hbox{--} & Radius of the CGM     \\ \rule{0pt}{3ex}

& $ M_{\rm cool}$ ($\rm M_{\odot}$)   & $10^{10}$ & $10^9$ \hbox{--} $10^{11}$  & Total cool gas mass in CGM   \\  
\hline
\multirow{4}{*}{cloud complex} & $ N_{\rm CC}$     & $10^3$  & $10^2$ \hbox{--} $10^4$ & Number of CCs  \\ \rule{0pt}{3ex}

  & $ R_{\rm min}$ (kpc)   & $10$ & \hbox{--} & Minimum distance of CC center from galactic center    \\ \rule{0pt}{3ex}
  
  & $ R_{\rm max}$ (kpc)   & $280$  & \hbox{--} & Maximum distance of CC center from galactic center   \\ \rule{0pt}{3ex}
  
 & $\alpha$ & $1$ & $0$ \hbox{--} $2$ & Power-law index of radial distance of CC center   \\ \hline \rule{0pt}{3ex}
 
\multirow{6}{*}{cloudlet} & $ r_{\rm min}$ (kpc)  & $0.1$ & \hbox{--} & Minimum distance of cloudlet center from CC center  \\ \rule{0pt}{3ex}

   & $ r_{\rm max}$ (kpc) & $10$  & $1$ \hbox{--} $20$ & Maximum distance of cloudlet center from CC center \\ \rule{0pt}{3ex}
   
   & $\beta$  & $0$ & $0$ \hbox{--} $2$  & Power-law index of radial distance of cloudlet from CC center \\ \rule{0pt}{3ex}
   
 & $ a_{\rm min}$ (kpc) & $0.01$ & $0.001$ \hbox{--} $0.1$  & Minimum semi-axes length of a cloudlet \\ \rule{0pt}{3ex}
 
 & $ a_{\rm max}$ (kpc) & $0.5$  & $0.1$ \hbox{--} $1$ & Maximum semi-axes length of a cloudlet \\ \rule{0pt}{3ex}
 
  & $\gamma$  & $1$  & $0$ \hbox{--} $2$   & Power-law index of semi-axes length of a cloudlet  \\ \hline 
  
\end{tabular}}
\caption{Model parameters, their fiducial values, tested range, and description for direct modeling of cloudlets in CCs spread throughout the CGM.}
\label{tab:params}
\end{table*}

\begin{figure}
    \centering
    \includegraphics[width=\columnwidth]{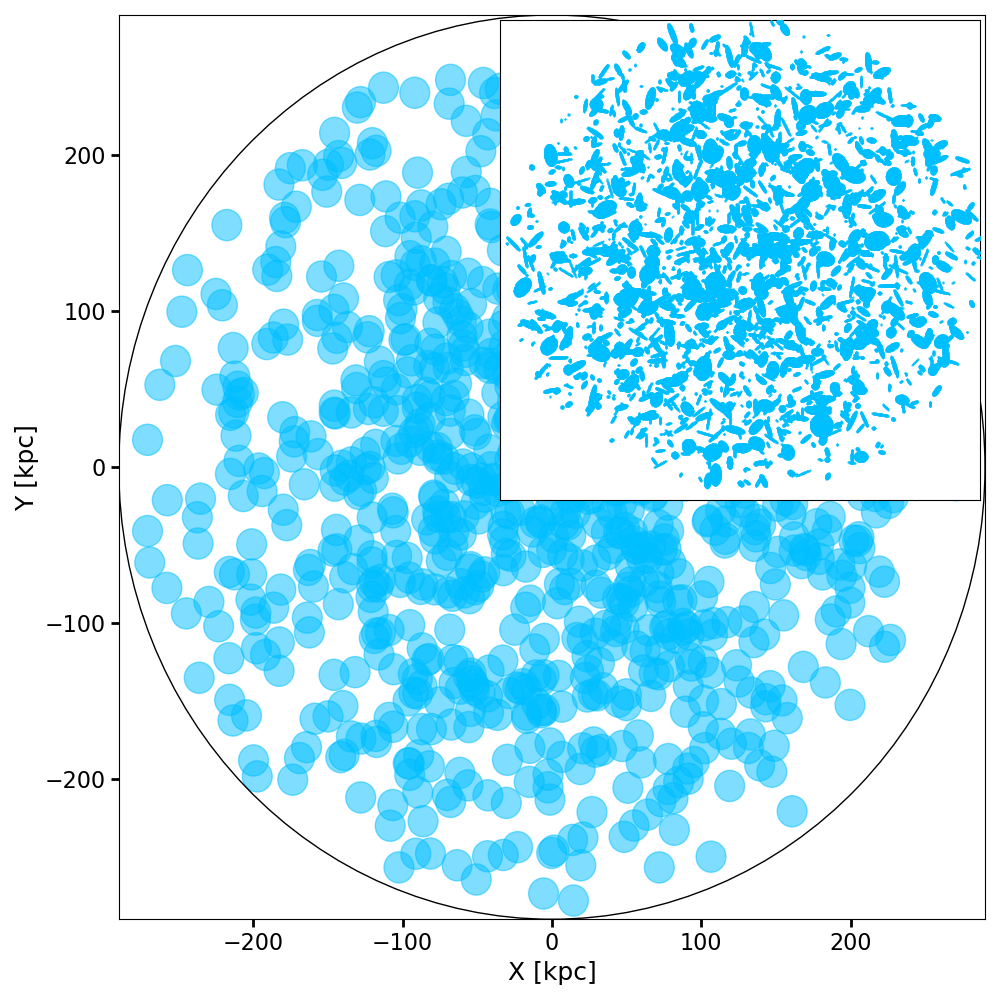}
    \caption{LOS projected distribution of CCs and cloudlets in the CGM for $M_{\rm cool}=10^{10} \, \rm  M_\odot$, $N_{\rm cc}=10^3$, and $\alpha=1$. Due to the power-law nature of the distribution of CCs and the projection effect, more CCs are found in the central region than in the outskirts. The inset shows the cloudlets within one of a CC for $\beta=0$ and $\gamma=1$ (fiducial parameters; see Table \ref{tab:params}).  In this fiducial case, $\approx 5 \times 10^8$ total cloudlets are generated, with $\approx 5\times 10^5$ cloudlets per CC. Note that we only show $2\times 10^3$ cloudlets in the inset to better highlight the variety of cloudlets with different sizes and orientations. Also, we do not explicitly show the distribution of cloudlets within all the CCs. The distribution of these cloudlets shows qualitative similarity to Figure 7 of \citealt{Ramesh2024} that shows a snapshot from a high-resolution simulation of a Milky Way-like CGM.}
    \label{fig:cgm_vis}
\end{figure}

\textit{Generation of CCs across the CGM:} We first generate the centers of  $N_{\rm CC}$ number of CCs with a power-law distribution in 
$R$ with an index $\alpha$, as done in Section \ref{sec:analytical} (see Eq. \ref{eq:pl_dist}). To determine the 3D coordinates of the centers of each CC, we follow the procedure described in section \ref{sec:MCMC}. The lower and upper limits of radial (galactocentric) distances of the centers of these CCs are $R_{\rm min}=10$ kpc and $R_{\rm max}=280$ kpc, respectively. For the fiducial case, we adopt $N_{\rm CC}=10^3$ and $\alpha=1$ (as in our fiducial \emph{mCC} model), but also check the variation with $N_{\rm CC}=10^{2,4}$ and $\alpha=0,2$ (see Table \ref{tab:params}). 

\textit{Location of Cloudlets:} After generating all the CCs, we populate cool cloudlets in each of them until the total cloudlet mass reaches the total cool gas mass $M_{\rm cool}$ (our fiducial value is $10^{10} M_\odot$). As in our {\it mCC} model, we follow the same cool gas number density profile (see section \ref{sec:cool_density})\footnote{We use the physical or the local density in this section. The average or the global density concept does not apply here due to the non-mist limit.} To obtain the mass of each cloudlet and the column density. We use a power-law distribution of index $\beta$ for the distribution of cloudlets within a CC as a function of radial distance from the center of each CC. This is similar to what we did for distributing CCs across the CGM. The lower and upper limits of distances of the centers of cloudlets from a CC center are $r_{\rm min}$ and $r_{\rm max}$, respectively. To determine the 3D location of the centers of each cloudlet, we follow the same procedure as adopted for determining the centers of CCs in the CGM. For the fiducial case, we adopt $M_{\rm cool}=10^{10} \, \rm M_{\odot}$, $\beta=0$, $r_{\rm max}=10$ kpc, $r_{\rm min}=0.1$ kpc. We also check the variation in observables due to variation in these parameters (see Table \ref{tab:params}).

\textit{Shape and size of cloudlets:} We assume the cloudlets to be ellipsoidal since they are expected to be non-spherical in general. The ellipsoidal cloudlets have three semi-axes and three rotation angles as parameters. We generate the semi-axes from a power-law distribution with index $\gamma$. The lower and upper limits of the semi-axes are $a_{\rm min}$ and $a_{\rm max}$ respectively. The three rotation angles are generated from a uniform distribution between $0$ and $2\pi$. For our fiducial case, we adopt $\gamma=1$, $a_{\rm min}=0.01$ kpc, and $a_{\rm max}=0.5$ kpc. We vary these parameters to check their impact on the observables.

We assign 3D velocity to each cloudlet following the same method as done for misty CCs in section \ref{sec:analytical}. The velocity fields are generated on a $600^3$ grid across the entire CGM. Table \ref{tab:params} lists all the parameters, their fiducial values, and the variations that we tried. 
We allow cloudlets to overlap as ensuring non-overlapping cloudlets is very expensive, and it has a minimal impact on the results (see section 2.2.1 in \citetalias{Hummels2024}). The total number of cloudlets generated will depend on various parameters. Decreasing $a_{\rm min}$, $a_{\rm max}$, and increasing $\gamma$, $M_{\rm cool}$, while holding other parameters fixed, will result in more cloudlets.

Once we generate all the CCs and cloudlets, we shoot $10^4$ sightlines (uniformly spaced in $\log_{10} R_{\perp}$) into the CGM. 
Fig. \ref{fig:cgm_vis} shows the LOS projected distribution of cool gas cloudlets in the CGM for the fiducial parameters. The circular regions show the CCs within which ellipsoidal cloudlets are distributed. There are more CCs in the central region than in the outskirts due to the power-law distribution in $r$ ($\alpha=1$) and the projection effect.
The inset shows the LOS projected distribution of cloudlets in one of the CCs. We only show $2\times 10^3$ cloudlets (total $\approx 5 \times 10^5$ cloudlets per CC) to better highlight the range in sizes and orientations of the ellipsoidal cloudlets.

Next, we calculate the number of cloudlets intersected and the intersected lengths along all the $10^4$ LOSs. We then calculate MgII column density using the length intersected through individual cloudlets along a sightline.
To determine the total MgII EW along a sightline, we first generate the total absorption profile due to all the intersected cloudlets along a LOS.  
We then calculate the area under the total absorption profile as the total EW along a LOS (as done in section \ref{sec:analytical} for misty CC). Note that simply adding EWs from individual intersected cloudlets results in a higher total EW. We assume the temperature of the cool gas to be $10^4$ K, corresponding to the Doppler $b$ parameter of $2.6$ km s$^{-1}$ for MgII. In this section, we do not consider turbulent broadening across a cloudlet. The turbulent broadening across the $\sim$ pc scale cloudlet is typically less than or equal to thermal broadening, so it does not significantly affect our EW and covering fraction results.

\subsection{Comparing with observations}

\begin{figure*}
    \centering
    \includegraphics[width=\textwidth]{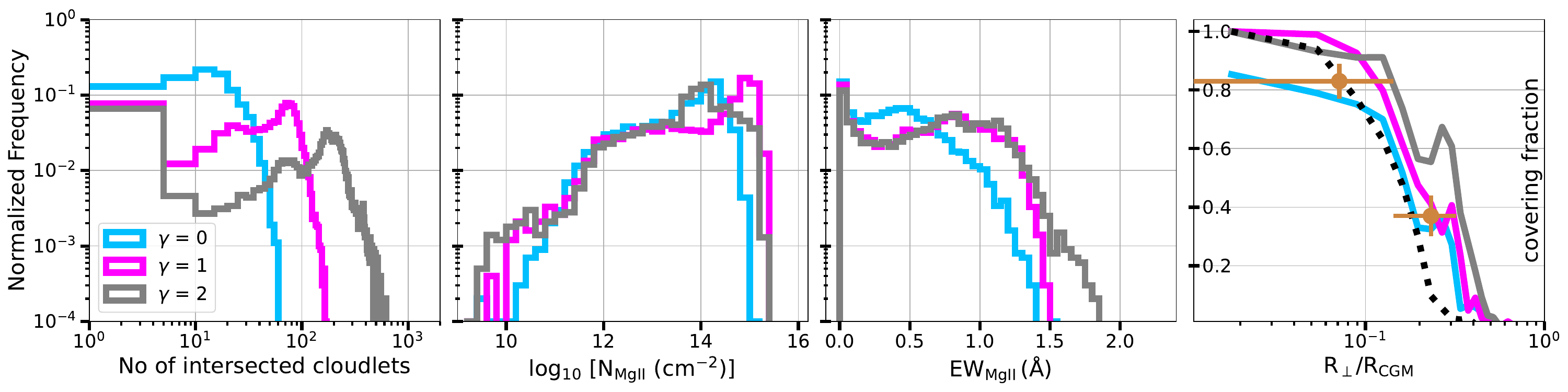}
    \includegraphics[width=\textwidth]{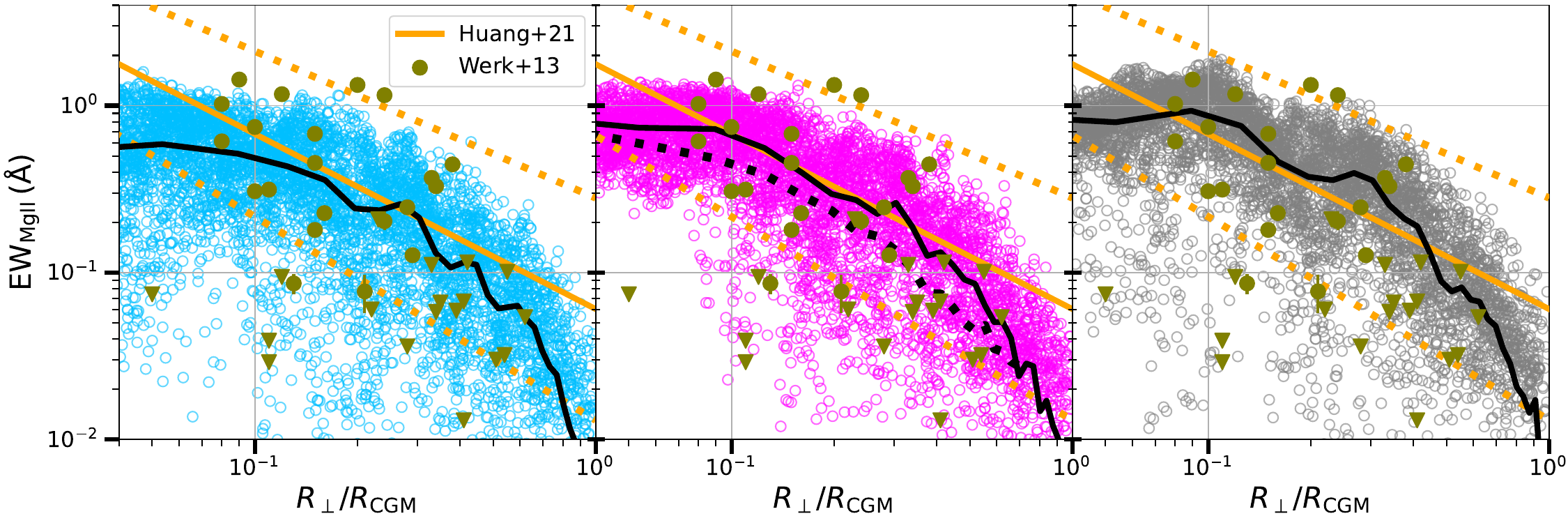}
    \caption{The \textit{top panel} shows the normalized histogram of the number of intersected cloudlets, $\log_{10} N_{\rm MgII}$, and MgII EW along $10^4$ LOSs (empty sightlines are not shown for number of intersected cloudlets and column density although they are taken into account for normalization). The blue, magenta, and grey colors correspond to $\gamma=0,1,2$, the power-law index of the semi-axes distribution of ellipsoidal cloudlets. Note that the larger the $\gamma$, the larger the number of cloudlets, keeping other parameters fixed. The total number of cloudlets generated for $\gamma=0,1,2$ is $\sim 5 \times 10^6, 5 \times 10^7, 10^9$ respectively. The rightmost panel shows the MgII covering fraction as a function of normalized impact parameter for an EW threshold of $0.3$ \r{A}. The data points are from \citealt{Huang2021} while the dotted black line is from our misty CC model (see Fig. \ref{fig:cov_eq}) with fiducial parameters. In the \textit{bottom panel}, the colored scatter points show the total EW along all the LOSs as a function of the normalized impact parameter for the different values of $\gamma$, respectively. The color of the points is consistent with the top panel. The solid black lines show the mean EW, while the dotted black line in the middle panel shows the mean EW for our fiducial misty CC model (see top panel of Fig. \ref{fig:cov_eq}). The solid orange and dashed lines show the best-fit relation and $1\sigma$ uncertainty from \citealt{Huang2021} while data points in green are from the COS-Halos survey (\citealt{Werk2013}). The observed EW trend and covering fraction match better for the $\gamma=1$ case.}
    \label{fig:fid_run}
\end{figure*}

\begin{figure*}
    \centering
    \includegraphics[width=\textwidth]{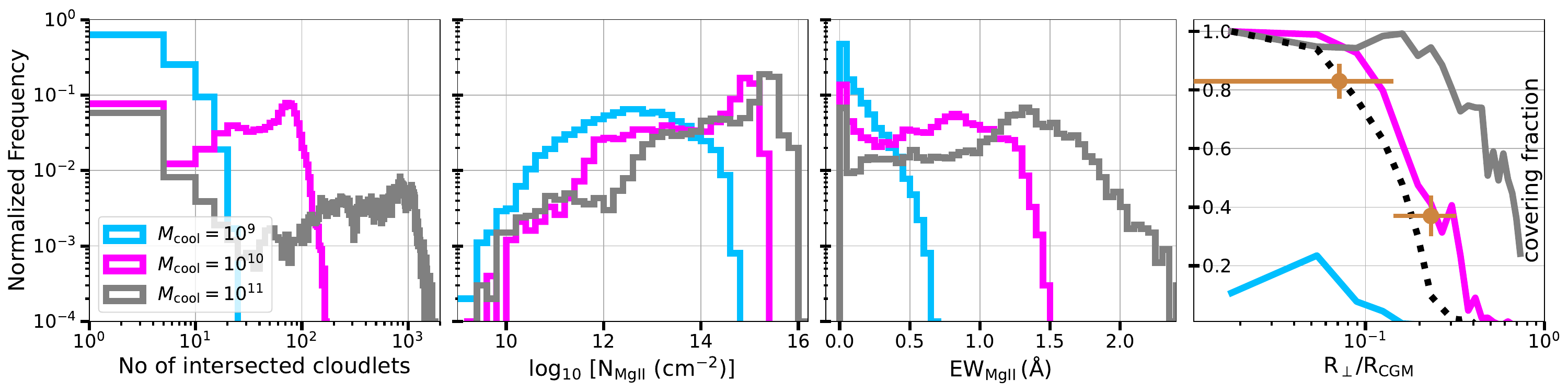}
    \includegraphics[width=\textwidth]{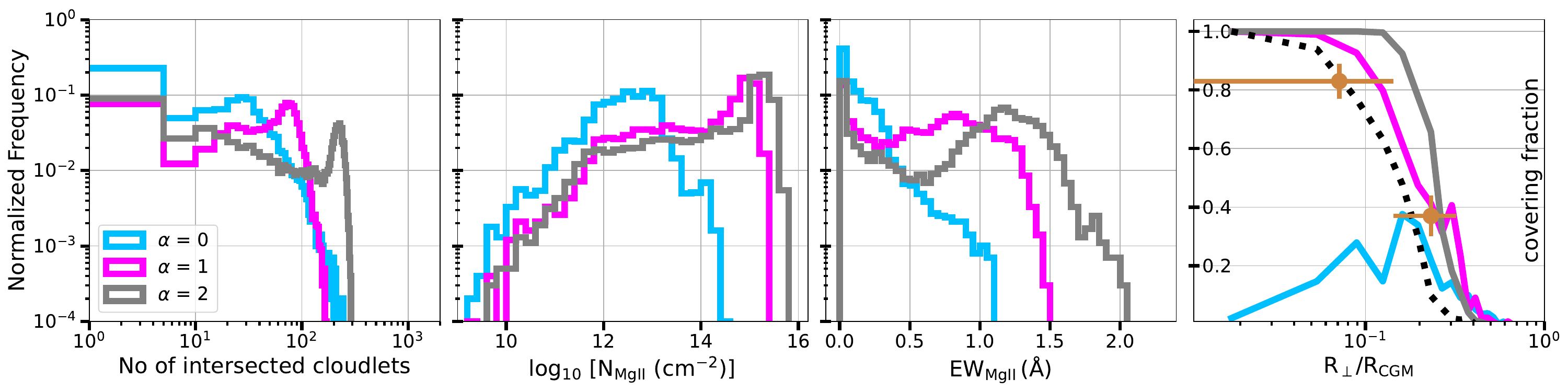}
    \includegraphics[width=\textwidth]{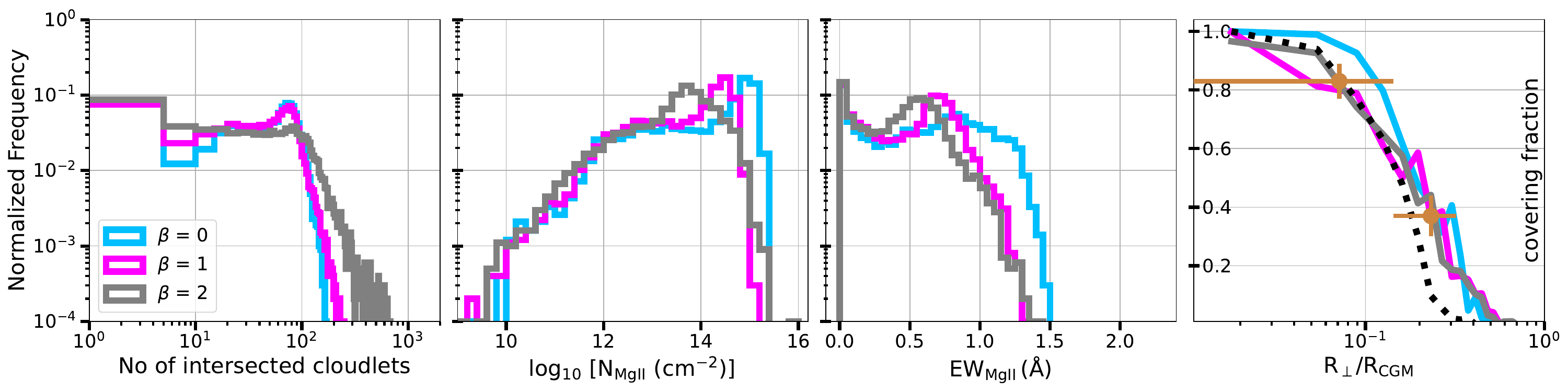}
    \includegraphics[width=\textwidth]{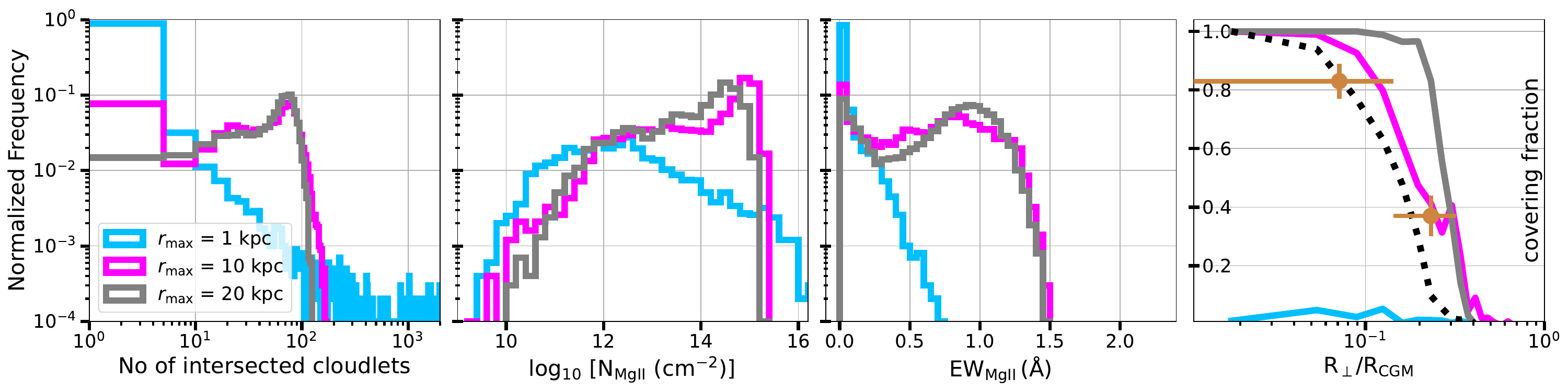}
    \includegraphics[width=\textwidth]{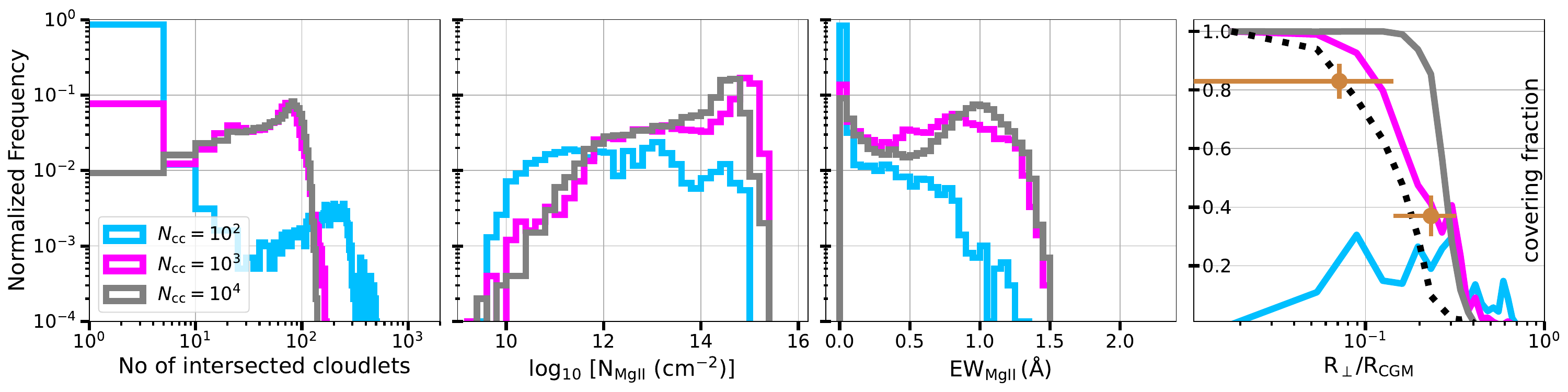}
    \caption{This figure shows the histograms of the number of intersected cloudlets,  MgII column density, MgII EW, and MgII covering fraction (same as the top left panels in Figure \ref{fig:fid_run}) to show variation with different parameters. We vary a single set of parameters for each row and hold the other parameters fixed to their fiducial values.}
    \label{fig:param1_var}
\end{figure*}

Figure \ref{fig:fid_run} shows the various observable statistics with the variation in $\gamma$, the power-law index of cloudlets' semi-axis length distribution (size distribution). We fix the other parameters to the fiducial values (see Table \ref{tab:params}). In the \textit{top panel}, we show the normalized histogram of the number of intersected cloudlets, $\log_{10} N_{\rm MgII}$, and $\rm EW_{MgII}$ along $10^4$ LOSs (empty sightlines are not shown for number of intersected cloudlets and column density although they are considered in normalization). Note that we quote total column density and EW along a LOS. The blue, magenta, and grey lines show the variation with $\gamma=0,1,2$, respectively. As $\gamma$ increases, the size distribution becomes steeper and therefore a larger number of cloudlets (smaller in size) are generated to keep the same cool gas mass. The total number of cloudlets generated for $\gamma=0,1,2$ is $\sim 5\times 10^6, 5\times 10^7, 10^9$ respectively. Due to the larger number of cloudlets, the number of intersected cloudlets is also high (top left panel), which results in a higher column density and thereby a higher equivalent width. The maximum EW is higher for higher values of $\gamma$, because adding EWs of a larger number of kinematic components results in larger EW along a LOS due to the flat part of the curve of growth (Fig. \ref{fig:eq_col_den}). The rightmost top panel shows the MgII covering fraction as a function of the normalized impact parameter for an EW threshold of $0.3$ \r{A}. The data points are from \citealt{Huang2021}. The dotted black line shows the covering fraction from the misty CC model (section \ref{sec:analytical}; see Fig. \ref{fig:cov_eq}). The covering fraction increases with increasing $\gamma$ (larger number of cloudlets) as EW is larger for larger values of $\gamma$ (see EW distribution).

In the \textit{bottom panels} of Fig. \ref{fig:fid_run}, we show the MgII EW distribution as a function of the normalized impact parameter for $10^4$ LOSs. The scatter points are the total EW values along each LOS (empty sightlines are not shown). The blue (left panel), magenta (middle panel), and grey (right panel) colors show the EW values for $\gamma=0,1,2$, respectively. 
The green data points are from the COS-Halos survey (\citealt{Werk2013}). The solid and dotted orange lines show the best-fit EW-impact parameter relation and $1\sigma$ scatter of the observed data from \citealt{Huang2021}.
The solid black line shows the mean EW profile in each panel, while the dotted black line in the middle panel shows the average EW value from the fiducial misty CC model. As we see from left to right, the mean EW increases with increasing $\gamma$ (larger number of cloudlets).
It shows that $\gamma=1$ with other parameters set to their fiducial values matches the observed range. 
Note that the $\gamma=1$ model with other parameters fixed to their fiducial value generates $\approx 5\times 10^7$ cloudlets while the $\gamma=2$ case generates $\sim10^9$ cloudlets, which is computationally expensive, and therefore we choose $\gamma=1$ as our fiducial value even though $\gamma=2$ also closely matches the observations. The fiducial number of CCs is $10^3$, which means that each CC has $\approx 5 \times 10^4$ cloudlets. This is less than the number of cloudlets $\sim 10^7$ generated in a single CC for achieving the mist limit (especially in velocity space; see Fig. \ref{fig:line_blended}) in section \ref{sec:turb}. Therefore, the cloudlets are not completely in the mist limit within a CC for these fiducial parameters.
To attain the mist limit within each $10^3$ CC, one needs to generate a total of $\sim 10^{10}$ cloudlets in the CGM, which is computationally prohibitive. Therefore, we are limited to generating and analyzing $\approx 10^8$ cloudlets in the entire CGM, which corresponds to $\sim 10^5$ cloudlets in each CC. 
Although our CCs are not entirely in the mist limit in this section for most parameter combinations in Table \ref{tab:params}, some of the key observables like the distribution and the covering fraction appear reasonably converged (see Fig. \ref{fig:fid_run}).

In Figure \ref{fig:param1_var}, we show the results similar to the top panels of Figure \ref{fig:fid_run} by varying cool gas mass ($M_{\rm cool}$), power-law index of radial distribution of CCs ($\alpha$), power-law index of distribution of cloudlets within CC ($\beta$), the size of CC ($r_{\rm max}$), and number of CCs ($N_{\rm CC}$). While showing the variation with a parameter, we fix the other parameters to their fiducial values. 

The first row in Fig. \ref{fig:param1_var} shows the variation with $M_{\rm cool}=10^9$ \hbox{--} $10^{11} \, M_\odot$. As the cool gas mass increases, the number of cloudlets increases, leading to more intersected cloudlets along LOSs, which in turn raises the MgII column density, EW, and covering fraction. The dotted black line shows the covering fraction from the misty CC model. The covering fraction values are better matched with the $10^{10} M_\odot$ model. The second row shows the variation with $\alpha$, the power-law index of the radial distribution of CCs. The number of cloudlets is the same for all the values of $\alpha$. As $\alpha$ increases, the CCs become more concentrated toward the inner regions of the CGM. This leads to a higher number of intersected cloudlets along the sightlines in the center, resulting in a larger column density, EW, and covering fraction. Our fiducial case with $\alpha=1$ (magenta line) matches the covering fraction better than other values of $\alpha$. 

The third row in Fig. \ref{fig:param1_var} shows the variation with $\beta$, the power-law index for the radial distribution of cloudlets within a CC. Even in this case, the number of cloudlets is the same. A larger value of $\beta$ suggests that the cloudlets are more concentrated towards the center of a CC. This results in more intersected cloudlets along a LOS passing close to the center of the CC. The column density does not show significant variation with $\beta$, while EW decreases with increasing $\beta$ due to the close packing of cloudlets in velocity space. The blending of absorption lines from cloudlets results in fewer kinematic components (see section \ref{sec:turb}), which results in a lower EW than having more kinematic components. The covering fraction values predicted from all values of $\beta$ are nearly the same. The fourth row shows the variation with $r_{\rm max}$, the CC size. For larger CC, the cloudlets are more dispersed, while for smaller $r_{\rm max}$, the cloudlets are tightly packed in a CC. The smaller CC results in a higher number of intersected cloudlets and a larger column density. The EW shows an opposite trend with the size of CC, as cloudlets are closely packed in velocity space for a much smaller size CC, which results in smaller EW values due to fewer kinematic components along a LOS, as discussed previously. The column density and EW values are very similar for $10$ and $20$ kpc CCs, whereas the covering fraction better matches the $10$ kpc CC model. 

The fifth row shows the variation with the number of CCs ($N_{\rm CC}$). The total number of cloudlets is roughly the same for all three cases, with a larger number of cloudlets per CC in the case of a smaller number of CCs and vice versa. The number of cloudlets intersected is larger for a smaller number of CCs, while the column density and EW are larger for a larger number of CCs due to the higher probability of intersection of more CCs along LOSs. However, the covering fraction is well matched by $10^3$ CCs in the CGM. Encouragingly, on comparing with observations, the fiducial parameters from the \emph{mCC} model and first principles modeling of cloudlets and CCs are the same (namely, $R_{\rm CC} = 10 $ kpc, $N_{\rm CC} = 10^3$, $\alpha = 1$, $M_{\rm cool} = 10^{10} M_\odot$).

\section{Discussion}
\label{sec:discussion}

Now we discuss the various implications of our models in modeling and understanding the multiphase CGM revealed by multi-wavelength observations. Our models and other similar approaches (e.g., \citetalias{Dutta2024, Hummels2024}, \citealt{Yang2025}) for the cool CGM go beyond the traditional hydrostatic models describing the hot CGM. The primary aim of these models is to provide robust physical insights rather than precisely matching the quasar absorption data (which are still biased by spectra quality and analysis methods).

\subsection{Utility of CC-like approaches to model multiphase CGM}

While an extreme idealization, CCs are useful building blocks of a multiphase CGM. Unlike the complete mist limit, CCs can give a non-unity covering fraction of cool and warm gas as probed by quasar sightlines. Moreover, the statistics of column densities and their covering fractions can constrain the mass of the cool gas, typical sizes of CCs, their numbers, and their occurrence with the distance from the center.

Most importantly, since the number of cloudlets in a CC is expected to be much larger than the number of CCs filling the CGM, it is computationally efficient to model CCs analytically in the mist limit and then populate their Monte Carlo realization to compare the post-processed column density and equivalent width statistics with observational inferences.

Our work demonstrates that the observationally inferred distribution of MgII column density and equivalent width can be reproduced by the misty CC model with several cloudlets along each LOS across a single CC. This translates into a cloudlet size of $\lesssim$ pc, which is beyond the reach of even the highest resolution galaxy formation simulations (e.g., see \citealt{Ramesh2024}). A small cloudlet size is also required for the cool CGM gas, in the form of cloudlets, to be suspended in long-lived CCs (see section \ref{sec:vterm}); large enough cloudlets will simply precipitate towards the galactic center.

\subsubsection{Comparison with \citetalias{Dutta2024}}
A large variation in EW and column density at the same distance from the galactic center implies that the misty cool gas does not fill the entire CGM uniformly, as assumed by \citetalias{Dutta2024}. A more accurate description is in terms of misty CCs that fill the CGM sporadically. Such a model can naturally explain the large variation in column density along different sightlines (see Fig. \ref{fig:col_var}). Unlike \citetalias{Dutta2024}, here we explore the distribution of direct observables such as EWs. With different levels of modeling of CCs and cloudlets within them, we explore the connections between the {\it misty CGM} model and the more computationally intensive models like {\it CloudFlex}.

\subsubsection{Comparison with \citetalias{Hummels2024}}

\citetalias{Hummels2024} computes the statistical properties of CCs from first principles by populating them with cloudlets with a range of masses (assuming a power-law distribution in the cloudlet mass, $dN/dm_{\rm cl} \propto m_{\rm cl}^{-\delta}$). For a constant density cloudlet, this corresponds to a size distribution of $dN/dr_{\rm cl} \propto r_{\rm cl}^{-3 \delta + 2}$ ($r_{\rm cl}^{-4}$ for their fiducial $\delta=2$; corresponds to many more small cloudlets compared to the big ones). The mass/volume distribution of cloudlets $dV/dr_{\rm cl} \propto r_{\rm cl}^3 dN/dr_{\rm cl} \propto r_{\rm cl}^{5 - 3 \delta}$; the area distribution $dA/dr_{\rm cl} \propto r_{\rm cl}^2 dN/dr_{\rm cl} \propto r_{\rm cl}^{4 - 3 \delta}$. Similarly, the column density distribution $dN_{\rm cool}/dr_{\rm cl} \propto r_{\rm cl} dN/dr_{\rm cl}$. For \citetalias{Hummels2024}'s fiducial value of $\delta = 2$, the mass/volume, area, and column densities (even more so than mass) are all dominated by the smallest cloudlets. For power-law distributions, the average properties are typically dominated by either the higher or the lower (as in the present case) cutoff. This justifies our misty-CC model in which CCs are made up of infinitesimal cloudlets (section \ref{sec:analytical}) that provide a computationally inexpensive way to produce observables from just thousands of CCs that can be compared efficiently with observations.

\citetalias{Hummels2024} carried out a comprehensive study of the variation of various parameters of their model of cloudlets within CCs. However, they were limited by computational cost in generating an enormous number of small cloudlets for their models approaching the mist limit. They found that the cool gas mass, the smallest cloudlet mass, the cool gas density, and the CC size most affected the column density and EW distributions. The variation of other parameters within a reasonable range, such as the cloudlet size distribution, their radial distribution within CC, and turbulent velocity parameters, had a relatively minor impact. Their comprehensive exploration allows us to focus on the complementary aspects in our work. In addition to the misty CC model (section \ref{sec:analytical}), we study the observational properties of CCs populated with tiny (up to 0.1 pc) cloudlets and their implications for the realistic modeling of the multiphase CGM. In particular, we study how the cloudlets within a CC start to overlap increasingly in velocity space with a decreasing cloudlet size (section \ref{sec:turb}). This further motivates our misty CC model, in which we approximate the spread in LOS velocity of numerous cloudlets with a single turbulent broadening parameter (see Fig. \ref{fig:line_blended}). Further, we model ellipsoidal (as opposed to spherical in \textit{CloudFlex}) cloudlets from first principles and populate the entire CGM with these. This approach is even more expensive than \citetalias{Hummels2024}'s approach but produces results largely consistent with them.

\subsection{Modeling intermediate temperature gas}
\label{sec:intermediate}

The primary focus of this paper is on modeling the cool gas. In principle, the same setup can also be applied to the intermediate temperature (warm) gas with few modifications. The warm gas is also expected to reside in clumpy structures like CCs, but with a larger volume fraction as compared to cool gas because of its lower (physical) density. To model the warm gas, we consider the misty cool CCs in the CGM. Each cool CC is assumed to have not only cool cloudlets but also warm and hot gas inside it. The warm gas is expected to envelop the cloudlets within a CC (see Fig. 8 in \citealt{Armillotta2017}). The cool cloudlet volume fraction within a CC is small ($\sim 0.016$; see Eq. \ref{eq:fV_cloudlet}). Being 10-30 times lower in density, the warm gas (traced by OVI) will occupy a larger volume by a similar factor, assuming a similar mass in warm and cool gas (a similar cool and warm gas mass is suggested by Table 3 in \citetalias{Dutta2024}). This implies a volume fraction of the warm phase in a CC $\sim 20-50$\%; the majority of volume, even within a CC, is in the hot phase.
We assume our CCs to be misty even in the warm phase with the same size as cool CCs and the same total cool and warm gas masses, shoot lines of sight through the CGM, and compute the column density of OVI using the photo (PIE) + collisional (CIE) ionization equilibrium OVI fraction at $10^{5.5}$ K and redshift of $0.2$ (the fraction of OVI is roughly constant $\approx 0.24$ over wide a density range).
Note that OVI is a good tracer of $\sim 10^{5.5}$ K warm gas in the CGM under CIE. There can, however, be a significant amount of OVI produced by the lower temperature gas ($\lesssim 10^5$ K) under PIE depending on the strength of the radiation field (\citealt{Werk2016}).

\begin{figure}
    \centering
    \includegraphics[width=\columnwidth]{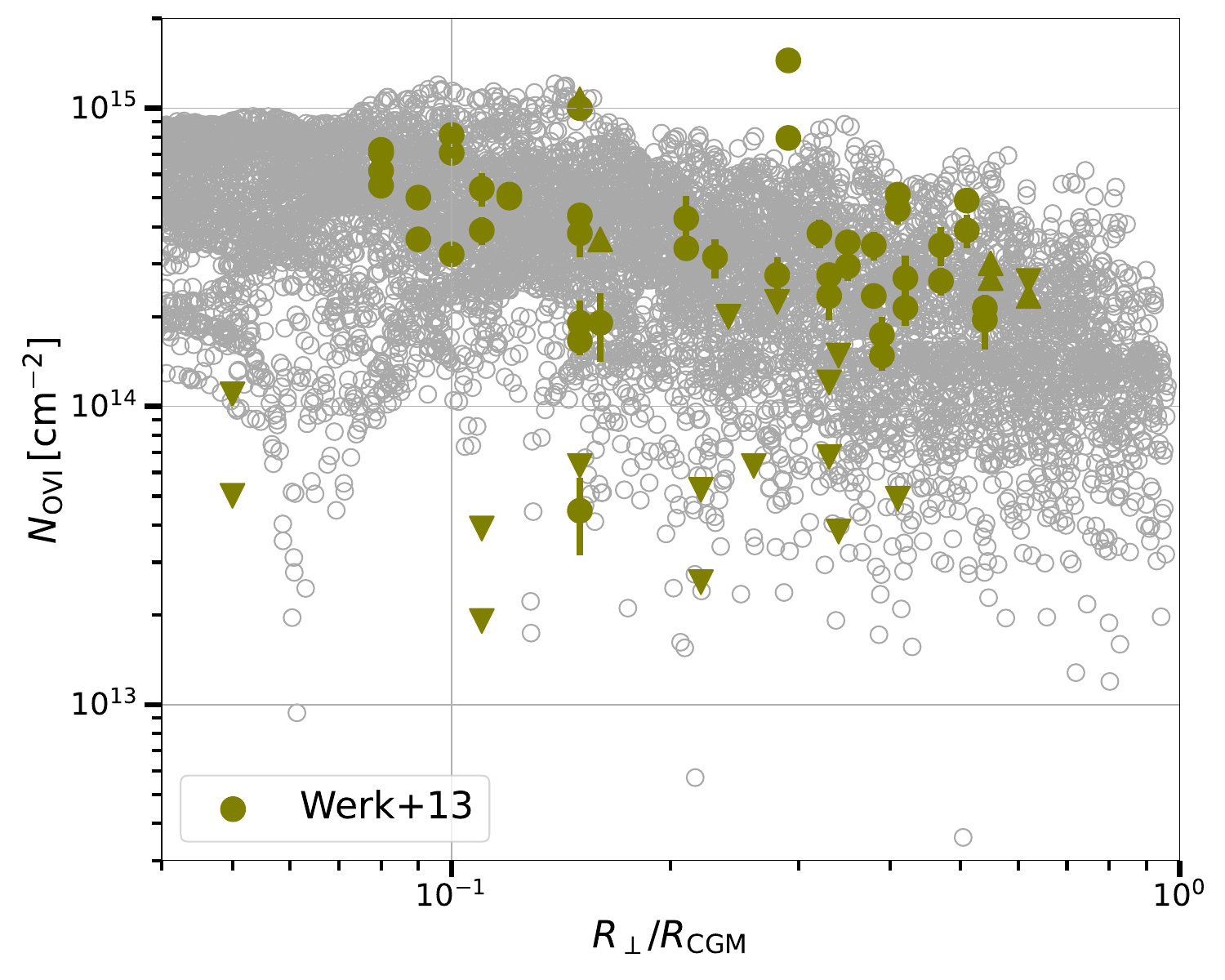}
    \caption{Column density distribution of OVI along $10^4$ sightlines, as predicted by our fiducial CC model with a simple prescription for warm gas (see section \ref{sec:intermediate}). Moreover, we assume the OVI ion fraction from photo+collisional ionization equilibrium at $10^{5.5}$ K, and assume similar cool and warm gas masses. The observed data points are from the COS-Halos (\citealt{Werk2013}) survey. The spread of OVI columns from our Monte Carlo model matches well with the observed spread.}
    \label{fig:col_OVI}
\end{figure}

Figure \ref{fig:col_OVI} shows the column density distribution of OVI tracing the warm ($10^{5.5}$ K) gas along $10^4$ sightlines (empty sightlines are not shown). The observational data points in green color are from the COS-Halos (\citealt{Werk2013}) survey. Predictions from our simple Misty CC model agree well with the observed distribution of OVI column density, including the upper and lower limits. This simple exercise shows the predictive power, flexibility, and computational efficiency of our Misty CC model for studying the multiphase CGM. Just as the combination of the covering fraction and the average column density provides an unbiased estimate of the cool CGM mass, the same observables for OVI can be applied easily to estimate the warm CGM mass.
However, warm gas may not always be associated with cool clouds, and there can be extended structures possibly due to the various formation channels like cosmological accretion (\citealt{McQuinn2018}), and galactic processes (\citealt{Shapiro1976}), etc. Therefore, this simple modeling of warm gas in the form of spherical cloud complexes does not reflect the expected diverse morphology of warm gas in the CGM.

\subsection{Astrophysical Implications}

The majority of the baryonic mass in the universe is in the diffuse intergalactic and circumgalactic medium. While the IGM dominates the global baryonic mass budget of the local Universe, even within galactic halos, the majority of baryons reside in the diffuse CGM and not in the dense interstellar medium and stars. Being major matter reservoirs, the CGM controls star formation in galaxies across cosmological timescales.

Fundamental questions in CGM research are regarding the fraction of baryons in the CGM as a function of halo mass, its spatial extent, and the distribution of the CGM mass across different temperature phases. While direct observations of the volume-filling hot CGM are rare (being volume-filling and close to hydrostatic, the hot CGM is easier to model), the cool and warm phases are routinely probed through quasar absorption lines. However, since cool and warm CGM phases are not volume-filling and occur in the form of clouds, the inferred column densities show a large scatter even in a uniform galaxy sample, which prevents us from drawing robust conclusions about their mass budget.

We show that a large variation in the column density of the cool/warm ions is natural in a CGM in which these phases are not spread uniformly throughout the CGM, but are instead confined to thousands of $\sim 10$ kpc CCs spread sporadically throughout the CGM (see Figs. \ref{fig:col_var}, \& \ref{fig:col_OVI}). Moreover, the average column density and the area covering fraction in a uniform galaxy+quasar sample can constrain the mass of the cool and warm phases of the CGM, as illustrated in Fig. \ref{fig:col_cov_frac}.

\subsubsection{Inferring cool CGM mass from quasar absorption}

One of the ultimate goals of our (and similar) models is to obtain a robust estimate of the cool/warm CGM mass from observations. Fig. \ref{fig:col_cov_frac} shows that a combination of the area-averaged MgII column density and the area covering fraction provides a robust proxy for the mass in cool CGM. Both ion column density and covering fraction can be inferred from observational samples of galaxies probed by quasar sightlines. However, the primary observable in these studies is the transmitted flux spectrum, which at high resolution can have many components separated by small centroid velocity shifts for the same sightline. Each of these resolved components (depending on the spectral resolution of the spectrograph) can be modeled as a misty and turbulent CC along the LOS, each composed of numerous cloudlets. Physically distinct CCs may coincide in velocity space, and similarly, cloudlets within a CC may be distinct in velocity (e.g., Fig. \ref{fig:line_fit}). Since the column density essentially counts all the ions along the LOS, this complication does not affect the cool CGM mass estimate. 

There are different approaches to analyzing absorption spectra. The most reliable technique with high spectral resolution ($\lesssim 10$ km s$^{-1}$) is to fit the absorption profile with a number of absorption components with Voigt profiles. The best fit Voigt profile can give us the turbulent broadening parameter $b_{\rm turb}$ (particularly in the presence of multiple ions tracing the same phase that coincide kinematically) and column density for each component. We can add the column densities of all components along the LOS to obtain the total column density along it. For low-resolution spectra, the Voigt profile fitting of components fails, and most studies employ the AOD (apparent optical depth) method (\citealt{Savage1991}). Under this, we obtain the optical depth spectrum and convert it into a column density spectrum (assuming that the absorption lines are not saturated). The column density spectrum is integrated to give the total ion column along the LOS. This popular method may underestimate the column density because of saturation and unresolved substructure. Section \ref{sec:turb} essentially shows that both Voigt profile fitting and even fitting a single turbulent broadened misty CC model (implying that AOD method will also work well; large cloudlets without much internal turbulence can become saturated, reducing the accuracy of the AOD/\emph{mCC} method) give a good estimate of the true column density of cloudlets along a LOS even with extreme blending in velocity due to thousands of cloudlets along the LOS.

Fig. \ref{fig:col_cov_frac}, which plots the covering fraction versus the average column density, shows that a combination of these two observables provides a good estimate of the cool CGM mass. The column density along different sightlines can give us the average column density that should be plotted on the y-axis of Fig. \ref{fig:col_cov_frac}.
All spectrographs have a finite sensitivity and will inevitably miss out on weak absorption components, giving us a lower limit on the LOS column density. Similarly, saturated lines can give us a lower limit on the column density (e.g., see COS-Halos markers in Fig. \ref{fig:variation_an}). We calculate the covering fraction on the x-axis of Fig. \ref{fig:col_cov_frac} as the fraction of LOSs with EW larger than a small enough value (say $0.3$ \r{A}) so that the bulk of the EW distribution is sampled by observations. If these conditions are satisfied, Fig. \ref{fig:col_cov_frac} provides a robust way to estimate the cool gas mass in the diffuse CGM.

\subsubsection{Quasar absorption lines versus other CGM probes}

Emission is weighted toward the highest density in the CGM, so it cannot typically probe the diffuse outer CGM (except through stacking; e.g., \citealt{Zhang2019}), which holds most of the CGM mass (e.g., \citealt{Nielsen2024}). Thus, emission gives a biased estimate of the CGM mass. FRBs are sensitive to the total number of electrons along the LOS, including the contribution of IGM, FRB host galaxy, and the Milky Way (e.g., \citealt{Macquart2020,Connor2025}), and the CGM contribution is typically subdominant. Moreover, while FRBs may be an unbiased probe of total CGM mass, one cannot separate the contribution from different temperature phases. 

Similarly, the thermal Sunyaev-Zeldovich (tSZ) effect probes the integrated pressure along the LOS and, like FRBs, cannot distinguish between different CGM phases. Moreover, the signal is too small to be detected in individual galaxies and requires stacking of similar mass galaxies (e.g., \citealt{Singh2015,Bregman2022,Das2023}). The kinetic SZ signal is proportional to the product of the LOS velocity of the galaxy halo (determining which requires large galaxy surveys) and the electron column density, but stacking is necessary to detect the signal averaged over the whole CGM (e.g., see \citealt{Schaan2021}). 

Therefore, quasar absorption line studies are the only means to study the various phases of the CGM. 
With an increasingly large number of background quasar-galaxy pairs, improving spatial and spectral resolution, and availability of a range of ions tracing a range of temperatures, quasar absorption lines have a unique position as the probe of the multiphase CGM. However, the modeling of the cool/warm phase is nontrivial, and more accurate models, going beyond the models presented in this work, are necessary to derive unbiased physical properties from observations.

Carefully combining the results from different observational probes from the CGM while being mindful of biases inherent in different techniques is the best way of obtaining robust physical constraints on the physical properties of the CGM. For this, we need robust phenomenological models of the multiphase CGM that can make testable predictions. While simulations of the CGM are steadily increasing in resolution and the included physics (\citealt{Ramesh2024}), we are still some way off from getting robust predictions from them because of insufficient spatial/mass resolution (we need a resolution better than a few pc in the cool phase to obtain converged results for MgII EW distribution; e.g., see section \ref{sec:turb} \& Figs. \ref{fig:line_blended} \& \ref{fig:fid_run}) and because of ambiguities related to feedback and sub-grid physics. The gap between expensive cosmological simulations and high-resolution observations can be filled by novel phenomenological models such as ours.

\subsection{Caveats \& future directions} 

The observational distribution of EWs and column densities of cool ion tracers in the CGM suggests a patchy distribution of cool gas in cloud complexes rather than uniformly throughout the CGM, as assumed in \citetalias{Dutta2024}. But our assumption of spherical CCs with a constant cloudlet size, a uniform distribution of cloudlets within a CC, a fixed turbulent broadening across a CC across all sightlines, etc., are extreme idealization. 
In our `advanced' model, we have varied the size, mass, and radial distribution of CCs and shown that the trends are the same for both models. We have also considered the ellipsoidal cloudlets to roughly mimic the true shape of cloudlets in the CGM in Sec. \ref{sec:realistic}. The distribution of cloudlets within a CC does not affect the results significantly (third row in Fig. \ref{fig:param1_var}). Consideration of the broadening based on the intersected length across a CC also does not significantly impact the EW distribution and covering fraction, since broadening is proportional to the cube root of the length scale.
However, the column density and EW distribution with a range of CC models (e.g., see Fig. \ref{fig:col_var},\ref{fig:cov_eq}) follow observations, highlighting the utility and computational tractability of the misty CC model. We assumed spherical CCs, but they are likely to have a head-tail structure. Moreover, with a range of cloudlets, the large ones may not be dragged enough (e.g., see Eq. \ref{eq:cloudlet_vt}) to remain in a turbulent CC but can instead precipitate towards the center replenishing the ISM, as required by the star formation histories over cosmological timescales (e.g., see \citealt{Bera2023}). 

Our misty CC model can easily include an intrinsic scatter (\citealt{Yang2025}) within a CC as well for different phases. Moreover, similar to \citetalias{Dutta2024}, which populates the whole CGM with a log-normal temperature (or 2-D density temperature) distribution, we can easily extend our misty CCs to have a range of temperatures and densities (which can also be adjusted according to the ambient pressure profile). These temperature and density distributions and the multiphase turbulence properties of CCs can be calibrated with detailed simulations of multiphase CGM patches (e.g., \citealt{Mohapatra2022b}). These models can also accommodate warm/cool gas within CCs that is mainly supported by nonthermal pressure of cosmic rays or magnetic fields, as suggested by some simulations (e.g., \citealt{Sharma2010,Nelson2020,Butsky2020}). The biggest advantages of the \emph{mCC} model presented in section \ref{sec:analytical} are their flexibility and much lower computational cost relative to models that model the cloudlets from first principles (e.g., \citetalias{Hummels2024}, our models in section \ref{sec:realistic}).

Misty CCs may also provide a useful recipe for modeling unresolved multiphase gas phenomenologically in galaxy simulations that typically lack sufficient resolution to resolve realistic CGM cloudlets (e.g., see \citealt{Weinberger2023,Butsky2024} for recent works along these lines).
We can populate CCs aware of their origins in different parts of the CGM, such as the high-pressure galactic outflows, low-metallicity IGM filaments, galactic fountains, and high-velocity clouds, etc., and produce observables to match more sophisticated observational samples that test anisotropies of CGM absorbers relative to the galaxy's minor axis and distance from the galactic center. The low computational cost, realism, and flexibility of our \emph{mCC} model allow us to adapt it to produce observable distributions for a range of model variations.

\section{Summary}
\label{sec:summary}
Here we provide a summary of the salient aspects of this paper.

\begin{enumerate}

\item We present the misty cloud-complex ({\it mCC}) model to understand and match the observed distribution of column densities, equivalent widths (EWs), and covering fractions of cool ions such as MgII (Fig. \ref{fig:col_var}, \ref{fig:cov_eq}) and warm ions like OVI (Fig. \ref{fig:col_OVI}). This model further expands and explores the connections between the misty CGM models of \citealt{Dutta2024} and the {\it CloudFlex} model presented by \citealt{Hummels2024}. The former model assumes that the cool/warm gas is uniformly spread throughout the CGM, producing a smooth variation of cool gas column density. The {\it CloudFlex} model assumes the CGM cool gas to be spread across cloud complexes (CCs) and quantifies the distribution of column densities and equivalent widths across sightlines through these CCs, which are populated by $\sim 10^6$ cloudlets (it is computationally difficult to increase the number of cloudlets per CC much beyond this). Like \textit{CloudFlex}, our misty CC model also assumes the cool gas to be spread over CCs but models the cool cloudlets analytically in the mist limit (analogous to \citetalias{Dutta2024}). This makes our model analytically tractable and computationally inexpensive compared to {\it CloudFlex} (which has a hard time reaching the mist limit due to computational limitations). With only $\sim 10^3$ CCs spread over the CGM, our {\it mCC} model can satisfactorily reproduce the observational spread of MgII EWs and other observables (e.g., see Figs. \ref{fig:col_var} \& \ref{fig:cov_eq}). 

\item Beyond our `basic' misty CC model, where we assume fixed CC size and mass, we consider more realistic mass, size, and radial distribution of CCs in the CGM in our `advanced' model, motivated by observations and simulations. Both models show similar qualitative features, but the scatter in the column density is larger for the `advanced' model due to size and mass variation of CCs, which results in a larger number of CCs than in the `basic' model.

\item We demonstrate that the combination of the average CGM column density and covering fraction provides a robust estimate of the cool CGM mass (see Fig. \ref{fig:col_cov_frac}). The observed distribution of the inferred MgII column density and EW from the COS-Halos survey (\citealt{Werk2013}; Milky Way-like galaxies) suggests a cool CGM mass $\sim 10^{10} M_\odot$ (see Fig. \ref{fig:col_var}, \ref{fig:cov_eq}). Further, about $\sim 10^3$ CCs each of $\sim 10$ kpc seems to match the observed distribution of MgII column density and equivalent width well. Encouragingly, these parameters are consistent with our mCC model and with the first principles modeling of cloudlets across the CGM.

\item We quantify the mist limit within a CC for the parameters quantifying a CC (see Appendix \ref{sec:mist_limit}): the volume fraction of cool gas within CC ($f_V$) and the number of cloudlets ($N_{\rm cl}$).
We find that the mist limit, for which a typical sightline encounters at least one cloudlet, starts to be approached for $N_{\rm cl} \gtrsim f_V^{-2}$. Further, we show that $\sim$ pc scale cloudlets within CCs produce extreme blending in absorption spectra, but the total absorption spectrum requires only a few components that can be well fit by the {\it mCC} ansatz for the scaling of turbulent broadening with the CC size (see Fig. \ref{fig:line_blended}) We also show that it is impossible to unambiguously infer the 3D information of cool cloud components from absorption spectra (see Fig. \ref{fig:line_fit}).

\item In section \ref{sec:realistic}, we directly generate cloudlets in all CCs spread throughout the CGM. We confirm that it is computationally challenging to approach the mist limit for a small number ($\lesssim 10^6$) of cloudlets within a CC. The observed MgII EW distribution and covering fraction (COS-Halos sample; \citealt{Werk2013}) match with our direct modeling of cloudlets for $\sim 10^3$ CCs with a total cool CGM mass of $\sim 10^{10} M_\odot$ and a typical size $\sim 10$ kpc, distributed radially with a power-law with index $\alpha=1$ (see Fig. \ref{fig:fid_run} and \ref{fig:param1_var}).
 
\end{enumerate}

To conclude, this paper bridges the gap between quasar absorption studies, simpler phenomenological modeling of cool CGM, and the first-principles cosmological galaxy formation simulations that typically severely lack the resolution to resolve cool cloudlets. The flexibility and ease in creating various observables from our simple, robust, but realistic model will help improve the modeling of the cool CGM. 

\section{Acknowledgements}
We thank the anonymous referees for their valuable comments and suggestions, which have greatly enhanced both the content and presentation of this paper.
We also thank Jasjeet Singh Bagla, Rajeshwari Dutta, Cameron Hummels, and Yakov Faerman for the useful discussions. This work was performed in part at the Aspen Center for Physics (ACP), which is supported by National Science Foundation grant PHY-2210452 and a grant from the Simons Foundation (1161654, Troyer). PS and AD acknowledge the lively discussions at ACP (in particular with Yakov Faerman and Cameron Hummels) that shaped some of the key ideas presented in this paper. Research of AD was supported by the Prime Minister's Research Fellowship (PMRF) from the Ministry of Education (MoE), Govt. of India. 

\section{Data Availability}
\label{sec:data_avail}

We have made all the codes and data used in this paper public in the {\it GitHub} repo \href{https://github.com/mukeshrri/mCC}{{\tt mCC}}\footnote{\href{https://github.com/mukeshrri/mCC}{https://github.com/mukeshrri/mCC}}. 
Any other data associated with this article is available upon reasonable request to the corresponding author.



\bibliographystyle{mnras}
\bibliography{ref} 


\appendix

\section{The mist limit}
\label{sec:mist_limit}
It is important to recall the meaning of the mist limit within a CC. Here we extend the discussion in section 4.2 of \citetalias{Dutta2024}, which uses cubical cloudlets (geometry only introduces order unity factors without altering the scaling relations with the parameters) of uniform size. The parameters for such a CC, also assumed to be a cubical, are the volume filling fraction $f_V$, the number of cloudlets $N_{\rm cl}$, and the CC size $L$. The column density of the cool gas in the mist limit is given by $n_{\rm cool} L f_V$, where $n_{\rm cool}$ is the physical number density of the cool gas. The area covering fraction is unity provided the average number of cloudlets along a sightline $f_V^{2/3} N_{\rm cl}^{1/3}$ exceeds unity, that is, when $N_{\rm cl} \gtrsim f_V^{-2}$. 

\begin{figure}
    \centering
    \includegraphics[width=\columnwidth]{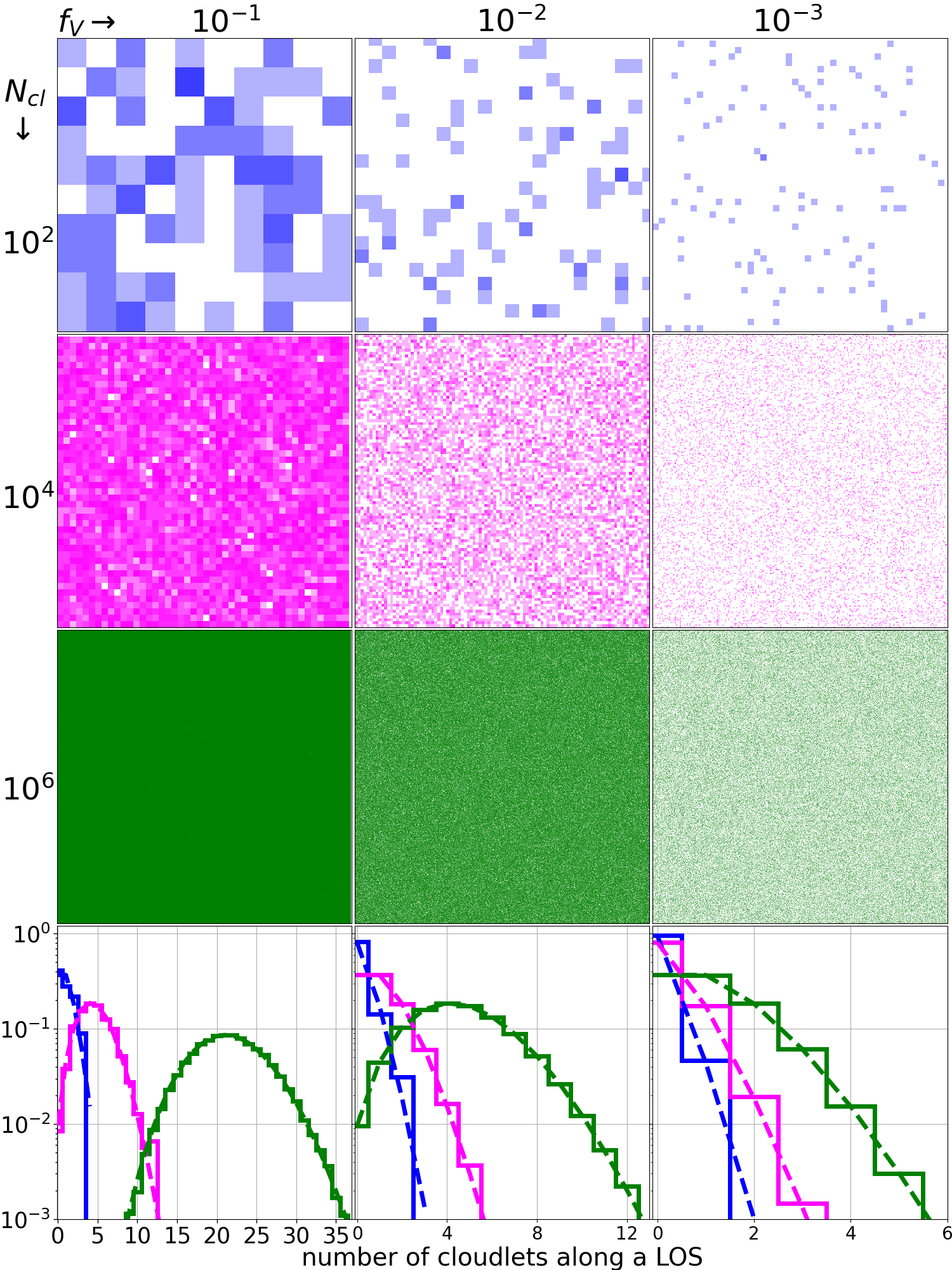}
    \caption{Projected cloudlet distribution (top 3$\times$3 panels) in a cubical volume with $N_{\rm cl}$ cubical cloudlets uniformly covering a volume fraction $f_V$. The size of vacancies equals the cloudlet size. As expected from simple estimates, the area covering fraction becomes of order unity when $N_{\rm cl} \gtrsim f_V^{-2}$. The panels farther from the diagonal (toward the bottom-left) are deeper into the mist limit. The bottom panels show the normalized histogram of the number of cloudlets along different sightlines. The histogram is very well fit by a Poisson distribution (shown by dashed lines) with a mean $f_V^{2/3} N_{\rm cl}^{1/3}$. Note that the number of sightlines with zero and one cloudlet becomes comparable for $N_{\rm cl} \sim f_V^{-2}$, corresponding to the approach toward the mist limit with a unity area covering fraction.}
    \label{fig:cubical}
\end{figure}

To test the above scaling borrowed from \citetalias{Dutta2024}, we carry out a simple Monte Carlo experiment (see Figure \ref{fig:cubical}). We divide a uniform cubical volume $L^3$ into equal cubical cells of size $l = L f_V^{1/3} N_{\rm cl}^{-1/3}$ of our cubical cloudlets. $N_{\rm cl}$ cloudlets, which occupy a volume fraction $f_V$ of the available volume, fill these available cells with uniform probability. In this arrangement, either the cell is fully occupied or unoccupied (there is no partial occupation). The number distribution of cloudlets along a LOS is Poisson, with a mean $f_V^{2/3} N_{\rm cl}^{1/3}$ and a standard deviation equaling the square root of the mean (as expected for a Poisson distribution). A Poisson distribution is expected for a uniform cloudlet distribution. These mist estimates provide a useful benchmark to compare with more sophisticated models.

\subsection{Terminal velocity of cloudlets within a CC}
\label{sec:vterm}

We can estimate the terminal velocity of a cloudlet within a CC by balancing the gravitational force and the drag force experienced by a cloudlet. The standard expression in the Rayleigh/turbulent drag regime is given by 
\begin{eqnarray}
    \nonumber
    & v_{\rm term}^{\rm cl} = \sqrt{\frac{4 \chi r_{\rm cl}}{3 C_D R_{\rm CGM}}} v_{c} \\
    &\approx 48~{\rm km~s}^{-1}\left( \frac{\chi}{100} \right)^{1/2} \left(\frac{r_{\rm cl}/R_{\rm CGM}}{4\times 10^{-4}} \right)^{1/2} \left(\frac{v_c}{220~\rm km s^{-1}} \right),
    \label{eq:cloudlet_vt}
\end{eqnarray}
where $\chi \approx 100$ is the density contrast between the cool and hot CGM phases, $C_D \sim 1$ is the drag coefficient, and $v_c \approx 220$ km s$^{-1}$ is the circular velocity of a Milky Way-like galaxy. For $r_{\rm cl} = 100$ pc (best resolution achievable by current cosmological galaxy formation simulations) and $R_{\rm CGM} = 280$ kpc, the above expression gives a terminal velocity $\sim 48$ km s$^{-1}$, higher than the turbulent velocity across a CC ($\sim 35$ km s$^{-1}$; see section \ref{sec:EW}). For cloudlets to be suspended in the CGM, and for mist description to be appropriate (as motivated by observations), the cloudlet terminal velocity should be $\lesssim 35$ km s$^{-1}$ (turbulent velocity within a CC), or equivalently, the cloudlet size should be $\lesssim$ a few pc! This simple argument emphasizes the importance of phenomenological models like ours, given the computational intractability of modeling realistic suspended CGM cloudlets from first principles.

\section{MgII ion fraction estimates}
\label{app:con_factor}

\begin{figure}
    \centering
    \includegraphics[width=\columnwidth]{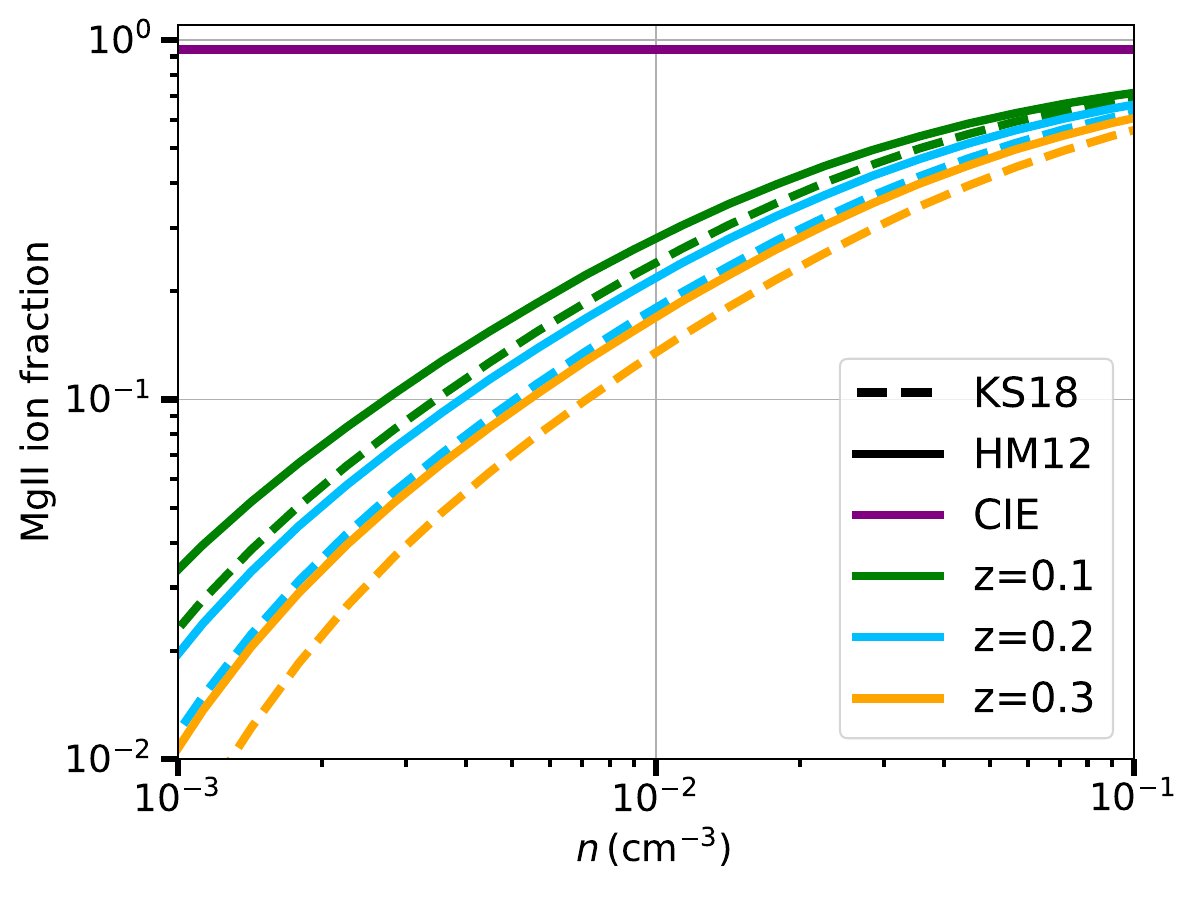}
    \caption{MgII ion fraction as a function of total number density at $10^4$ K. The purple line shows the MgII fraction for the Collisional Ionization Equilibrium, while the green, blue, and orange colors show the values for Photoionization Equilibrium at the redshifts of $0.1$, $0.2$, and $0.3$, respectively. The solid and dashed colored line shows the ion fraction with the HM12 (\citealt{Haardt2012}) and KS18 (\citealt{Khaire2019}) background radiation field. We computed the ion fraction values using the radiative transfer code {\tt CLOUDY} (\citealt{Ferland2017}). MgII ion fraction values vary by $\sim 2$ orders of magnitude from $\approx 0.01 \hbox{--} 0.6$ for $n \approx 10^{-3} \hbox{-} 10^{-1} \, \rm cm^{-3}$ at $z \sim 0.2$.}
    \label{fig:ion_frac}
\end{figure}

The number density of MgII is given by
\begin{equation}
    n_{\rm MgII} = \mu X \left (\frac{Z}{Z_\odot} \right) \left( \frac{n_{\rm Mg}}{n_{\rm H}} \right)_\odot \left( \frac{n_{\rm MgII}}{n_{\rm Mg}} \right) n_{\rm cool},
\end{equation}
where $X=0.74$ (\citealt{Asplund2009}) is the hydrogen mass fraction, $\mu=0.6$ is the mean molecular weight, $Z=0.3 Z_\odot$ (\citealt{Prochaska2017}) is the metallicity of the gas, $(n_{\rm Mg}/n_{\rm H})_\odot=3.47\times10^{-5}$ (\citealt{Ferland2017,Holweger2001}) is the solar abundance of Mg, and ($n_{\rm MgII}/n_{\rm Mg}$) is the fraction of MgII. Figure \ref{fig:ion_frac} shows the MgII ion fraction as a function of the total number density. The purple line shows the ion fraction in Collisional Ionization Equilibrium (CIE). The green, blue, and orange lines show the ion fraction at three different redshifts in Photoionization Equilibrium (PIE).

\section{Standard deviation estimates}
\label{app:std}

Just as we can calculate the average number of CCs along a LOS and the average path length along a CC, we can also calculate the standard deviation of these quantities.

For average CC density, which is only a function of the distance $r$ and independent of $\theta$ and $\phi$, the distribution of CCs in a radial shell is expected to be Poisson, with the variance equal to the mean number of clouds. For the number of CCs encountered along a LOS, this means that the variance in the number of CCs along a LOS equals the average number of CCs. Thus, the standard deviation in the number of CCs along a LOS is simply the square root of the average number of clouds encountered.

The deviation in the length along a single CC is given by
$\sigma_L = \sqrt{\langle L^2 \rangle - \langle L \rangle^2 }$, with 
\begin{equation}
    \langle L^2 \rangle = \frac{\int_0^{R_{\rm CC}} 2\pi b db \, 4(R_{\rm CC}^2 - b^2)}{\int_0^{R_{\rm CC}} 2\pi b db} = 2 R_{\rm CC}^2.
\end{equation}
Combining this, and Eq. \ref{eq:mean_l} gives, $\sigma_L = \sqrt{2} R_{\rm CC}/3$.

Both the number of clouds encountered along a LOS and the path length through a CC are random variables. Thus, the variation in column density along a LOS will have contributions due to both of these random variables. Adding all the deviations in quadrature (and assuming $N_{\rm CC, LOS}$ and $L$ to be independent random variables), we can write

\begin{equation}
\sigma_N^2 = \sigma_{N_{\rm CC, LOS}}^2 \langle L \rangle^2 + \langle N_{\rm CC, LOS} \rangle^2 \sigma_L^2 + \sigma_{N_{\rm CC, LOS}}^2 \sigma_L^2.
\label{eq:err_n}
\end{equation}

Here, the column density $N = \sum (n_{\rm ion})_j L_j = N_{\rm CC, LOS} L n_{\rm ion} $ is a sum over CCs along the LOS. Recall that $\langle N \rangle = \langle N_{\rm CC, LOS} \rangle \langle L \rangle \langle n_{\rm ion} \rangle $, $\sigma_{N_{\rm CC, LOS}}/\langle N_{\rm CC, LOS} \rangle = 1/\sqrt{\langle N_{\rm CC, LOS} \rangle}$ and $\sigma_L/\langle L \rangle = \sqrt{2}/4$.
These estimates are used to show (using dotted lines) the expected deviation from average column density in the \textit{bottom panel} of Fig. \ref{fig:variation_an}.

\bsp	
\label{lastpage}
\end{document}